\documentclass{article}

 \usepackage[preprint]{neurips_2026}

\usepackage[utf8]{inputenc} 
\usepackage[T1]{fontenc}    
\usepackage{hyperref}       
\usepackage{url}            
\usepackage{booktabs}       
\usepackage{amsfonts}       
\usepackage{nicefrac}       
\usepackage{microtype}      
\usepackage{xcolor}         

\usepackage{times}
\usepackage{latexsym}

\usepackage[T1]{fontenc}
\usepackage{url}

\usepackage{breakurl}
\usepackage{fancybox}
\usepackage[most]{tcolorbox}
\usepackage{nicefrac}       
\usepackage{microtype}      
\usepackage{colortbl}
\usepackage[dvipsnames]{xcolor}
\usepackage{enumitem}
\usepackage{makecell}
\usepackage{float}
\usepackage{placeins}
\usepackage{lineno}
\usepackage{epsfig}
\usepackage{comment}
\usepackage{multirow}
\usepackage{adjustbox}
\usepackage{textcomp}
\usepackage{xspace}
\usepackage{setspace}
\usepackage{kantlipsum, wrapfig,graphicx}
\usepackage[rightcaption]{sidecap}
\usepackage{soul}
\usepackage{kotex}
\usepackage{CJKutf8}
\usepackage{pifont}
\usepackage{acronym}
\usepackage{amsmath}
\usepackage{booktabs}
\usepackage{amssymb}
\usepackage[capitalize,noabbrev]{cleveref}
\usepackage{wrapfig}
\usepackage{amsmath}
\usepackage{amssymb}
\usepackage{algorithm}
\usepackage{algorithmic}
\usepackage{enumitem}
\usepackage{xcolor}
\usepackage{sidecap}

\usepackage{amsmath}
\usepackage{enumitem}

\newtcolorbox{codebox}{
  breakable,
  colback=gray!5,
  colframe=gray!50,
  boxrule=0.5pt,
  arc=2pt
}

\newtcolorbox{promptbox}{
  breakable,
  colback=blue!5,
  colframe=blue!40,
  boxrule=0.5pt,
  arc=2pt
}

\newcommand{\pos}[1]{\textcolor{green!60!black}{(#1)}}

\usepackage{graphicx}
\usepackage{subcaption}
\usepackage{caption}

\newcommand{\greencheck}{{\color[HTML]{009901}\ding{51}}}
\newcommand{\redcross}{{\color[HTML]{CB0000}\ding{55}}}
\newcommand{\codeline}[1]{\textcolor{blue}{\texttt{#1}}}

\title{WebUIProof: Benchmarking WebUI Code Generators with UI-Agent Execution Harness}

\author{~~Yun-Yun Tsai$^1$\thanks{Work done in internship at Meta},~~Yuning Mao$^2$,~~Shiqi Wang$^2$,~~Junfeng Yang$^1$,~~Sinong Wang$^2$\\ \\
$^1$Columbia University, $^2$Meta SuperIntelligence Labs\\ \\
{\tt\small $^1$\{yunyuntsai,junfeng\}@cs.columbia.edu,} \\{\tt\small $^2$\{yuningm,tcwangshiqi,sinongwang\}@meta.com} \\
}

\begin{document}

\maketitle

\begin{abstract}
Evaluating WebUI code generation at scale is difficult: outputs may compile and look plausible yet fail under user interaction, and prior benchmarks largely rely on free-form prompts with static checks (build success, screenshots) that miss functional correctness. We introduce \textbf{WebUIProof}, an execution-oriented benchmark that provides structured specifications and dense, executable interaction tests for WebUI generation across two task families: general WebUIs (e.g., dashboards, game, interactive tools) and 3D interactive simulation (e.g., particle/galaxy systems, physics dynamics). WebUIProof includes a UI-agent harness that runs executable interaction tests in a headless browser using an iterative plan–act–observe loop: it locates DOM elements, performs actions, observes resulting UI/DOM changes, and checks the specified assertions. We evaluate across eight commercial LLMs and observe frequent failures on interaction-based requirements even when pages render successfully, especially on 3D simulation interfaces. Finally, we show the UI-agent harness can provide outcome-level training signals. Training compact models (e.g., Qwen2.5 14B and MIMO 7B) with RL rewards derived from executable interaction tests improves functional completion while reducing build failures.
Project link: \url{https://github.com/yunyuntsai/webuiproof-benchmark}
\end{abstract}

\section{Introduction}

With LLMs increasingly integrate into the software development cycles, reliable evaluation for code generation becomes critical, motivating the benchmarks for real-world GitHub issues 
~\cite{jimenez2023swebench}, programming competitions
~\cite{li2022alphacode}, multilingual coding tasks~\cite{zheng2023humanevalx}, and interview coding problems~\cite{hendrycks2021apps}.  
Frontend code generation (WebUI generation) differs fundamentally from backend code synthesis and remains comparatively underexplored. The correctness depends not only on whether code compiles, but on what the UI renders and how it behaves under interaction.

Existing WebUI benchmarks only partially address this challenge. WebDev-Arena~\cite{zheng2024webdevarena} provides a large set of web prompts, but instructions are largely free-form and evaluation relies on human preference rather than executable functional verification. WebGen-Bench~\cite{liu2024webgenbench} introduces test cases for web prompts, but many requirements stated in instructions are left unchecked, resulting in incomplete coverage of instruction-following. More broadly, many benchmarks~\cite{xiao2025designbench, zhang2025artifactsbench, reacteval2024} depend on static checks such as build success, lint/type checks, or screenshots, which are insufficient to validate interactive behavior.

\begin{SCfigure}[0.7][t]
    \centering
    \includegraphics[width=0.52\textwidth]{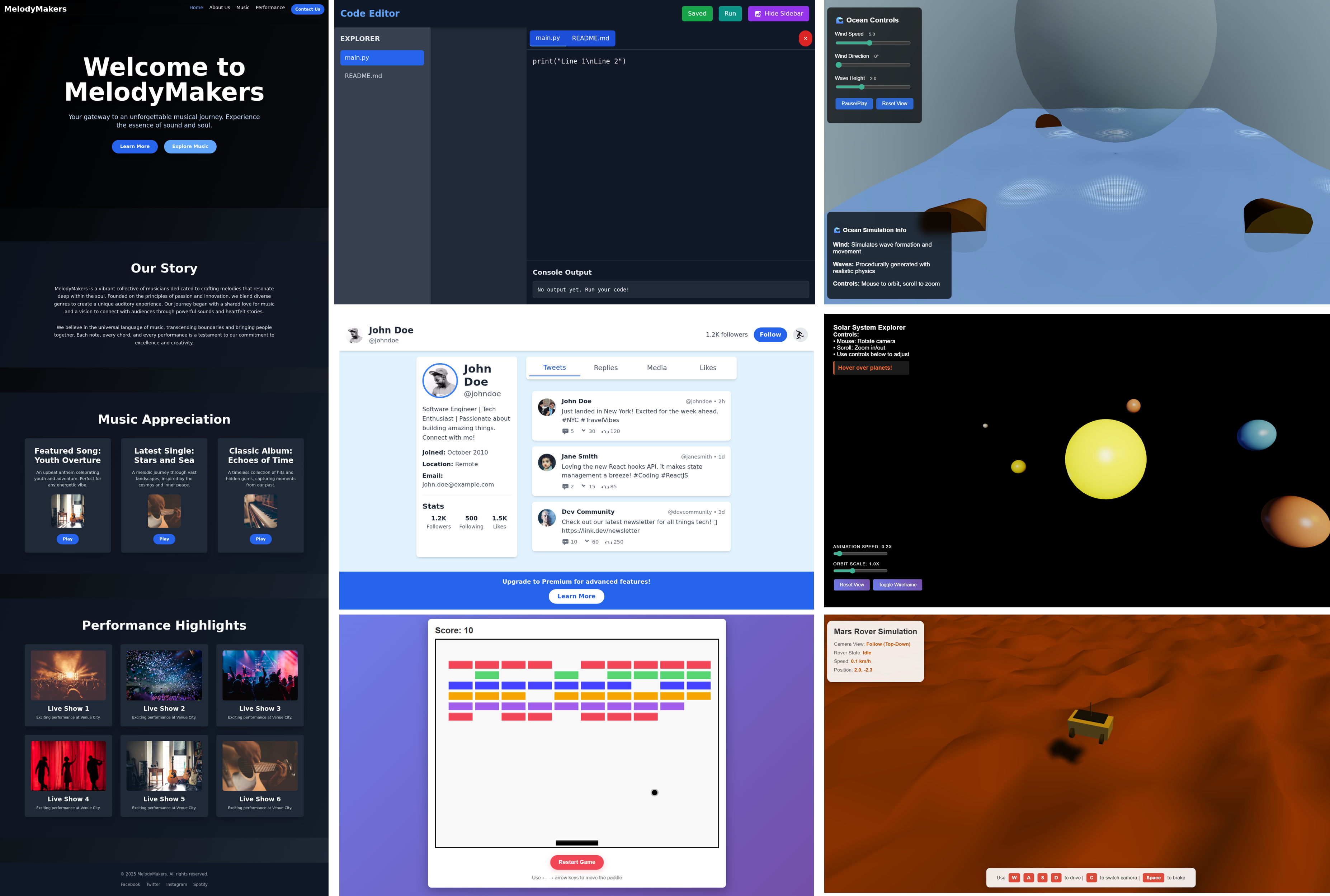}
    \caption{Examples of WebUI outputs in \textbf{WebUIProof}, spanning general applications (landing pages, code IDE, social/profile pages, games) and 3D simulation interfaces (physics dynamics, planetary scenes). These examples illustrate the benchmark’s emphasis on both visual fidelity and interaction-driven functionality. Each task is evaluated with executable interaction tests that check user actions (click/type/drag) and the resulting UI behavior.}
    \label{fig:webuiproof_teaser}
\end{SCfigure}

We introduce \textbf{WebUIProof}, an execution-oriented benchmark designed to measure functional, interactive correctness of WebUI code generation. WebUIProof replaces free-form prompts with structured specifications (website summary and UI element requirements) and pairs them with executable interaction tests that drive the UI agent to verify expected behavior.
WebUIProof contains 219 tasks (149 general + 70 3D) paired with ~3k executable interaction tests, enabling functional evaluation at much higher granularity than prompt-only benchmarks. The tasks span two families that cover standard UI workflows and richer interactive behaviors, including dashboards/games/tools and interactive 3D simulations such as physics dynamics and 3D scenes (see Fig.~\ref{fig:webuiproof_teaser}).
Our evaluation protocol is automated and reproducible, with a fixed UI-agent execution harness that runs these tests in a headless browser by iteratively planning the next interaction, executing browser actions (e.g., click, select, type, scroll), observing DOM/visual state changes, and verifying test assertions. 

While WebUIProof is primarily a benchmark contribution, the harness also provides outcome-level training signals by self-observing the rendered UI in training rollouts and deriving interactive tests from it. We therefore include \textbf{VisRL} (Visual-guided Reinforcement Learning), an execution-grounded RL approach that optimizes a compact model using composite rewards from build/runtime execution, visual quality, and UI-agent execution test success. 
In addition, we adopt several training strategies, such as multi-temperature sampling and fast-gate filtering, to improve rollout efficiency and stabilize learning under costly executable evaluation.

Our contributions are as follows:
\vspace{-3mm}
\begin{enumerate}[leftmargin=*]

\item We introduce \textbf{WebUIProof}, an execution-oriented benchmark for WebUI code generation that pairs \emph{structured specifications} with \emph{executable interaction tests} and a \emph{fixed UI-agent evaluation protocol} for functional correctness. It contains \textbf{219} tasks spanning (i) \textbf{general WebUIs} (e.g., dashboards, games, interactive tools) and (ii) \textbf{3D simulation WebUIs} (e.g., physics-driven interactive scenes).

\item We provide an agent-assisted construction pipeline for producing structured specs. and element-aligned executable tests at scale (Sec.~\ref{sec:method}).

\item With our executable UI-agent protocol (DOM grounding + interaction execution + assertions), we benchmark \emph{eight} commercial LLMs and show that many outputs \emph{build/render} yet fail interaction requirements, especially on 3D simulation interfaces. (Sec.~\ref{sec:experiment}).

\item  We include \textbf{VisRL} as an execution-grounded RL approach enabled by the execution harness with composite rewards. Training compact model such as \emph{Qwen2.5 14B Instruct and MIMO 7B} yield significant improvement of accuracy on general WebUI (\textbf {+15.7\%} gain) and 3D simulation tasks (\textbf{+4.9\%} gain) over SFT. We provide comprehensive ablations on evaluator robustness, benchmark quality analysis, and learning strategies (Sec.~\ref{sec:ablation}).

\end{enumerate}


\section{Related Works}

\paragraph{Benchmark for Frontend Code Generation.}
Prior agentic SWE systems (e.g., SWE-Agent~\cite{he2024sweagent}, SWE-bench~\cite{jimenez2023swebench}, SWE-Gym~\cite{yang2024swegym}, StackEval~\cite{zhang2024stackeval}) focus on repository-level editing and bug fixing, whereas we study end-to-end \emph{WebUI generation} evaluated through buildability, visual fidelity, and UI-agent-tested interactivity.
FrontendBench~\cite{zhu2025frontendbench} first standardizes the pipeline for frontend code generation as: (1.) LLMs receive instructions for generating code, (2.) render the code in a controlled sandbox environment, (3.) automate test scripts, and let the agent provide feedback loops.
WebDev-Arena~\cite{zheng2024webdevarena} is a large-scale benchmark that contains over 10k instruction prompts collected from online sources. Models are evaluated through pairwise comparisons with human votes determining preference. 
While valuable for capturing human preference, the instructions are loosely structured and lack automated test cases.
DesignBench~\cite{xiao2025designbench} evaluates framework-based web development (e.g., React/Vue/Angular) across generation, editing, and repair. 
The key difference in their setting is the \emph{evaluation signal}, which scores outputs based on screenshot-level visual similarity and code modification similarity via Jaccard similarity.
In particular, it does not employ a UI agent to locate DOM elements and verify behavioral assertions via scripted test cases.
WebGen-Bench~\cite{liu2024webgenbench} introduces web agents that take prompt instructions to generate a frontend project with full repository, and evaluate the website with executable web tasks in the sandbox build environment 
In contrast, our \textbf{WebUIProof} provides \emph{structured specifications} (website summary, functional descriptions, and required UI elements) paired with \emph{executable test cases}, and evaluates generations by a UI agent that interacts with the rendered page via DOM-level element localization and action execution (click/type/navigate) to verify test assertions. Table~\ref{tab:compare_benchmark} provides a detailed comparison of benchmarks for WebUI code generation.


\begin{figure}[t]
    \centering
    \includegraphics[width=0.85\textwidth]{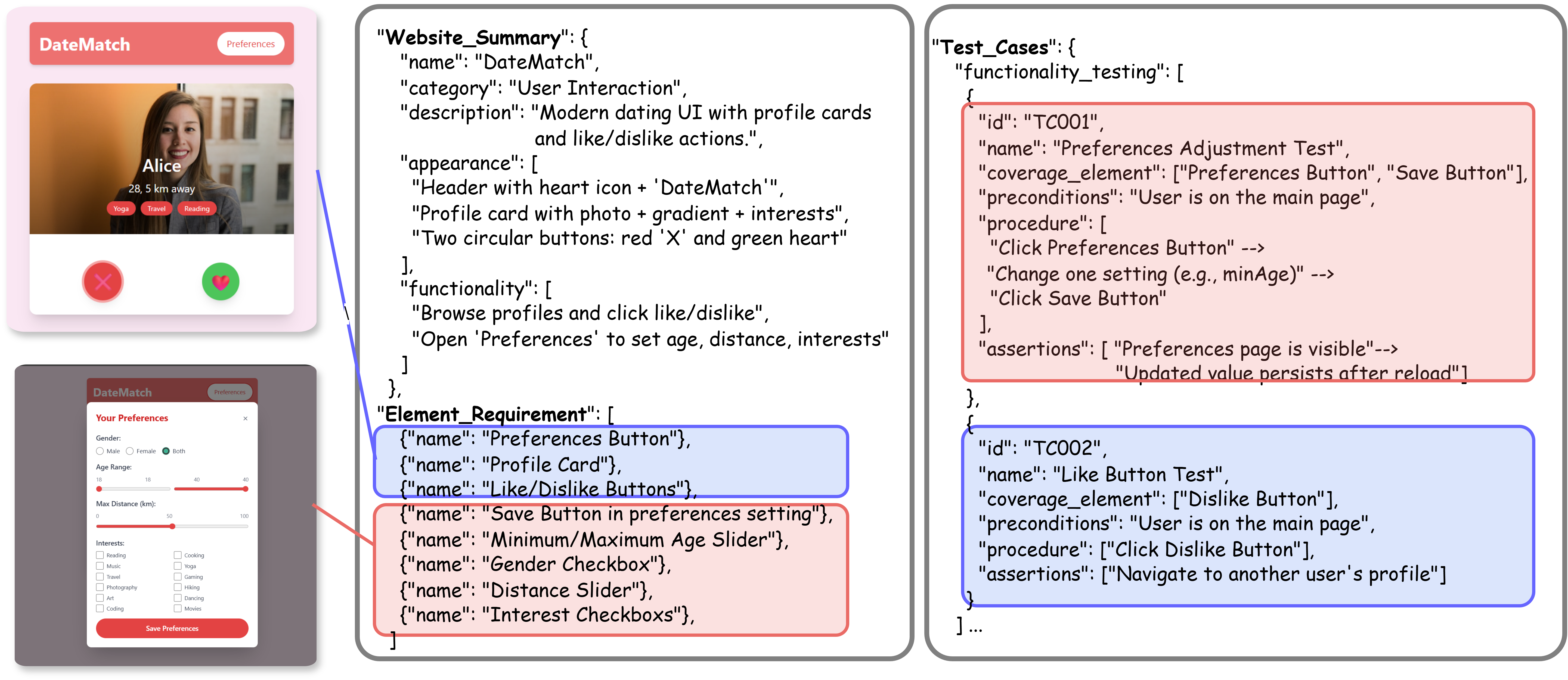}
    \caption{Benchmark instance format of \textbf{WebUIProof}. We show a structured spec. with paired test cases. Our benchmark construction pipeline first lets the UI agent collect a sequence of visual snapshots and HTML/DOM structure from the given seed webpage. Based on observations from webpages, the UI agent provides a website summary and test case recommendations. The \textcolor{red}{red} and \textcolor{blue}{blue} boxes illustrate element-to-test alignment: a subset of entries in $\textit{"element\_requirement"}$ is grouped and paired with executable test cases, making element-level coverage explicit.}
    \label{fig:webuiproof_sample}
    \vspace{-6mm}
\end{figure}

\paragraph{Web-based Testing with UI Agents.}
WebUI evaluation commonly relies on headless browsers such as Playwright\footnote{\url{https://playwright.dev}}
and Selenium WebDriver\footnote{\url{https://www.selenium.dev/}}, which supports reliable automated testing
via DOM inspection, event simulation, and screenshot capture. Building on this, recent UI agents, such as
WebVoyager~\cite{he2024webvoyager}, combine HTML and screenshots for end-to-end reasoning and action.
In addition, WebArena~\cite{zhou2023webarena} and BrowserGym~\cite{raman2023browsergym} provide standardized environments for
reproducible web-agent evaluation. Related benchmarks such as WebBench~\cite{webbench2025} and
VisualWebBench~\cite{liu2024visualwebbench} assesses UI-agent capability on webpage browsing, understanding,
and grounding.
In our work, we implement our evaluator based on WebVoyager for both general
WebUI generation and 3D simulation.

\begin{table}[t]

\caption{Benchmarks comparison for WebUI code generation.
We compare the benchmark based on instruction number, form, type, length, and output scope.
\textit{UI Topic} indicates general WebUIs (dashboards/tools/games) or additionally includes \textit{3D simulation interfaces} (classical mechanics/particles simulation). \textbf{UI Agent Eval.} denotes whether the evaluation involves a UI agent that executes test suites. Test coverage rate (\textit{TCR}) measures element-level coverage. Our \textbf{WebUIProof} provides diverse UI topics (general+3D) with rich test suites and high TCR.}
\label{tab:compare_benchmark}
\scriptsize
\centering
\setlength{\tabcolsep}{2pt}
\renewcommand{\arraystretch}{1}
\begin{tabular}{l|ccccccccc}
\toprule
   \textbf{Benchmark}           & \#Instr. &\begin{tabular}[c]{@{}c@{}}Instr. \\ Form\end{tabular} & \begin{tabular}[c]{@{}c@{}}Instr. \\ Type\end{tabular}  &  \begin{tabular}[c]{@{}c@{}}Instr. \\ Length\end{tabular}    &\begin{tabular}[c]{@{}c@{}}Output \\ Scope\end{tabular} &\begin{tabular}[c]{@{}c@{}}UI \\ Topic\end{tabular} & \begin{tabular}[c]{@{}c@{}}UI Agent \\ Eval.\end{tabular} & \#Tests  & TCR(\%)  \\ \midrule
DesignBench~\cite{xiao2025designbench} & 900    & NL prompts & Gen./Edit/Repair & $\sim$80-150     & Single file (HTML) & General                 & \redcross         & 0           & 0\%                               \\
WebDev-Arena~\cite{zheng2024webdevarena} & $>$10k     & NL prompts & Gen. &   $\sim$60    &  Repo (App)  & General               &   \redcross      & 0             & 0\%                                     \\
FrontendBench~\cite{zhu2025frontendbench} & 148    & NL prompts  & Gen. &   $\sim$50-100    & Single file (HTML)    & General               & \greencheck           & 148            & $\sim$90\%                               \\
WebGen-Bench~\cite{liu2024webgenbench} & 101   & NL prompts  & Gen.  & $\sim$40-70    & Repo (App)      & General              & \greencheck             & $\sim$600           & 49.24\%                                          \\ \midrule
\textbf{WebUIProof (Ours)}        & 219   &Structured Spec. & Gen. & $\sim$270-550      & Repo (App)    & General+3D           &  \greencheck            & $\sim$3k            & 97.23\%                                     \\ \bottomrule
\end{tabular}
\end{table}

\section{Method}
\label{sec:method}

This section describes \textbf{WebUIProof}, including (1) the benchmark overview, (2) the UI-agent evaluation protocol used for scoring, and (3) the agent-assisted benchmark construction pipeline. Separately, we also present \textbf{VisRL} as a training approach to illustrate how the executable harness can provide outcome-level learning signals.


\subsection{WebUIProof Benchmark Overview}
\label{sec:benchmark_overview}

\paragraph{Task Families and Scope.}
WebUIProof contains 219 tasks spanning two families: \textbf{(i) General WebUIs} (149) and \textbf{(ii) 3D simulation WebUIs} (70). General tasks cover five types of common web applications, such as dashboards and data analytics tools, social profile interfaces, and interactive games; 3D simulation tasks cover six types of interactive Three.js-style scenes, such as physics dynamics, fluid waves simulation, and astro particles visualizations. (Appendix~\ref{app:b_1-5_uicategory} Fig.~\ref{fig:webui-category-distribution} shows the category distribution). 

\paragraph{Data Sources.}
We collect seed prompts from WebGen-Bench~\cite{liu2024webgenbench} and WebDev-Arena~\cite{zheng2024webdevarena}, and we generate seed webpages with these prompts to drive agent-assisted structured specification and test generation.
The 3D simulation WebUIs are built from Three.js\footnote{\url{https://threejs.org/}} example pages and similar public demo scenes, focusing on browser-based WebGL simulations with interactive user controls (See Appendix~\ref{app:b_1_datasource}).

\paragraph{Benchmark Instance Format.}
Fig.~\ref{fig:webuiproof_sample} illustrates the benchmark instance format of WebUIProof. Each task is packaged as \textit{(website\_summary, element\_requirement, test\_cases)}. The \textit{website\_summary} describes appearance and functionality; \textit{element\_requirement} lists required UI elements with stable identifiers and locator hints; and \textit{test\_cases} are executable interaction tests with \textit{preconditions}, \textit{procedure}, and \textit{assertions}. We group tests into three semantic types: \textit{Functionality}, \textit{Data-display}, and \textit{Design-validation} tests. (Appendix~\ref{app:b_1-5_uicategory} Fig.~\ref{fig:webui-testcase-distribution} shows the test case distribution).

\subsection{UI-Agent Evaluation Protocol}
\label{sec:evaluator}

WebUIProof uses a fixed UI-agent execution harness to score functional correctness via executable interaction tests. For each test case, the agent follows an iterative \textit{Plan}$\rightarrow$\textit{Action}$\rightarrow$\textit{Observation} loop. Specifically, it observes DOM and visual state changes,  thinks and plans the interaction for the next step, executes browser actions (e.g., click/type/hover/drag/scroll, etc), and checks the test assertions. The final score aggregates test outcomes into \texttt{Full/Partial/No} completion based on passed assertions. This protocol is applied identically to all models and tasks during the evaluation.

\subsection{Benchmark Construction Pipeline}
\label{sec:webuiproof}

We construct \textbf{WebUIProof} with an agent-assisted pipeline that produces \emph{structured specifications} paired with \emph{executable test cases}. The pipeline is designed to (i) scale to diverse webpages, (ii) maximize test coverage over UI elements and behaviors, and (iii) minimize manual annotation. Each benchmark instance is packaged as:
\textit{(website\_summary, element\_requirement, test\_cases)}.

\noindent\textbf{Phase I: Seeding from static webpages.}
Given a target webpage, we first collect (1) a set of visual snapshots (multiple views if available) and (2) the HTML/DOM structure of the page. We prompt a UI agent to produce a \textbf{website summary} that includes appearance and functionality, and to propose a small set of \textbf{seed test cases} that cover common workflows (e.g., search/filter, navigation, form submission). This stage provides basic test cases for UI agent to start the interactive exploration.

\noindent\textbf{Phase II: Interactive exploration and UI elements retrieval.} The dynamic states, such as hidden panels, modal dialogs, pagination, and hover states, may exist but are underexplored. We therefore run the agent on the webpage with basic test cases for exploration. Specifically, the agent executes an iterative
\textit{Plan}$\rightarrow$\textit{Action}$\rightarrow$\textit{Observation} loop: (1.) \textit{Plan:} The agent receives the test cases and plans the execution (e.g., open settings, trigger sorting, expand menus). (2.) \textit{Action:} The agent makes decisions on the next action (e.g., click/type/hover/drag/scroll).
(3.) \textit{Observation:} The agent records state changes and extracts structured information: (i) newly discovered UI elements, (ii) updated DOM attributes, (iii) textual/UI feedback, (iv) intermediate task progress, and (v) test cases recommendation. We ask the LLMs to provide the test cases in three perspectives, including \textit{function accessibility}, \textit{data display correctness}, and \textit{UI design or style consistency}.

From these loop trajectories, we incrementally update our structured spec.:
\textit{(i) element\_requirement}, a canonical list of required UI elements with stable identifiers and locator hints (text, role, class/id, hierarchical path); and \textit{(ii) test\_cases}, expanded to reflect newly discovered states.

\noindent \textbf{Phase III: Test case synthesis and post-processing.}
After each agent loop, we parse the interaction trajectory and extract candidate test cases from step-wise observations.
Each case is formatted with: \textit{test\_name, coverage\_element, preconditions, procedure, and assertions}.
We then perform element-level coverage checking to identify UI elements that never appear in interactions or assertions.
For any uncovered elements, we prompt the UI agent to generate additional test cases.
Finally, we de-duplicate near-identical cases by merging variants that differ only in constants (e.g., slider values or dropdown options).
The resulting instance is stored as a JSON file containing the structured specification and executable test suite. In Appendix~\ref{app:b_2_datacollection} and \ref{app:b_3_datasample}, we provide more detail about collection pipeline and samples for structured specification.

\begin{SCfigure}[0.50][t]
    \centering
    \includegraphics[width=0.58\textwidth]{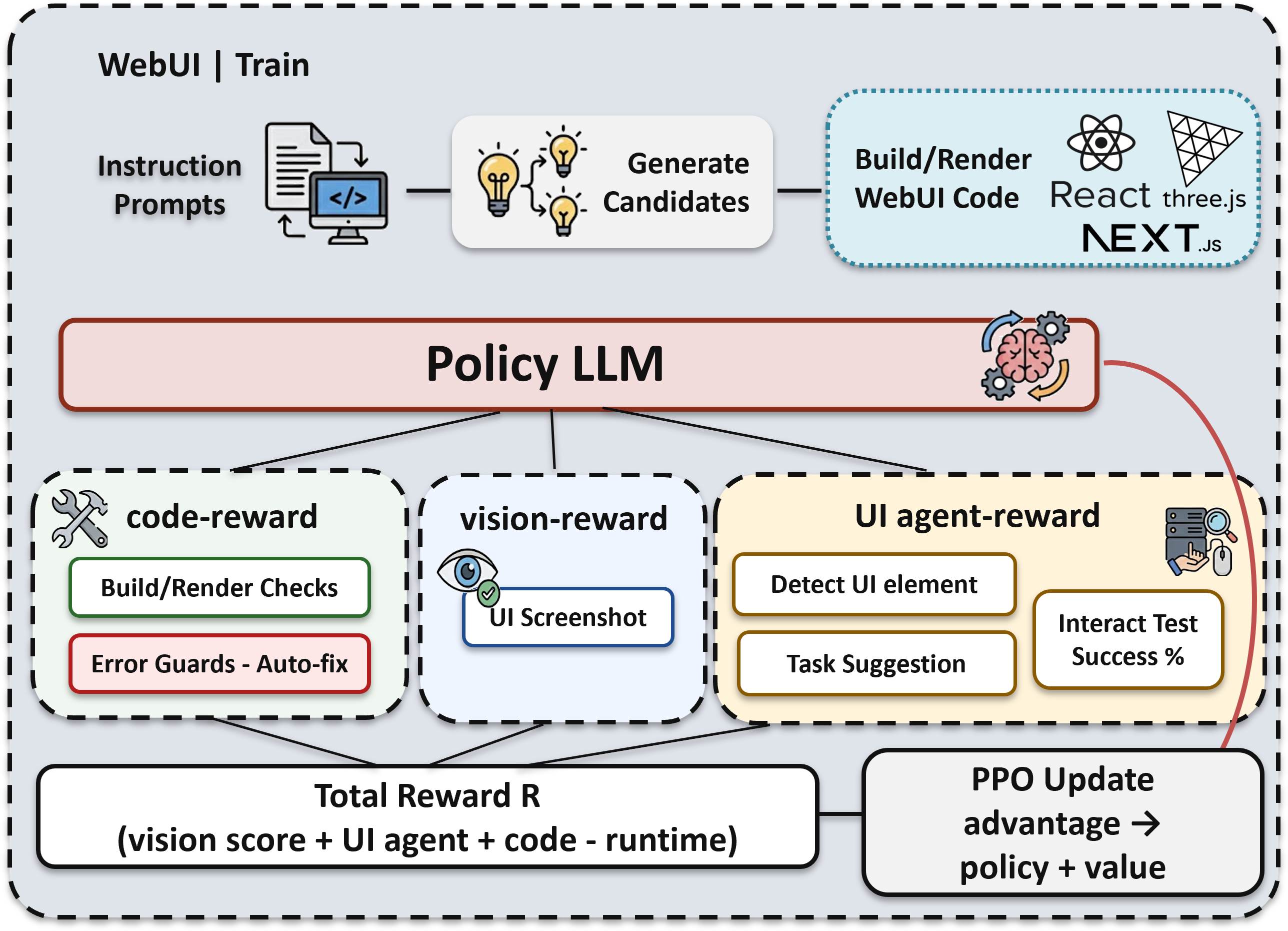}
    \caption{\textbf{VisRL Overview}: Given instruction prompts, the Policy LLM generates UI code candidates, which are built and rendered to produce screenshots; rewards combination (i) code validity with error fixing loop, (ii) vision-based screenshot scoring, and (iii) UI-agent interaction tests (element detection, task proposal, and execution success), and are used to update the policy via PPO.}
    \label{fig:webui-flow}
\end{SCfigure}

\vspace{-2mm}

\subsection{VisRL: RL Training with Executable Harness}
\label{sec:visrl}

While WebUIProof is primarily an evaluation benchmark, the executable UI-agent harness can provide outcome-level training signals. We include \textbf{VisRL} (Visual-guided Reinforcement Learning) as a new training approach that optimizes a compact model using rewards computed from multiple outcomes: (i) build/runtime execution, (ii) rendered visual quality, and (iii) UI-agent interaction-test success. Concretely, given generated code $c$ and a task specification $p$, we build and render the project and compute a composite reward
$r(c,p)=\alpha_v r_v(I(c))+\alpha_u r_u(s(c),\mathcal{T}(p))+\alpha_m r_m(c)$,
where $r_v$ scores rendered screenshots, $r_u$ is the interaction-test pass rate, and $r_m$ is a lightweight analyzer penalty/bonus for common failure patterns. Full PPO details are provided in Appendix~\ref{app:c_1_seq_ppo}~\ref{app:c_2_reward_definition},~\ref{app:c_3_additional_learning}.

\noindent\textbf{Design Insight.}
We observe that \emph{single-signal rewards are insufficient} in WebUI generation. Optimizing interaction tests only tends to produce minimal but functional UIs, while optimizing visual scores only improves visual appearance quality without guaranteeing correct behavior. Thus, combining build/runtime, visual, and interaction signals yields the most balanced improvements across functionality and UI appearance, We provide comprehensive studies on reward decomposition and weight sensitivity in Appendix~\ref{app:d_2_reward_decompose_analysis}, \ref{app:d_3_alpha_sensitivity_analysis}. 

\noindent\textbf{Training Stability.}
Because rewards are only observed after build/render/test, we adopt two lightweight strategies to stabilize and reduce rollout cost: \textbf{multi-temperature sampling} to increase candidate diversity under a fixed best-of-$N$ budget, and \textbf{fast-gate filtering} to skip obviously invalid generations before expensive execution. Ablations in Sec.~\ref{sec:learn_strategies_visrl} show both components matter for accuracy and build failure reduction.

\section{Experiment}
\label{sec:experiment}

\begin{table*}[t]
\scriptsize
\centering
\setlength{\tabcolsep}{2pt} 
\renewcommand{\arraystretch}{1} 
\caption{\textbf{General WebUI generation on WebUIProof.}
We report the \textbf{build failure rate} (\% of generations that fail to build/render), and the \textbf{UI-agent verified task outcome} among \emph{build-successful} generations: \textbf{No/Partial/Full} completion rate (\%). Overall \textbf{Accuracy} is computed as
$\mathrm{Acc}=\frac{\#\mathrm{Full}+0.5\cdot \#\mathrm{Partial}}{N}\times 100$,
where build failures are counted as \textbf{No} completion. 
Rows are grouped into eight proprietary (API) models and two self-trained models, including Qwen2.5 14B Instruct and MIMO 7B under three training variants: SFT, PPO w/ vision agent, and VisRL (ours).}

\begin{tabular}{llccccc}
\hline
\begin{tabular}[c]{@{}l@{}}Model \\ Source\end{tabular} & \begin{tabular}[c]{@{}l@{}}Model Name / \\ Training Variant\end{tabular}  & \begin{tabular}[c]{@{}c@{}}Build Failure \\ Rate (\%)\end{tabular} & \multicolumn{3}{c}{\begin{tabular}[c]{@{}c@{}}Build Success -- \\ UI-Agent Verified Task Outcome (\%)\end{tabular}}                                                                  & \begin{tabular}[c]{@{}c@{}} Accuracy (\%)\\(Full + 0.5$\cdot$Partial)\end{tabular} \\ \cline{4-6}
                                                      &                                                       &                                                               & \begin{tabular}[c]{@{}c@{}}No \\ Completion\end{tabular} & \begin{tabular}[c]{@{}c@{}}Partial \\ Completion\end{tabular} & \begin{tabular}[c]{@{}c@{}}Full\\ Completion\end{tabular} &                                                          \\ \hline
\multirow{8}{*}{Proprietary (API)}                            & Claude Sonnet 4~\cite{Anthropic2025Sonnet4}                                       & 30.37                                                         & 12.59                                                    & 22.10                                                      & 35.05                                                     & 46.10                                                    \\
                                                      & Gemini 2.5 Pro~\cite{GoogleDeepMind2025Gemini2_5Pro}                                        & 32.21                                                         & 15.02                                                    & 20.38                                                      & 32.67                                                     & 42.86                                                    \\
                                                      & Qwen 3 Coder 480B~\cite{Alibaba2025Qwen3Coder}                                    & 33.05                                                         & 13.92                                                    & 20.76                                                      & 31.62                                                     & 41.99                                                    \\
                                                      & DeepSeek R1~\cite{DeepSeek2025R1}                                          & 32.89                                                         & 14.91                                                    & 20.51                                                      & 31.65                                                     & 41.90                                                    \\
                                                      & GPT 4.1~\cite{OpenAI2025GPT4_1}                                               & 36.58                                                         & 11.67                                                    & 20.30                                                      & 31.25                                                     & 41.40                                                    \\
                                                      & Llama 4 Maverick~\cite{MetaAI2025Llama4Maverick}                                      & 46.64                                                         & 2.53                                                     & 18.72                                                      & 31.41                                                     & 40.77                                                    \\
                                                      & Kimi K2~\cite{MoonshotAI2025KimiK2}                                               & 46.31                                                         & 8.70                                                     & 11.94                                                      & 33.33                                                     & 39.30                                                    \\
                                                      & Llama 3 405B~\cite{MetaAI2024Llama3_405B}                                         & 47.32                                                         & 10.13                                                    & 14.18                                                      & 29.10                                                     & 36.19                                                    \\ \hline
\multirow{6}{*}{Self-Trained}                         & Qwen 2.5 14B~\cite{qwen2.5} SFT                              & 56.25                                                         & 10.97                                                    & 11.84                                                      & 20.66                                                     & 26.08                                                    \\
                                                      & + PPO w/ Vision agent                               & 41.42  \pos{-14.83}                                                           & 13.96                                                    & 11.65                                                      & 32.97                                                     & 38.79 \pos{+12.71}                                                     \\
                                                      & + \textbf{VisRL (Ours)}                            & \textbf{36.10  \pos{-20.15}}                                                          & 13.71                                                    & 15.13                                                      & 34.23                                                     & \textbf{41.80  \pos{+15.72}}                                                    \\ \cmidrule{2-7}
                                                      & MIMO 7B~\cite{coreteam2025mimounlockingreasoningpotential} SFT                              & 58.71                                                         & 12.15                                                    & 7.33                                                    & 20.2                                                   & 23.8                                                    \\
                                                      & + PPO w/ Vision agent                               & 51.44 \pos{-7.27}                                                           &   15.33                                                 &   11.52                                                 & 21.3                                                   &  27.06 \pos{+3.26}                                                     \\
                                                      & + \textbf{VisRL (Ours)}                            & \textbf{40.22 \pos{-18.49}}                                                          & 17.43                                                  &   15.71                                                  &  25.63                                               & \textbf{33.43 \pos{+9.63}}                                                    \\
                                                      \hline

\end{tabular}
\label{tab:general_main}
\end{table*}

\vspace{-3mm}

\subsection{Experimental Setting}
\paragraph{Model.} We evaluate eight state-of-the-art LLMs spanning different architectures, scales, and training paradigms.
Claude Sonnet 4~\cite{Anthropic2025Sonnet4} and Gemini 2.5 Pro~\cite{GoogleDeepMind2025Gemini2_5Pro} represent strong closed-source reasoning-oriented models with multimodal capabilities.
Qwen 3 Coder 480B Instruct\footnote{\url{https://huggingface.co/Qwen/Qwen3-Coder-480B-A35B-Instruct}}~\cite{Alibaba2025Qwen3Coder} and Llama 4 Maverick~\cite{MetaAI2025Llama4Maverick} / Llama 3 405B~\cite{MetaAI2024Llama3_405B} are open-source instruction-tuned models optimized for code synthesis and general reasoning.
DeepSeek R1~\cite{DeepSeek2025R1} is an open-weight model emphasizing step-by-step “reasoning-first” training.
Moonshot Kimi K2~\cite{MoonshotAI2025KimiK2} is a high-efficiency proprietary model designed for long-context understanding and structured output.
Finally, GPT 4.1~\cite{OpenAI2025GPT4_1} serves as a strong general-purpose baseline.
Furthermore, we use two compact models Qwen 2.5 14B Instruct\footnote{\url{https://huggingface.co/Qwen/Qwen2.5-14B-Instruct}}~\cite{qwen2.5, qwen2} and MIMO 7B~\cite{coreteam2025mimounlockingreasoningpotential} as the base model for training. We compare the results of VisRL with two training variants (SFT baseline and PPO with Vision agent).

\paragraph{UI-agent Evaluator Setup.}
We evaluate functional correctness using a fixed UI-agent execution harness implemented with Playwright webdriver and a vision-language model (\textit{Qwen2.5 VL 72B}) as the evaluator. The evaluator runs in a headless browser with a 1280$\times$800 viewport, a per-task step budget of 15, and supports the actions \{click, type, select, scroll, drag, hover, wait\}. Each task is scored by executing its test suite and aggregating outcomes into \texttt{Full/Partial/No} completion. We keep this evaluator \emph{frozen and identical} across all models and methods. Details of the execution environment, prompts, and failure handling are provided in Appendix~\ref{app:a_main_reproduce}. We additionally validate evaluator robustness via expanded human--agent agreement (N=100 across multiple model families) and step-budget sensitivity analysis (5/10/15 steps) with failure-mode attribution (see Sec.~\ref{sec:evaluator_robustness}).

\noindent\textbf{Training Details of VisRL.} We use Qwen 2.5 14B Instruct~\cite{qwen2.5} as our base model. We first launch the SFT training to establish a cold-start checkpoint. After that, we launch the PPO learning with LoRA adapters ($r=16$, $\alpha=32$ for general WebUI task), batch size 24, learning rate $2\times10^{-7}$, and vLLM for off-policy generation with periodic adapter reloads. We use 4 H100-80 GB NVIDIA GPUs for training. We compare VisRL with two training variant, including pure SFT and PPO with vision agent. The baseline, PPO with vision agent, leverages the agent to score the WebUI screenshot with vision rubric (see Appendix~\ref{app:c_6_visionrubric}). 
We provide implementation details, including training hyperparameters, hardware resources, system configuration, and training infrastructure ( see Appendix~\ref{app:c_0_visrl_implementation}). In Appendix~\ref{app:d_1_training_curve}, we show the training curve analysis.

\vspace{-1mm}

\noindent\textbf{Training Data.} For general WebUI task, we leverage the 6000 training set from webgen-bench~\cite{liu2024webgenbench} to do the SFT and PPO training. We de-duplicate the training data and also avoid using data that requires a login page first to facilitate our training process.
For the 3D simulation task, we construct our own \texttt{Three.js} training dataset by scraping the official
\texttt{threejs.org} examples. Using the \texttt{files.json} index, we crawl each demo and parse its HTML to extract
metadata (title, description, tags), which we convert into instruction prompts. The resulting dataset covers a wide
range of real-world \texttt{Three.js} patterns (e.g., animation, interactive controls, external assets, visual effects)
and contains 1600+ prompts.

\begin{table*}[t]
\scriptsize
\centering
\setlength{\tabcolsep}{2pt} 
\renewcommand{\arraystretch}{1} 
\caption{\textbf{3D simulation WebUI generation on WebUIProof.}
Same protocol as Table~\ref{tab:general_main}: we report build failure rate and UI-agent verified task outcomes (\%), including \textbf{No/Partial/Full} completion rate, and compute Accuracy,
counting build failures as \textbf{No} completion. Results compare eight proprietary (API) models and self-trained models, including \textbf{Qwen2.5 14B Instruct} and \textbf{MIMO 7B} under three training variants: SFT baseline, PPO w/ vision agent, and VisRL (ours).}

\begin{tabular}{llccccc}
\hline
\begin{tabular}[c]{@{}l@{}}Model \\ Source\end{tabular} & \begin{tabular}[c]{@{}l@{}}Model Name / \\ Training Variant\end{tabular}  & \begin{tabular}[c]{@{}c@{}}Build Failure \\ Rate (\%)\end{tabular} & \multicolumn{3}{c}{\begin{tabular}[c]{@{}c@{}}Build Success -- \\ UI-Agent Verified Task Outcome (\%)\end{tabular}}                                                                  & \begin{tabular}[c]{@{}c@{}} Accuracy (\%)\\(Full + 0.5$\cdot$Partial)\end{tabular} \\ \cline{4-6}
                                                      &                                                       &                                                               & \begin{tabular}[c]{@{}c@{}}No \\ Completion\end{tabular} & \begin{tabular}[c]{@{}c@{}}Partial \\ Completion\end{tabular} & \begin{tabular}[c]{@{}c@{}}Full\\ Completion\end{tabular} &                                                          \\ \hline
\multirow{8}{*}{Proprietary (API)}                            & Claude Sonnet 4~\cite{Anthropic2025Sonnet4}                                       & 0.00                                                        & 49.66                                                    & 20.21                                                      & 30.14                                                     & 40.24                                                    \\
                                           & Qwen 3 Coder 480B~\cite{Alibaba2025Qwen3Coder}                                                   & 1.43                                                         & 51.37                                                 & 17.12                                                      & 29.11                                                   & 37.67                                                    \\
                                                     & Gemini 2.5 Pro~\cite{GoogleDeepMind2025Gemini2_5Pro}                                & 0.00                                                         & 54.79                                                    & 18.49                                                     & 26.71                                                   & 35.96                                                    \\
                                                       & Kimi K2~\cite{MoonshotAI2025KimiK2}                                               & 0.00                                                         & 54.79                                                     & 20.21                                                    & 25.00                                                     & 35.10                                                    \\
                                                          & Llama 4 Maverick~\cite{MetaAI2025Llama4Maverick}                                      & 0.00                                                        & 59.93                                                    & 15.75                                                      & 24.43                                                    & 32.19                                                    \\
                                                      & DeepSeek R1~\cite{DeepSeek2025R1}                                          & 11.43                                                         & 49.66                                                    & 17.12                                                      & 21.23                                                     & 29.79                                                    \\
                                                      & GPT 4.1~\cite{OpenAI2025GPT4_1}                                               & 2.85                                                         & 65.41                                                  & 15.75                                                     & 15.07                                                   & 22.95                                                  \\

                                                      & Llama 3 405B~\cite{MetaAI2024Llama3_405B}                                         & 2.86                                                         & 69.86                                                    & 13.70                                                    & 12.67                                                    & 19.52                                                    \\ \hline
\multirow{6}{*}{Self-Trained}                        &  Qwen 2.5 14B~\cite{qwen2.5} SFT                              & 0.00                                                       & 55.68                                                 & 22.94                                                     & 20.97                                                    & 32.44                                                   \\
                                                      & + PPO w/ Vision agent                               & 0.00                                                          & 51.00                                                   & 26.37                                                    & 23.28                                                 & 36.47 \pos{+4.03}                                                     \\
                                                      & \textbf{+ VisRL (Ours)}                            &   0.00                                                          & 52.24                                                   & 19.31                                                     & 27.73                                                     & \textbf{37.38 \pos{+4.94}}                                                    \\ \cmidrule{2-7}
         & MIMO 7B~\cite{coreteam2025mimounlockingreasoningpotential} SFT                              &    4.77                                                    &     71.24                         &      13.31                                           &     10.54                                              &        17.20                                          \\
                                                      & + PPO w/ Vision agent                               &  2.86 \pos{-1.91}                                                         &    70.30                                               &  14.86                                                 &   11.32                                         & 18.75 \pos{+1.55}                                                     \\
                                                      & \textbf{+ VisRL (Ours)}                            &        \textbf{0.00 \pos{-4.77}}                      &   69.20                                           &    18.86                                            &     10.67                    & \textbf{ 20.10 \pos{+2.9}}                                                    \\ \hline
\end{tabular}
\label{tab:sim3d_main}
\vspace{-3mm}
\end{table*}

\subsection{Experimental Results}
\label{sec:experimentresults}


Tables~\ref{tab:general_main} and~\ref{tab:sim3d_main} report WebUI generation performance on WebUIProof under the same \emph{fixed, executable} UI-agent evaluation protocol across eight commercial models. We report (i) build failure rate (cannot build/render), and (ii) UI-agent verified task outcomes among build-successful generations (\textsc{No}/\textsc{Partial}/\textsc{Full}), summarized as $\mathrm{Acc}=(\#\textsc{Full}+0.5\cdot \#\textsc{Partial})/N$ with build failures counted as \textsc{No}.
On General WebUI (Table~\ref{tab:general_main}), Claude Sonnet 4 performs best (\textbf{46.10\%} acc.) but still fails to build in \textbf{30.37\%} of cases. The second-best models, Gemini 2.5 Pro (42.86\% acc. and 32.21\% build failure rate) and Qwen 3 Coder 480B (41.99\% acc. and 33.05\% build failure rate), form a tight cluster around 42\% accuracy with similarly high build failures.
On 3D simulation WebUIs (Table~\ref{tab:sim3d_main}), Claude Sonnet 4 achieves \textbf{40.24\%} accuracy with 0\% build failure, with Qwen 3 Coder 480B close behind at 37.67\% (1.43\% build failure), indicating that even frontier models leave substantial headroom on both of these interactive interfaces under executable UI-agent tests.

We also report the results for compact open-sourced models trained under VisRL. On General WebUI, \textbf{VisRL} improves Qwen 2.5 14B Instruct from 26.08\% to 41.80\% accuracy while reducing build failures drop by 20.15\%, and improves MIMO 7B from 23.80\% to 33.43\% while reducing build failures by 18.49\% (Table~\ref{tab:general_main}). On 3D simulation WebUIs, VisRL improves Qwen 2.5 14B from 32.44\% ro 37.38\% (build failures remain 0\%), and improves MIMO 7B from 17.12\% to 20.10\% while reducing build failures by 4.77 points (Table~\ref{tab:sim3d_main}). These results illustrate that the UI-agent harness provides a learnable signal for training compact baselines.
Appendix~\ref{app:e_1_additional_results} provides per-category accuracy breakdowns (Tables~\ref{tab:cat_acc_generalui_apimodel},~\ref{tab:cat_acc_generalui_selftrained},~\ref{tab:cat_acc_sim3d_apimodel},~\ref{tab:cat_acc_sim3d_selftrained}).




\section{Ablation Studies}
\label{sec:ablation}

\subsection{Evaluator Reliability and Failure Modes}
\label{sec:evaluator_robustness}

Since WebUIProof relies on a UI-agent execution harness to run interaction tests, we quantify evaluator reliability by reporting (i) human-agent agreement on the \textsc{Full}/\textsc{Partial}/\textsc{No} labels across multiple model families, and (ii) sensitivity to the evaluator step budget with failure attribution.

\textbf{$\bullet$ Human--agent agreement across model families.}
We annotate $N=100$ General WebUI samples and compare the UI agent's labels against human judgments for three model families (Claude Sonnet 4, Gemini 2.5 Pro, and self-trained Qwen 2.5 14B with VisRL). Table~\ref{tab:ha_agreement_compact} reports agreement rates and confusion matrices. Overall agreement is consistently above $\sim$80\%, suggesting that the evaluator provides a reasonably stable signal across different generation styles.

\textbf{$\bullet$ Step budget sensitivity and failure attribution.}
Evaluator errors can arise from limited interaction budget or brittle element grounding. To characterize this, we vary the per-task step budget (5/10/15) on the same $N=100$ samples and attribute disagreements to concrete failure modes: timeout/step-limit, interaction failure (action executed but UI does not respond as expected), and element-locator failure (target cannot be grounded). Table~\ref{tab:step_sensitivity_100} shows agreement increases with larger budgets while timeouts decrease, indicating that a meaningful portion of disagreement is due to resource/grounding limitations rather than systematic scoring bias.

\begin{table}[t]
\centering
\scriptsize

\begin{minipage}[t]{0.53\linewidth}
\centering
\setlength{\tabcolsep}{3pt}
\renewcommand{\arraystretch}{1}
\captionof{table}{Human--agent agreement (N=100 on General WebUI). We report overall agreement rate, full agreement breakdown (N--N, P--P, Y--Y), and disagreement breakdown (Adjacent v.s. Far). "Adjacent" (Adj.) counts No$\leftrightarrow$Partial and Partial$\leftrightarrow$Yes confusions;  "Far" counts No$\leftrightarrow$Yes confusions.}
\label{tab:ha_agreement_compact}
\begin{tabular}{l c ccc cc}
\toprule
& & \multicolumn{3}{c}{\textbf{Full Agree.}} & \multicolumn{2}{c}{\textbf{Disagree.}} \\
\cmidrule(lr){3-5}\cmidrule(lr){6-7}
\textbf{Model Families} &
\textbf{\begin{tabular}[c]{@{}c@{}}Total\\Agree.\end{tabular}} &
\textbf{N--N} & \textbf{P--P} & \textbf{Y--Y} &
\textbf{Adj.} & \textbf{Far} \\
\midrule
Claude Sonnet 4 & 86 & 18 & 30 & 38 & 13 & 1 \\
Gemini 2.5 Pro  & 82 & 14 & 31 & 37 & 15 & 3 \\
Qwen2.5 14B (VisRL) & 85 & 18 & 32 & 35 & 10 & 5 \\
\bottomrule
\end{tabular}
\end{minipage}\hfill
\begin{minipage}[t]{0.44\linewidth}
\centering
\setlength{\tabcolsep}{2pt}
\renewcommand{\arraystretch}{1}
\captionof{table}{Step budget sensitivity of the UI-agent evaluator (N=100, General WebUI). Increasing the step budget improves human--agent agreement and reduces timeouts; remaining errors are primarily interaction and element-locator failures.}
\label{tab:step_sensitivity_100}
\begin{tabular}{c c c c c}
\toprule
& & \multicolumn{3}{c}{\textbf{Evaluator Errors}} \\
\cmidrule(lr){3-5}
\textbf{\# Step} &
\textbf{\begin{tabular}[c]{@{}c@{}}Total\\Agree.\end{tabular}} &
\textbf{Timeout} &
\textbf{\begin{tabular}[c]{@{}c@{}}Interaction\\fail.\end{tabular}} &
\textbf{\begin{tabular}[c]{@{}c@{}}Element Loc.\\fail.\end{tabular}} \\
\midrule
5  & 80.0 & 6.7 & 10.0 & 3.3 \\
10 & 84.0 & 3.3 & 6.7  & 6.7 \\
15 & 86.7 & 0.0 & 4.3  & 9.0 \\
\bottomrule
\end{tabular}
\end{minipage}

\vspace{-2mm}
\end{table}

\subsection{Analysis on Benchmark Quality}

\begin{figure}[t]
  \centering
  \captionsetup{font=small,skip=4pt}

  \begin{minipage}[t]{0.53\textwidth}
    \vspace{0pt}
    \centering
    \includegraphics[width=\linewidth]{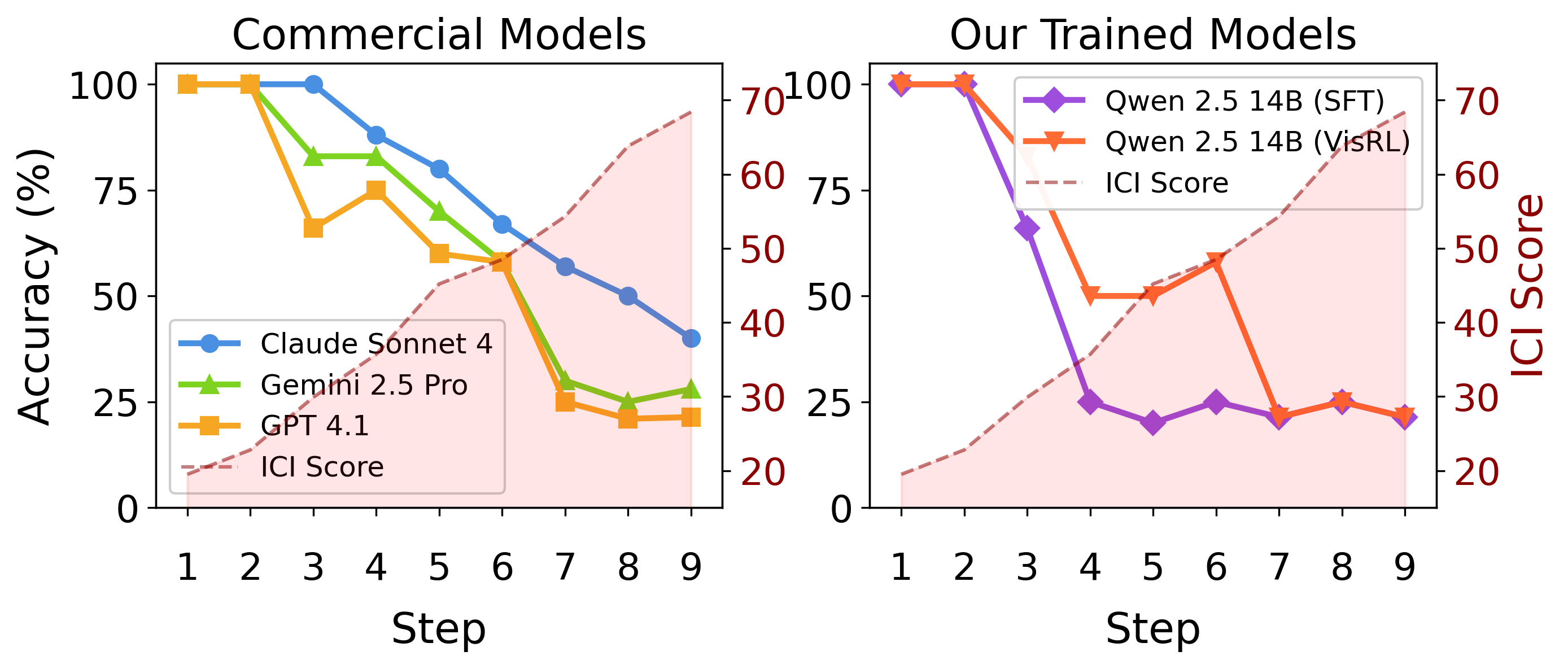}
    \caption{Sensitivity analysis of model robustness vs.\ instruction complexity. Commercial models exhibit superior accuracy retention (Claude Sonnet 4: 100\% to 40\%) compared to our trained models Qwen 2.5 14B (VisRL), which drops from 100\% to 21.4\% across 9 steps, with ICI scores increasing from 19.5 to 68.4 (\textcolor{red}{red} margin with dotted line).}
    \label{fig:ici_row}
  \end{minipage}\hfill
   \begin{minipage}[t]{0.44\textwidth}
    \vspace{0pt}
    \centering
    \includegraphics[width=\linewidth]{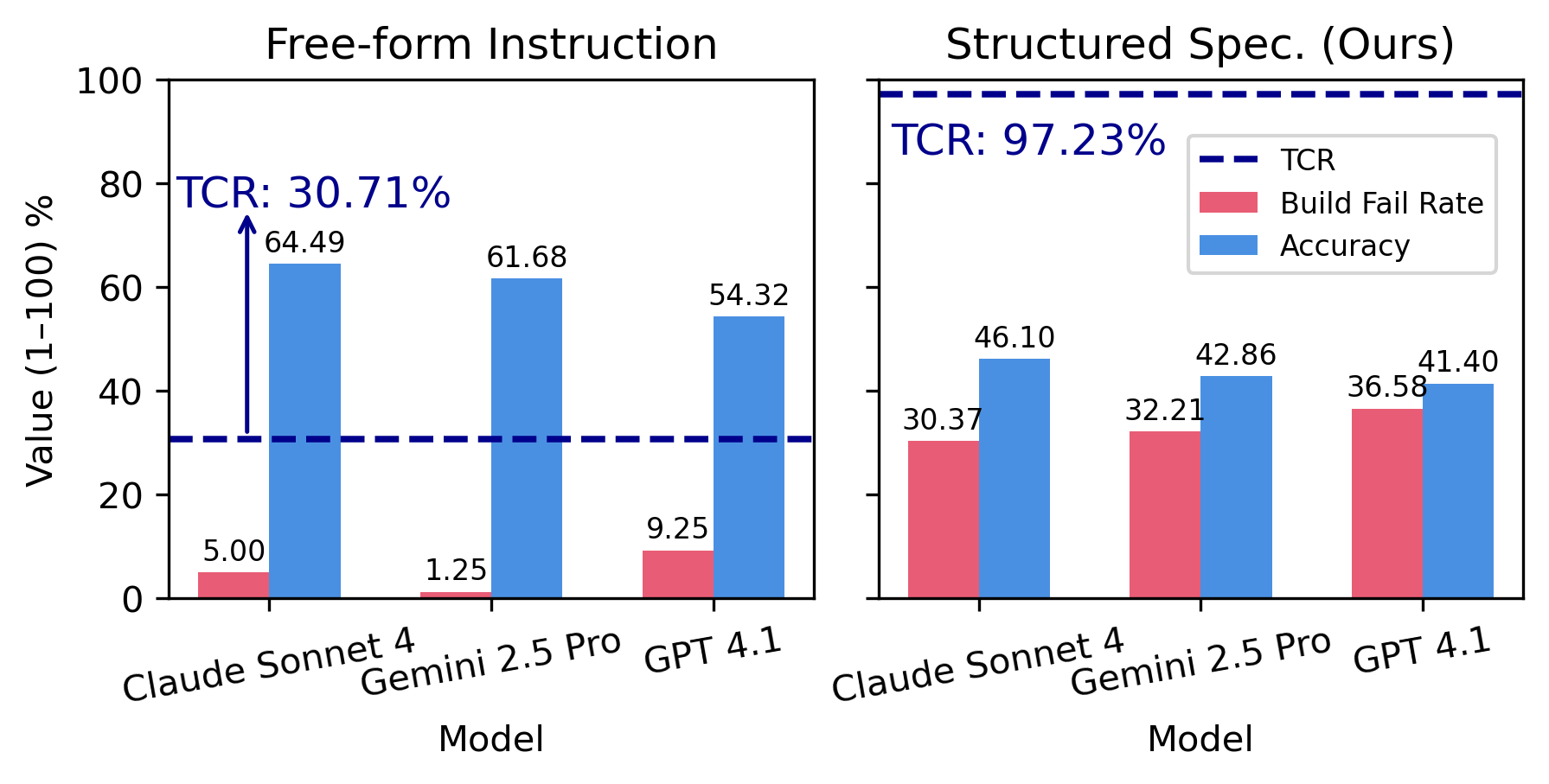}
    \caption{Benchmark quality analysis via test coverage. We compare \textbf{free-form} vs.\ \textbf{structured specifications} on General WebUI. Dashed lines indicate the average TCR. Structured specs. increase coverage (30.71\%$\rightarrow$97.23\%) and yield a more discriminative evaluation.}
    \label{fig:tcr_row}
  \end{minipage}
  \vspace{-3mm}
\end{figure}

\textbf{$\bullet$ Sensitivity to instruction complexity.} To quantify how functional requirements scale with instruction difficulty, we define an \textbf{instruction complexity index (ICI)} that captures more than prompt length. ICI aggregates two factors, including \emph{Size} $S$ (number of required elements and tests), \emph{Interactivity} $I$ (interaction verbs/actions):
$
\text{ICI}=0.5S+0.5I
$.
While increasing the step, we increment the instruction with more requirements. The generation of each step is built sequentially; that is, the model will generate the code for the current step based on the code generated from prior steps.
In Fig~\ref{fig:ici_row} (left side subgraph), we show the commercial models' performance. When increasing the step from 1 to 9, the ICI score increases up to 68 (red dotted line), and the accuracy of the three frontier models gradually drops. On the right side subgraph, we show the accuracy of our self-trained model, including baseline SFT and VisRL (PPO with UI agent). The trend is similar to the frontier models.

\textbf{$\bullet$ Test coverage rate of WebUIProof.}
We assess benchmark quality using TCR, the fraction of required UI elements exercised by the test suite. For each task, a UI element is counted as \emph{covered} if it is referenced by any test case as an interaction target or assertion.
We compare TCR between free-form instruction prompts collected from WebGen-Bench~\cite{liu2024webgenbench} and stuctured specs. from WebUIProof (ours). We report two metrics: model accuracy and build failure rate. High TCR paired with low accuracy indicates that the instruction provides dense, informative tests that remain challenging. Fig.~\ref{fig:tcr_row} shows TCR increases from 30.71\% (free-form) to 97.23\% (structured spec.); the resulting drop in accuracy suggests that WebUIProof is more challenging.

\begin{figure*}[t]
  \centering
  \captionsetup{font=small,skip=4pt}

  \begin{minipage}[t]{0.30\textwidth}
    \vspace{0pt}
    \centering
    \scriptsize
    \setlength{\tabcolsep}{4pt}
    \renewcommand{\arraystretch}{1.08}

    \captionof{table}{VisRL learning-strategy ablations. The upper table shows that MTS and FGF improve UI stability/efficiency; Lower table shows that multi-temperature outperforms single-temperature under the same best-of-$N$.}
    \label{tab:ablation_learnstrategy}

    \vspace{2pt}
    \begin{adjustbox}{max width=\linewidth}
    \begin{tabular}{lcc}
    \toprule
    Learning strategy & Build fail. (\%) & Acc. (\%) \\
    \midrule
    \textbf{VisRL (Ours)} & \textbf{36.10} & \textbf{41.80} \\
    w/o MTS ($\tau{=}0.7$) & 46.31 & 36.76 \\
    w/o FGF & 49.53 & 39.00 \\
    w/o MTS \& FGF & 50.32 & 32.12 \\
    \bottomrule
    \end{tabular}
    \end{adjustbox}

    \vspace{2pt}

    \begin{adjustbox}{max width=\linewidth}
    \begin{tabular}{lcc}
    \toprule
    Best-of-$N$ & $\tau$ & Acc. (\%) \\
    \midrule
    \textbf{$N{=}3$} & \textbf{$\{0.3,0.45,0.7\}$} & \textbf{41.80} \\
    $N{=}2$ & $\{0.3,0.7\}$ & 41.53 \\
    $N{=}2$ & $0.3$ & 39.00 \\
    $N{=}1$ & $0.7$ & 36.76 \\
    $N{=}1$ & $0.3$ & 35.28 \\
    \bottomrule
    \end{tabular}
    \end{adjustbox}
  \end{minipage} \hfill
   \begin{minipage}[t]{0.67\textwidth}
    \vspace{0pt}
    \centering
    \includegraphics[width=\linewidth]{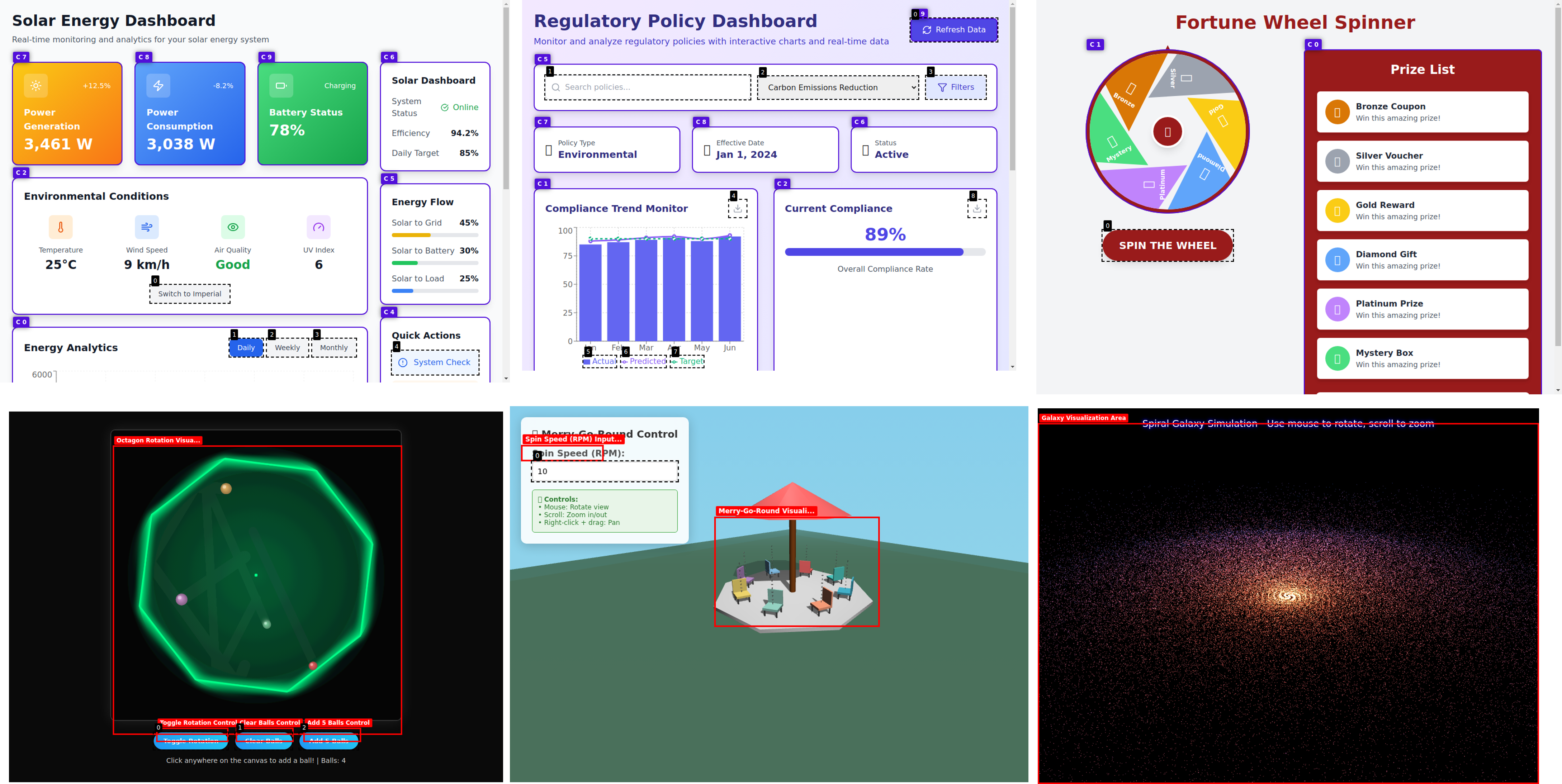}
    \caption{\textbf{UI-agent execution examples on WebUIProof.} Top: general WebUIs (e.g., interactive dashboards and widgets). Bottom: 3D simulation WebUIs (e.g., physics-driven interactive scenes). In all cases, the UI agent executes interaction tests by grounding to DOM/UI elements and performing actions (click/type/drag) to validate functional behavior.}
    \label{fig:webui_vis}
  \end{minipage}

  \vspace{-3mm}
\end{figure*}

\subsection{Analysis of Learning Strategies in VisRL}
\label{sec:learn_strategies_visrl}
We study learning strategies to improve rollout efficiency in VisRL. Table~\ref{tab:ablation_learnstrategy} provides two analyses: (i) the effect of multi-temperature candidate sampling (MTS) and fast-gate filtering (FGF).
(ii) the effect of sampling budget and temperature.
The upper sub-table shows that removing either MTS or FGF increases build failures and reduces overall accuracy, showing both are important for stable and efficient RL training. The lower sub-table analyzes the budget of best-of-$N$ sampling under different sampling numbers ($N=1,2,3$), with multi-or single temperature ($\tau$ indicates temperature set). For the same $N$, multi-temperature achieves higher accuracy than the same temperature, indicating that the gain comes from diversity, not just more sampling attempts.

\vspace{-2mm}

\subsection{Visualization of Tasks and UI-Agent Execution Traces}

Figure~\ref{fig:webui_vis} visualizes six WebUIProof tasks and the corresponding UI-agent execution traces. The top row shows general WebUI applications (e.g., dashboards and interactive widgets), while the bottom row shows 3D simulation interfaces (e.g., object collisions). The UI agent localizes task-relevant elements by cropping their bounding regions from the rendered page and marking them with colored boxes and index labels, making the executed interaction targets explicit. Additional visualization of UI-agent execution traces and task samples for both domains is provided in Appendix~\ref{app:e_2_additional_visualization}.




\section{Conclusion}

We introduce \textbf{WebUIProof}, an execution-oriented benchmark for WebUI code generation that pairs structured specifications with \emph{executable} interaction tests spanning general WebUIs and 3D simulation interfaces. WebUIProof includes a reproducible UI-agent harness for automated evaluation, enabling functional verification beyond build success or static visual checks. Using this protocol, we benchmark eight frontier models and find that interactive WebUI generation remains challenging in both domains, with substantial headroom on 3D tasks.
We also show that the same harness can provide outcome-level training signals by including \textbf{VisRL}, an execution-grounded RL appraoch that optimizes compact models with composite rewards from build/runtime execution, rendered visual quality, and UI-agent test success. VisRL improves both functional completion and reliability compared to SFT and vision-only baselines. Our ablation suggests that performance remains bottlenecked by complex instruction and multi-step requirements, motivating future work on reward assignment, richer executable signals, and robust evaluation agents for dynamic 3D WebUI.
Overall, our work shed light on the importance of outcome-grounded, executable evaluation and training signals for reliable code generation beyond static correctness and visual-only metrics.

{
\small

\bibliography{neurips_2026}
\bibliographystyle{plain}
}

\clearpage
\vspace*{-1em}
\begin{center}
{\Large \bfseries
Appendices
}
\end{center}
\vspace{0.4em}
\section*{Appendix Outline}
\definecolor{navy}{RGB}{10,35,80}
\noindent\textbf{\ref{app:a_main_reproduce}. Reproducibility and Evaluation Protocol}
\begin{itemize}
    \item \ref{app:a_1_exec_environment} Execution Environment (web framework and build process) 
    \item \ref{app:a_2_agent_evaluator} UI-agent Evaluator Setup (action space, step budget)
    \item \ref{app:a_3_prompt_for_codegen} System Prompt for Code Generation (general + 3D WebUI)
    \item \ref{app:a_4_failure_handling} Evaluation Metrics and Failure Handling
\end{itemize}

\noindent\textbf{\ref{app:b_main_benchmark}. WebUIProof Benchmark Construction}
\begin{itemize}
    \item \ref{app:b_1_datasource} Data sources and Task Scope (general + 3D WebUI)
    \item \ref{app:b_1-5_uicategory} WebUI Categories and Test Case Distribution
    \item \ref{app:b_2_datacollection} Data Collection Pipeline and Benchmark Collection Prompt
    \item \ref{app:b_3_datasample} Sample of Structured specification
\end{itemize}

\noindent\textbf{\ref{app:c_main_visrl}. VisRL: Training Details}
\begin{itemize}
    \item \ref{app:c_0_visrl_implementation} Implementation Details (training hyperparameters configuration, resource usage, training infrastructure)
    \item \ref{app:c_1_seq_ppo} Sequence-level Rollout and Probability
    \item \ref{app:c_2_reward_definition} Reward definition
    \item \ref{app:c_3_additional_learning} Additional learning strategies (multi-temperature sampling, fast gate filtering)
    \item \ref{app:c_4_pseudo_algo} Pseudo Algorithm of VisRL
    \item \ref{app:c_5_uiagent_exec} UI Agent Task Execution Process
    \item \ref{app:c_6_visionrubric} Vision Agent - Evaluation Rubric
\end{itemize}

\noindent\textbf{\ref{app:d_main_training_analysis}. Reward and Hyperparameter Analyses}
\begin{itemize}
   \item \ref{app:d_1_training_curve} Training Curve for Different PPO Strategy 
    \item \ref{app:d_2_reward_decompose_analysis} Reward Decomposition Analysis 
    \item \ref{app:d_3_alpha_sensitivity_analysis} Reward Weight Analysis - $\alpha$ Sensitivity and Trade-off Discussion
\end{itemize}

\noindent\textbf{E. Extended Results and Diagnostics}
\begin{itemize}
    \item \ref{app:e_1_additional_results} Additional Experiments (per-test category results of general WebUI and 3D WebUI)
    \item \ref{app:e_2_additional_visualization} Visualization of UI Agent Execution Traces
    \item \ref{app:e_3_moremore_visualization} More Visualization for general WebUI and 3D WebUI
    \item \ref{app:e_3_build_failure} Build Failure Taxonomy (error categories and examples)
\end{itemize}


\clearpage

\appendix

\section{Reproducibility and Evaluation Protocol}
\label{app:a_main_reproduce}
\subsection{Execution Environment}
In the main paper Sec.~\ref{sec:experiment}, we introduce the evaluation setup for WebUIProof, including model details, UI-agent evaluator setup. Here, we provide additional detail on the execution environment for evaluation.

\paragraph{Web Framework and Build Process}

We use React, Next.js, and Node.js to build the web app during the evaluation.
We transform the React codes generated from LLMs into a fully functional Next.js web application via a specific build process:

(1.) JSON code files parsing: At the beginning, the LLM generates codes and saves them in the JSON files. We extract TSX/React code from the JSON  payload (e.g., \codeline{payload.get("metadata", \{\}).get("code")}) 

\begin{wrapfigure}{r}{0.35\textwidth}
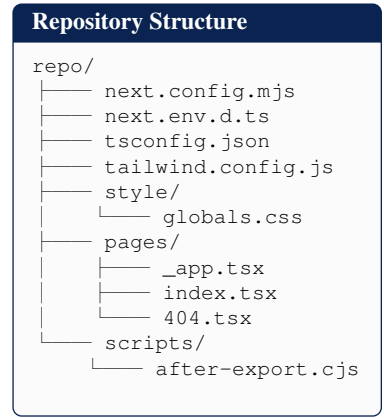

\vspace{-0.8em} 
\centering
\begin{tcolorbox}[
  title={Repository Structure},
  width=\linewidth,
  colback=gray!3,
  colframe=navy,
  boxrule=0.5pt,
  arc=3pt,
  left=4pt,right=4pt,top=3pt,bottom=3pt,
  fontupper=\ttfamily\footnotesize,  
  fonttitle=\bfseries\small          
]
\ttfamily\footnotesize
\begin{verbatim}
repo/
├── next.config.mjs
├── next.env.d.ts
├── tsconfig.json
├── tailwind.config.js
├── style/
│   └── globals.css
├── pages/
│   ├── _app.tsx
│   ├── index.tsx
│   └── 404.tsx
└── scripts/
    └── after-export.cjs

\end{verbatim}
\end{tcolorbox}
\vspace{-0.8em} 
\caption{Next.js project structure}
\label{fig:repo-structure-wrap}
\end{wrapfigure}

(2.) Code Sanitization: we find valid TSX code with \codeline{export default} using a regex search. (3) Web Project Scaffolding: we create a completely new Next.js project with \textit{npm} build commands, such as \codeline{npm init -y} for project initialization, \codeline{npm install react@18.3.1 next@15.5.4} for react/next.js installation, \codeline{npm install -D typescript @types/react @types/node tailwindcss@3 postcss} to setup libraries for UI style.
(4) Files Generation: After all the dependency installation, we create a repo structure for web app with several files (as shown in Fig.~\ref{fig:repo-structure-wrap}). The Next.js web app includes one file for dynamically inserting the LLM-generated code (\codeline{index.tsx}), and several fixed structure files with pre-built templates to ensure static export for deployment
(5) Build the Repo: after constructing the repo structure, we run \codeline{npm run build}, which triggers Next.js static export. The build script will finally output an \codeline{out/} folder with \codeline{index.html}, \codeline{404.html}, \codeline{\_next/} that contain all the Next.js static build assets (JavaScript bundles, compiled css files, build manifest files) for us to directly render in the browser using the command \codeline{npx serve out}. 


\label{app:a_1_exec_environment}
\subsection{UI-agent Evaluator Setup}
\label{app:a_2_agent_evaluator}
Our evaluation pipeline is built upon the existing UI agent evaluator Webvoyager~\cite{he2024webvoyager}. a vision-language web agent that automate browser interaction through a combination of visual percpetion and DOM analysis.
The web agent uses Selenium as a Webdriver with our own configured Chrome/Chromium browser (Chrome driver version 119.0.6045.105).

The agent operates through a number of iteration steps, where each step follows a thinking loop: (1.) Selenium takes UI screenshots and parses the DOM interactive web elements via executable JavaScript code snippets. (2.) LLMs reason from the visual information in the UI screenshots and provide "Thought" and "Action" responses. The actions are predefined under an action space (e.g., click/type/scroll/select/hover...) (see Table~\ref{tab:action_space}). LLMs will choose one action for each step, and the action is executed by Selenium. The UI state will update for the next iteration, and at the final step, we ask LLMs to assess task completion, including three answer options: "FULL", "NO", and "PARTIAL" completion. In our experiment, we set the maximum iteration step as 15. (Mentioned in Sec.~\ref{sec:method}.1 as well)

For visual understanding, the agent supports multiple vision-language models (VLMs), either via the API (e.g., GPT-4O from OpenAI), HuggingFace endpoints, or open-source VLMs hosted locally. In our evaluation, we chose to use the Qwen2.5-VL-72B\footnote{\url{https://huggingface.co/Qwen/Qwen2.5-VL-72B-Instruct}}, a powerful open-source VLM. 

\clearpage
\begin{table}[h]

\caption{Action space of UI agent}
\centering
\scriptsize
\setlength{\tabcolsep}{2pt} 
\renewcommand{\arraystretch}{1} 
\begin{tabular}{l!{\hspace{2pt}\vrule width 0.8pt\hspace{2pt}}l}
\toprule
\textbf{Action Type} & \textbf{Detail}                                                                                                                          \\ \midrule
Click                & Click any web element identified by its numerical label (e.g., Click {[}0{]})                                                            \\
Type                 & Clear textbox and input content (automatically hits ENTER) (e.g., Type {[}0{]})                                                          \\
Scroll               & Scroll specific element or entire window (e,g, Scroll WINDOW up / down)                                                                  \\
Hover                & Hover for tooltips/effects, with precise coordinate positioning for charts (e.g., Hover {[}0:75,50{]} for 75\% from left, 50\% from top) \\
Wait                 & Pause for 5 seconds (for loading processes)                                                                                              \\
GoBack               & Navigate to previous webpage                                                                                                             \\
Answer               & Provide final task result (terminates interaction) (e.g., FULL/PARTIAL/NO)                                                               \\ \bottomrule
\end{tabular}
\label{tab:action_space}
\end{table}

\subsection{System Prompt for Code Generation (General + 3D WebUI)}
\label{app:a_3_prompt_for_codegen} 

To ensure reproducibility and make our evaluation pipeline fully auditable, we disclose the exact system prompts used for our two task families: \textbf{(i) General WebUI generation} and \textbf{(ii) 3D WebUI simulation} (As shown in Fig.~\ref{fig:genui_prompt} and~\ref{fig:threejs_prompt}). These two settings require \emph{different code formats and runtime assumptions}, and we therefore use task-specific prompts to standardize model outputs. For \textbf{General WebUI} tasks (Fig.~\ref{fig:genui_prompt}), models are instructed to generate a \emph{single-file} React/TypeScript component compatible with a Next.js (pages router) scaffold, with strict output constraints (default export only, no surrounding text) and styling constraints (Tailwind classes without arbitrary values). In contrast, \textbf{3D simulation} tasks (Fig.~\ref{fig:threejs_prompt}) require a self-contained \emph{HTML} document that renders an interactive Three.js scene. The prompt fixes the Three.js version and enforces verified CDN imports to avoid dependency drift, and additionally emphasizes runtime robustness (error handling, interactivity checks) and output formatting (start from \texttt{<!DOCTYPE html>} and return HTML only). By explicitly specifying the target artifact (\texttt{pages/index.tsx} React code vs.\ standalone HTML), these prompts reduce formatting variance across models and ensure that downstream building, rendering, and UI-agent testing are applied consistently.

\begin{figure*}[h]
\centering
\begin{tcolorbox}[title=General WebUI Generation, colback=cyan!5!white,colframe=cyan!65!black,fonttitle=\bfseries,
  fontupper=\ttfamily\footnotesize,  
  fonttitle=\bfseries\small          
]

You are an expert frontend React engineer who is also a great UI/UX designer. Follow the instructions carefully. Think carefully step by step. 
\\

\begin{itemize}[leftmargin=1.6em, itemsep=0.25em, topsep=0.25em]
  \item Create a React component for whatever the user asked you to create and make sure it can run by itself by using a default export.
  \item Make sure the React app is interactive and functional by creating state when needed and having no required props.
  \item If you use any imports from React like \texttt{useState} or \texttt{useEffect}, import them directly. Use TypeScript as the language for the React component.
  \item Use Tailwind classes for styling. DO NOT USE ARBITRARY VALUES (e.g., \texttt{h-[600px]}). Use a consistent color palette.
  \item Specify and install ALL additional dependencies.
  \item Include all necessary code in one file. Do not touch project dependency files like \texttt{package.json}, \texttt{package-lock.json}, \texttt{requirements.txt}, etc.
  \item Use Tailwind margin and padding classes so components are spaced nicely.
  \item ONLY return the full React code starting with imports, nothing else. DO NOT start with \texttt{`typescript}, \texttt{`javascript}, \texttt{`tsx}, or \texttt{`}.
  \item ONLY if the user asks for a dashboard/graph/chart, you may import \texttt{recharts} (e.g., \texttt{import \{LineChart, XAxis, ...\} from "recharts"}). You may also use shadcn/ui charts (e.g., \texttt{ChartConfig}, \texttt{ChartContainer} from \texttt{@/components/ui/chart}).
  \item For placeholder images, use \texttt{<div className="bg-gray-200 border-2 border-dashed rounded-xl w-16 h-16" />}.
\end{itemize}
    You can use the following templates: \\
    
    nextjs-developer: "A Next.js 13+ app that reloads automatically. Using the pages router. All components must be included one file.". File: pages/index.tsx. Dependencies installed: nextjs@15.5.4, typescript, @types/node, @types/react, @types/react-dom, postcss, tailwindcss, shadcn. Port: 3000.

\end{tcolorbox}
\caption{System prompt for General WebUI task. Models generate React code generation with strict styling and output constraints.}
\label{fig:genui_prompt}
\end{figure*}

\clearpage

\begin{figure*}[t]
\centering
\begin{tcolorbox}[
title=3D WebUI Simulation,
  colback=cyan!5!white,
  colframe=cyan!65!black,
  boxrule=0.6pt,
  arc=3pt,
  left=6pt,right=6pt,top=6pt,bottom=6pt,
  fontupper=\ttfamily\footnotesize,
  fonttitle=\bfseries\small
]
You are an expert web developer specializing in creating interactive Three.js HTML simulations and visualizations. Follow these instructions carefully to ensure proper rendering:

\begin{itemize}[leftmargin=1.6em, itemsep=0.25em, topsep=0.25em]
  \item ALWAYS use the same Three.js version consistently across all imports.
  \item Use only verified CDN URLs. For example:
\end{itemize}

\begin{lstlisting}[
  basicstyle=\ttfamily\footnotesize,
  columns=fullflexible,
  breaklines=true,
  breakatwhitespace=true,
  numbers=none
]
<script src="https://cdn.jsdelivr.net/npm/three@0.140.0/build/three.min.js"></script>
<script src="https://cdn.jsdelivr.net/npm/three@0.140.0/examples/js/controls/
OrbitControls.js"></script>
\end{lstlisting}

\begin{itemize}[leftmargin=1.6em, itemsep=0.25em, topsep=0.25em]
  \item NEVER mix different Three.js versions in the same file.
  \item ALWAYS include comprehensive error handling with JavaScript \texttt{loader.load()}.
\end{itemize}

\noindent OUTPUT FORMAT:
\begin{itemize}[leftmargin=1.6em, itemsep=0.25em, topsep=0.25em]
  \item Return ONLY the complete HTML code.
  \item Do not include explanatory text before or after.
  \item Start directly with \texttt{<!DOCTYPE html>}.
  \item Ensure the simulation is functional and interactive.
  \item Test all interactive elements work properly.
\end{itemize}
\end{tcolorbox}
\caption{System prompt for Three.js HTML simulation generation.}
\label{fig:threejs_prompt}
\end{figure*}


\subsection{Evaluation Metrics and Failure Handling}
\label{app:a_4_failure_handling}
For each task, we categorize the outcome into four mutually exclusive cases:
\emph{(i) Full Success}, \emph{(ii) Partial Success}, \emph{(iii) No Success}, and \emph{(iv) Build Failure}.

\emph{Build Failure} covers build-time or launch-time failures (e.g., compilation errors, runtime crashes, or timeouts) where the generated project cannot be executed for UI-agent testing. 
Importantly, \textbf{we average Full/Partial/No Success over \emph{all} tasks} by treating build failure with no credit. As a result, the corresponding rates satisfy
$\text{Full} + \text{Partial} + \text{No Success} + \text{Build failure} \approx 100\%$.

\clearpage
\section{WebUIProof: Benchmark Construction}
\label{app:b_main_benchmark}

\subsection{Data Source and Task Scope}
\label{app:b_1_datasource}
\paragraph{General WebUI}
In the main paper, Sec.~\ref {sec:method}, we mention that the data collection process for structured specification requires seed pages. Here, we describe how we collect the seed pages. 
For the general WebUI task, we collect seed instructions mainly from two data sources: WebGen Bench~\cite{liu2024webgenbench} and WebDev-Arena~\cite{zheng2024webdevarena}. 
We use 101 instructions from the WebGenBench testset and 48 instructions from WebDev-Arena-10k as our seed instructions and generate seed webpages using DeepSeek R1 model.

This collection encompasses five main application domains, including creative media, business data, game sports, social interaction, and learning style. 
For example, 
social media platforms (reddit clone, social profile like Twitter, dating app), 
interactive gaming platforms (chess applications, 2048 games, wheel of fortune), 
chat interface (mail box with chat function)
sports analytics systems (basketball betting analysis, baseball simulation), business applications (credit repair lead generation, clinical office websites), 
data visualization tools (regulatory policy dashboards, stock report generators), 
and complex strategy games (text-based browser games with resource management). 

Our collected instructions emphasize both frontend user experience elements and backend functionality, requiring implementations that support real-time interactions, data visualization via charts and dashboards, user authentication systems, and responsive design patterns compatible with modern web development frameworks.
We show more sample visualization in \ref{app:e_3_moremore_visualization}

\paragraph{3D WebUI Simulation}

The 3D simulation test instruction dataset features advanced Three.js-based interactive visualizations and simulations derived from Three.js opensource website examples and community contributions. This specialized collection contains 70 sophisticated 3D programming challenges that span astronomical simulations (particle galaxy formations with spiral arms, black hole visualizations with event horizon effects), geometric animations (tesseract rotations with collision detection, merry-go-round mechanics with adjustable parameters), physics-based demonstrations (bouncing ball dynamics within complex 3D shapes), and interactive exploration systems that allow users to navigate through immersive 3D environments using JavaScript controls. 
Each instruction demands implementation of complex mathematical concepts including 3D transformations, particle systems, collision detection algorithms, and real-time rendering optimizations, while maintaining smooth 60fps performance and responsive user interaction capabilities. 

The tasks require understanding of WebGL fundamentals, Three.js scene graph management, camera controls, lighting systems, material properties, and animation loops, with many instructions specifically requesting single HTML file implementations that demonstrate advanced computer graphics concepts through browser-based 3D experiences. These instructions evaluate both technical proficiency in 3D programming and creative problem-solving abilities in translating complex visual concepts into interactive web-based simulations.
We show more sample visualization in \ref{app:e_3_moremore_visualization}

\clearpage

\subsection{WebUI Categories and Test Case Distribution}
\label{app:b_1-5_uicategory}

\begin{figure*}[h]
    \centering
    \includegraphics[width=0.96\textwidth]{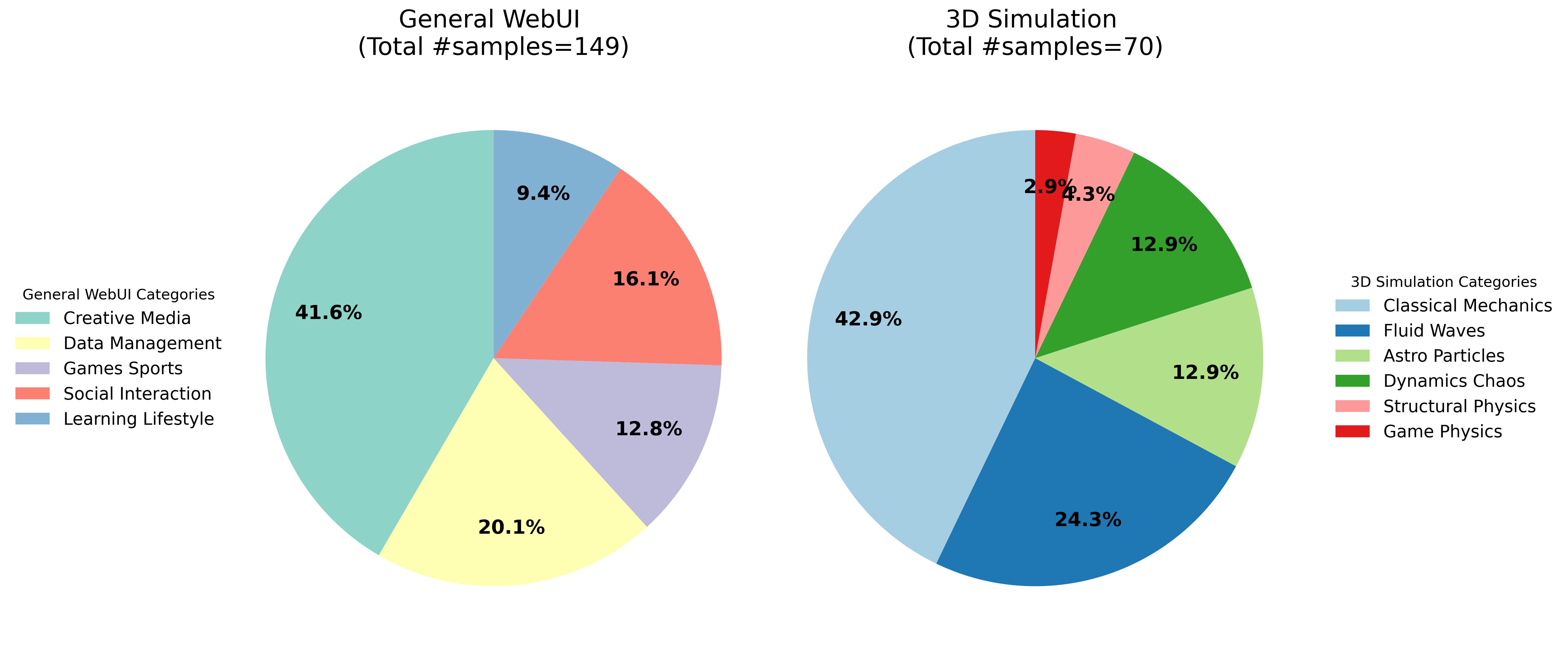}
    \caption{WebUI Category Distribution for WebUIProof.}
    \label{fig:webui-category-distribution}
\end{figure*}

In Fig.~\ref{fig:webui-category-distribution}, we provide the category distribution of two distinct WebUI tasks in WebUIProof: General WebUI applications (149 samples) and 3D Simulation applications (70 samples).
General WebUI is dominated by Creative Media applications (41.6\%), followed by Data Management (20.1\%) and Social Interaction (16.1\%). The remaining categories include Learning \& Lifestyle (12.8\%) and Games \& Sports (9.4\%).
3D Simulation shows a more balanced distribution with Classical Mechanics leading at 42.9\%, Fluid Waves at 24.3\%, and equal representation of Astro Particles and Dynamics Chaos (both 12.9\%). Structural Physics (4.3\%) and Game Physics (2.9\%) represent smaller segments.

\begin{figure*}[h]
    \centering
    \includegraphics[width=0.76\textwidth]{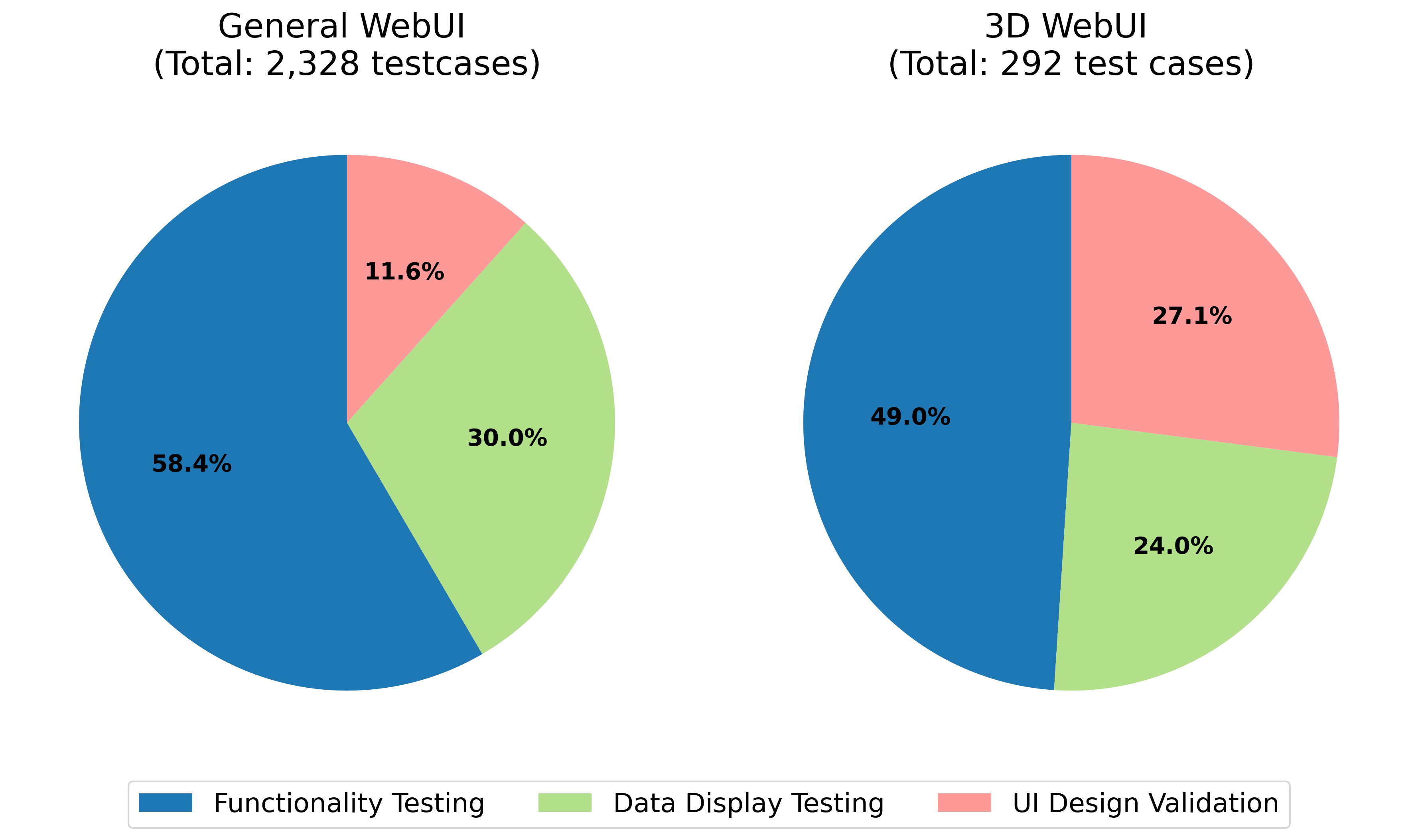}
    \caption{Test Case Distribution for WebUIProof.}
    \label{fig:webui-testcase-distribution}
\end{figure*}

In Fig.~\ref{fig:webui-testcase-distribution}, we further provide the test cases distribution on two main tasks, we categorize the test cases into three categories, including functionality testing, data display testing, and design validation testing. In the later section~\ref{app:e_1_additional_results}, we show additional experimental results on category accuracy for each test case across multiple proprietary models.

\clearpage
\subsection{Data Collection Pipeline and Detail Prompt}
\label{app:b_2_datacollection}
\begin{figure}[h]
    \centering
    \includegraphics[width=0.70\textwidth]{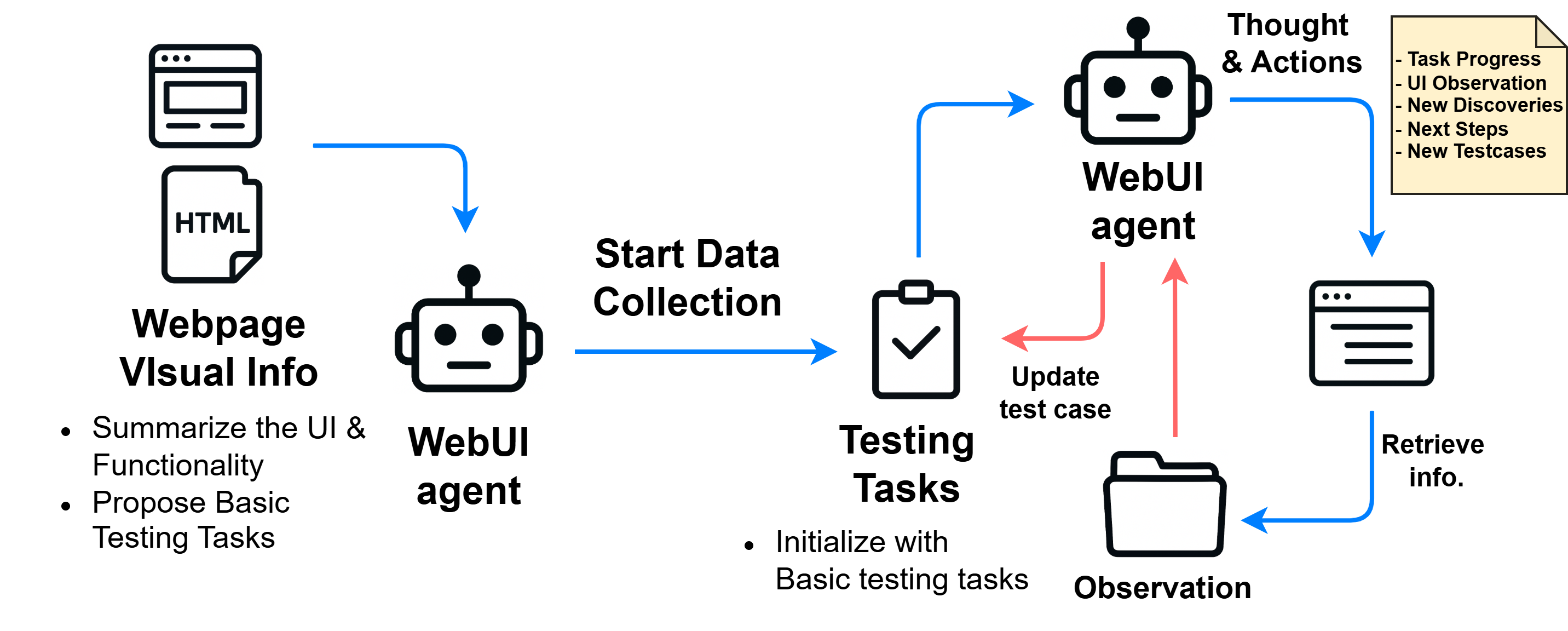}
    \caption{Benchmark Collection Pipeline.}
    \label{fig:webui-data-collection-flow}
\end{figure}

Although we have already described how we collect the data in the main paper, we would like to provide more detail, including the data collection flow, and the system prompt we provide to UI agent for WebUI exploration.
As Fig~\ref{fig:webui-data-collection-flow} shows, the UI agent receives the webpages screenshots and html file as input at the first step. Then it summarizes the UI functionality, and provides the basic test cases for later exploration. At the second step, it starts the thinking loop, and each loop it will provide the UI observation, newly discovered element, and test case recommendations based on the updated UI state at every iteration.

\begin{figure*}[h]
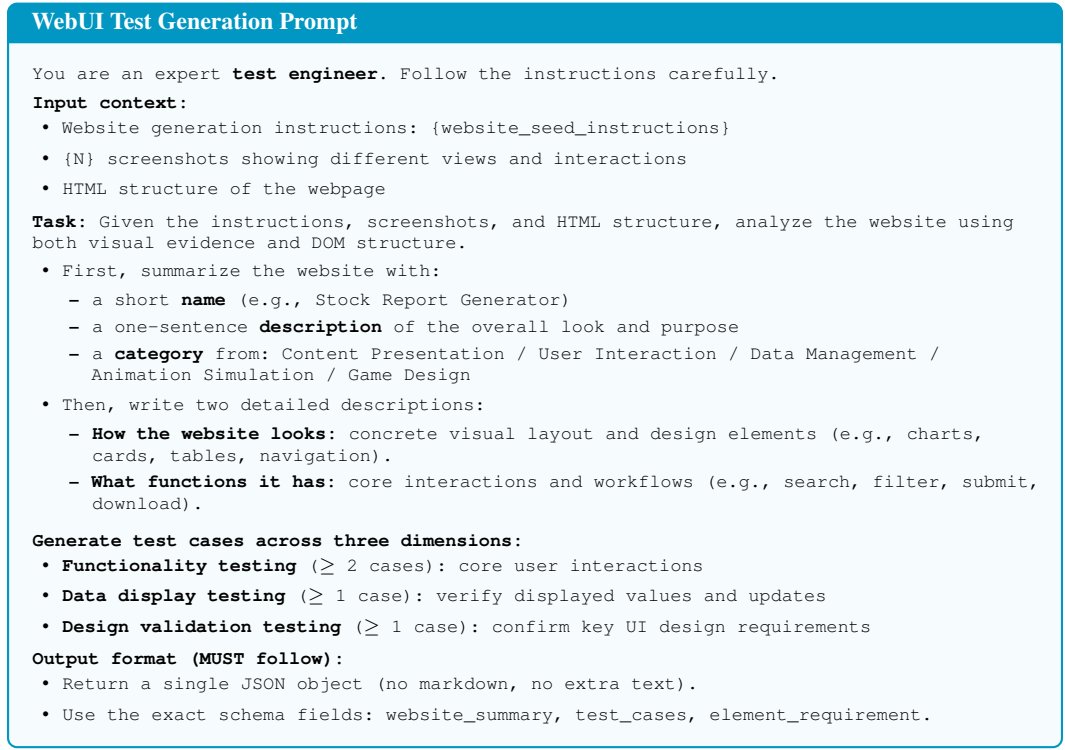

\centering
\begin{tcolorbox}[
  title={WebUI Test Generation Prompt},
  colback=cyan!3!white,
  colframe=cyan!75!black,
  coltitle=white,
  colbacktitle=cyan!75!black,
  boxrule=0.8pt,
  arc=3pt,
  left=6pt,right=6pt,top=6pt,bottom=6pt,
  fonttitle=\bfseries\small,
  fontupper=\ttfamily\scriptsize,
]

You are an expert \textbf{test engineer}. Follow the instructions carefully.

\smallskip
\textbf{Input context:}
\begin{itemize}[leftmargin=1.6em, itemsep=0.2em, topsep=0.2em]
  \item Website generation instructions: \texttt{\{website\_seed\_instructions\}}
  \item \texttt{\{N\}} screenshots showing different views and interactions
  \item HTML structure of the webpage
\end{itemize}

\smallskip
\textbf{Task:} Given the instructions, screenshots, and HTML structure, analyze the website using both visual evidence and DOM structure.

\begin{itemize}[leftmargin=1.6em, itemsep=0.2em, topsep=0.2em]
  \item First, summarize the website with:
  \begin{itemize}[leftmargin=1.6em, itemsep=0.15em, topsep=0.15em]
    \item a short \textbf{name} (e.g., \texttt{Stock Report Generator})
    \item a one-sentence \textbf{description} of the overall look and purpose
    \item a \textbf{category} from:
    \texttt{Content Presentation / User Interaction / Data Management / Animation Simulation / Game Design}
  \end{itemize}

  \item Then, write two detailed descriptions:
  \begin{itemize}[leftmargin=1.6em, itemsep=0.15em, topsep=0.15em]
    \item \textbf{How the website looks:} concrete visual layout and design elements (e.g., charts, cards, tables, navigation).
    \item \textbf{What functions it has:} core interactions and workflows (e.g., search, filter, submit, download).
  \end{itemize}
\end{itemize}

\smallskip
\textbf{Generate test cases across three dimensions:}
\begin{itemize}[leftmargin=1.6em, itemsep=0.2em, topsep=0.2em]
  \item \textbf{Functionality testing} (\(\geq\) 2 cases): core user interactions
  \item \textbf{Data display testing} (\(\geq\) 1 case): verify displayed values and updates
  \item \textbf{Design validation testing} (\(\geq\) 1 case): confirm key UI design requirements
\end{itemize}

\smallskip
\textbf{Output format (MUST follow):}
\begin{itemize}[leftmargin=1.6em, itemsep=0.2em, topsep=0.2em]
  \item Return a single JSON object (no markdown, no extra text).
  \item Use the exact schema fields: \texttt{website\_summary}, \texttt{test\_cases}, \texttt{element\_requirement}.
\end{itemize}

\end{tcolorbox}
\caption{Multimodal prompt (instructions + screenshots + HTML) used for WebUI test-case generation.}
\label{fig:benchmark_collect_prompt}
\end{figure*}

\paragraph{Benchmark Collection Prompt} 
In Fig.~\ref{fig:benchmark_collect_prompt}, we show the detail prompt that we provide to WebUI agent to let them explore the seed webpages. As the prompt shows, we give the agent three inputs, including website seed instruction, screenshots of webpages, and an HTML file. Model will provide a JSON file with certain schema.

\clearpage
\subsection{Sample of Structured Specification in WebUIProof}
\label{app:b_3_datasample}
We show two samples of structured specifications in our WebUIProof benchmark (the first one for General WebUI and the second one for 3D WebUI).  We totally have 219 structured specifications. During the evaluation, we provide the model with the specification and let it follow the structured instructions to generate WebUI code.
\newtcblisting{instructionbox}[1][]{%
  enhanced,
  breakable,
  colback=cyan!5!white,colframe=cyan!65!black,
  boxrule=0.8pt,
  arc=2mm,
  left=3mm,right=3mm,top=2mm,bottom=2mm,
  fonttitle=\bfseries,
  title=Instructions,
  listing only,
  listing options={
    basicstyle=\ttfamily\footnotesize,
    breaklines=true,
    breakatwhitespace=false,
    prebreak=,              
    postbreak=,             
    columns=fullflexible,
    keepspaces=true,
    showstringspaces=false,
    numbers=none
  },
  #1
}

\begin{instructionbox}[title={Structured Specification for General WebUI}]
{
  "website_summary": {
    "name": "Stock Report Generator",
    
    "description": "The website displays dynamic stock information and offers analytical tools with a clean, professional design.",
    
    "category": "Data Management",
    
    "appearance": "The website has a clear and organized layout with distinct sections. It features a navigation header with \"Stock Report Generator\" prominently displayed. Below this, there's a section for stock information which includes a live price update, volume, and market cap. The \"Historical Data\" section showcases price and volume over time with a line graph. Additionally, the \"Financial Metrics\" and \"Financial Summary\" tables present key financial indicators and revenue details in a tabular format. The \"Ownership Distribution\" offers a pie chart for visualizing investor distribution, while the \"Stock Performance\" shows a line graph of price trends. The footer includes a customizable report section with checkboxes for content selection (e.g., Overview, Financials) and a \"Download Report\" button. The background is white, and various components utilize navy blue for text and buttons, ensuring a consistent color scheme and modern appearance."
  },
  "test_cases": {
    "functionality_testing": [
      {
        "id": "TC001",
        "name": "Download Button Test",
        "objective": "To verify that the Download Button triggers the report download process, providing complete standardized or customized report data.",
        "coverage_element": "Download Button",
        "preconditions": "The page is loaded, and a report is ready to be downloaded.",
        "expected_results": "The downloading process visually starts, leading to successfully downloading a report file reflecting data on the UI.",
        "procedure": [
          "Click on the Download Button and confirm a download concern initiating accordingly."
        ]
      } // ... (the rest of test cases)] }
  "element_requirement": {
  "1. Company Selector : Button for selecting different companies, currently set to \"Apple Inc.\"", 
  "2. Summary Tab : Button that likely switches the display to a summary view", 
  "3. Download Button : Button likely for downloading the current or customized report", 
  "4. Stock Information : Card displaying current price, volume, and market cap data for Apple Inc.", 
  "5. Historical Data Card : Contains a historical data chart with price and volume over time", 
  "6. Financial Metrics Card : Displays key financial indicators", 
  "7. Financial Summary : Card with revenue and income data"
  }
}
\end{instructionbox}

\clearpage

\begin{instructionbox}[title={Structured Specification for 3D WebUI Simulation}]
{
  "website_summary": {
    "name": "3D Spiral Galaxy Simulation",
    
    "description": "The website features a dark-themed background to emphasize the 3D spiral galaxy simulation.",
    
    "category": "Astro Particles",
    
    "appearance": "At the top of the page, there is a centered text informing the user how to interact with the simulation: \"Spiral Galaxy Simulation - Use mouse to rotate, scroll to zoom\". The galaxy is displayed in full screen within a canvas, creating a visually immersive experience with thousands of small, colored particles representing stars forming spiral arms around a central core. The website allows users to interact with a 3D spiral galaxy simulation using mouse controls. Users can:\n- Rotate the view of the galaxy by clicking and dragging the mouse.\n- Zoom in and out using the mouse scroll wheel.\n- Experience a visualization of spiral galaxy formation with realistic star positioning and colors."
  },
  "test_cases": {
    "functionality_testing": [
      {
        "id": "TC001",
        "name": "Rotate Galaxy View",
        "objective": "To verify that the user can rotate the galaxy view using the mouse.",
        "preconditions": "The website is loaded and the galaxy is visible.",
        "expected_results": "The galaxy rotates smoothly and interactively in real-time with the mouse movements, allowing the user to explore different angles.",
        "procedure": [
          "Open the website in a supported web browser.",
          "Click and hold the left mouse button on the galaxy display.",
          "Drag the mouse to the left and right."
        ]
      }
      // ... (the rest of test cases)
    ]
  }
  "element_requirement": {
     "1. Canvas Element : Full-screen 3D rendering canvas displaying the spiral galaxy simulation with thousands of star particles",
      "2. Instructional Text : Centered white text at the top reading \"Spiral Galaxy Simulation - Use mouse to rotate, scroll to zoom\"",
      "3. Mouse Rotation Control : Interactive mouse click-and-drag functionality for rotating the galaxy view in real-time",
      "4. Mouse Scroll Zoom Control : Mouse wheel scroll functionality for zooming in and out of the galaxy simulation",
      "5. Star Particles : Thousands of colored particles representing stars with varying colors (blue-white, yellow-white, redder) and sizes",
      "6. Spiral Galaxy Structure : 3D particle system forming realistic spiral arms around a central galactic core",
      "7. Real-time Animation : Smooth, interactive rendering system maintaining 60fps performance during user interactions",
      "8. Camera Controls : Three.js camera system enabling 360-degree exploration of the galaxy from different angles",
      "9. Particle Color System : Randomized star coloring system creating realistic stellar color distribution throughout the galaxy"
  }
}
\end{instructionbox}

\clearpage




\clearpage

\section{VisRL: Training Detail}
\label{app:c_main_visrl}

We provide algorithm details and implementation-level details for our sequence-level PPO training, including reward computation, candidate selection, and efficiency heuristics.

\subsection{Implementation Detail}
\label{app:c_0_visrl_implementation}

\paragraph{$\bullet$ Cold Start from Supervised Fine-Tuning (SFT)}
We finetune Qwen2.5-14B-Instruct on a curated dataset of 6000 high-quality webUI generation samples. We use the 6k training set from WebGen-Bench as our seed instruction and generate code snippets using DeepSeek-R1. Our model is trained with distributed training across 4 NVIDIA H100 GPUs (80 GB each) with full parameter tuning. 
To handle long code sequences, we employ dynamic RoPE scaling with a factor of 2.0, extend the context window to 65536 tokens, and apply a chunking strategy that segments sequences into 1792-token windows with 128-token overlap. 
The model is trained under 8 epochs with a learning rate $8\times10^{-6}$ and an effective batch size of 256 (per-device batch size of 1 with 64 gradient accumulation steps across 4 GPUs). Other training hyperparameters, hardware configuration are shown in Table~\ref{tab:sft_train_config}. We apply chat template formatting where only assistant responses are supervised (prompt tokens masked with -100). Training converges in approximately 8 hours after eight training epochs.

\begin{table}[h]
\small
\centering
\caption{Supervised Fine-Tuning Configuration and Resource Usage}
\label{tab:sft_train_config}
\begin{adjustbox}{max width=\textwidth}
\begin{tabular}{cc}
\begin{tabular}{lc}
\toprule
\multicolumn{2}{c}{\textbf{Training Hyperparameters}} \\
\midrule
Base Model & Qwen2.5-14B-Instruct \\
Training Samples & 6,000 \\
Train/Val Split & 95\% / 5\% \\
Epochs & 8 \\
Learning Rate & $8 \times 10^{-6}$ \\
LR Schedule & Cosine (5\% warmup) \\
Optimizer & AdamW \\
Batch Size (per-device) & 1 \\
Gradient Accumulation & 64 \\
\textbf{Effective Batch Size} & \textbf{256} \\
Weight Decay & 0.07 \\
Gradient Clipping & 1.0 \\
Max Sequence Length & 65,536 \\
Chunk Length & 1,792 \\
Chunk Overlap & 128 \\
RoPE Scaling & Dynamic (2.0$\times$) \\
Save Frequency & Every 100 steps \\
\bottomrule
\end{tabular}
&
\begin{tabular}{lc}
\toprule
\multicolumn{2}{c}{\textbf{Hardware \& Resources}} \\
\midrule
GPUs & 4$\times$ H100 (80GB) \\
per GPU Memory Usage & $\sim$60GB \\
System RAM & 256GB \\
Storage & 1TB \\
Training Time & $\sim$8 hours \\
\midrule
\multicolumn{2}{c}{\textbf{System Configuration}} \\
\midrule
Training Framework & HuggingFace TRL \\
Precision & BF16 mixed \\
Parallelization & DeepSpeed ZeRO-2 \\
Attention & Flash Attention 2 \\
Gradient Ckpt. & Enabled \\
Distributed Training & DDP (4 GPUs) \\
Logging & WandB \\
\bottomrule
\end{tabular}
\end{tabular}
\end{adjustbox}
\end{table}

\begin{figure*}[h]
  \centering
    \includegraphics[width=\linewidth]{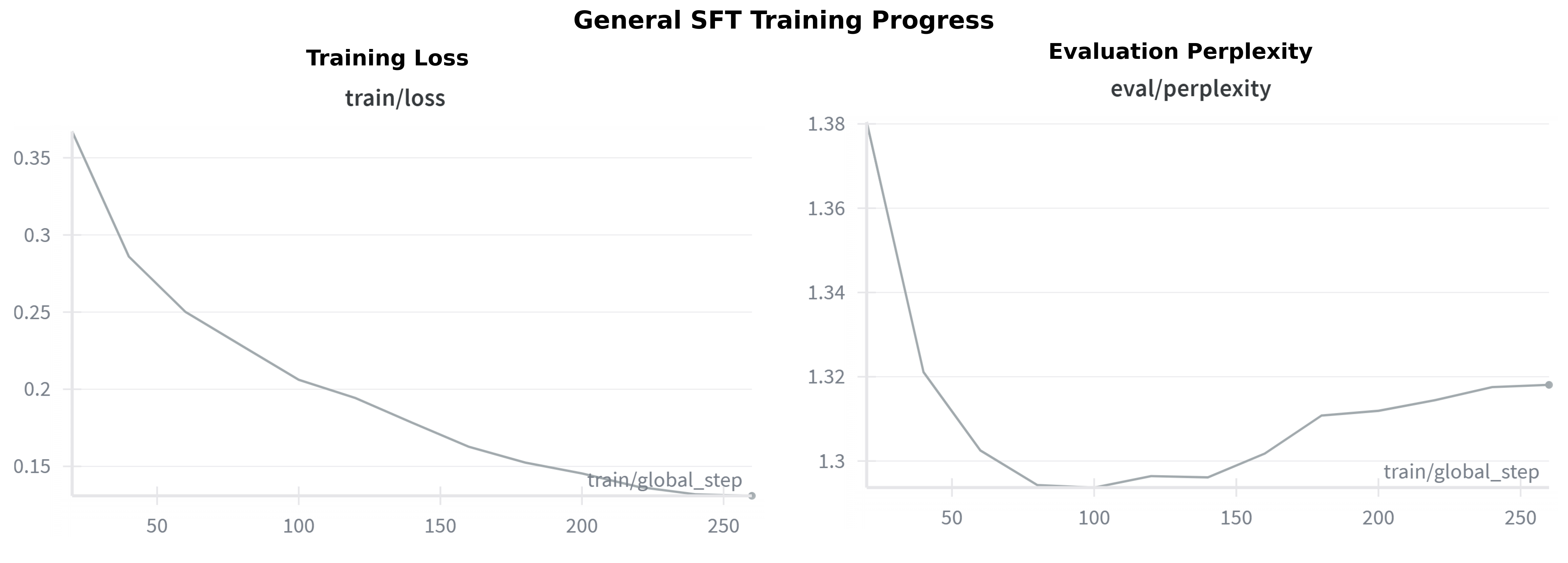}

  \caption{Visualization of SFT Training Curve}
  \label{fig:sft_train_curve}
  \vspace{-3mm}
\end{figure*}

\paragraph{$\bullet$ Training Curve of SFT}
In Fig.~\ref{fig:sft_train_curve}, we show the training curve of SFT for general WebUI. We finetune a Qwen 2.5 VL Instruct with the hyperparameter mentioned in Table~\ref{tab:sft_train_config}.

\clearpage

\paragraph{$\bullet$ VisRL Training and Implementaiton Detail}

\begin{figure}[h]
    \centering
    \includegraphics[width=0.99\textwidth]{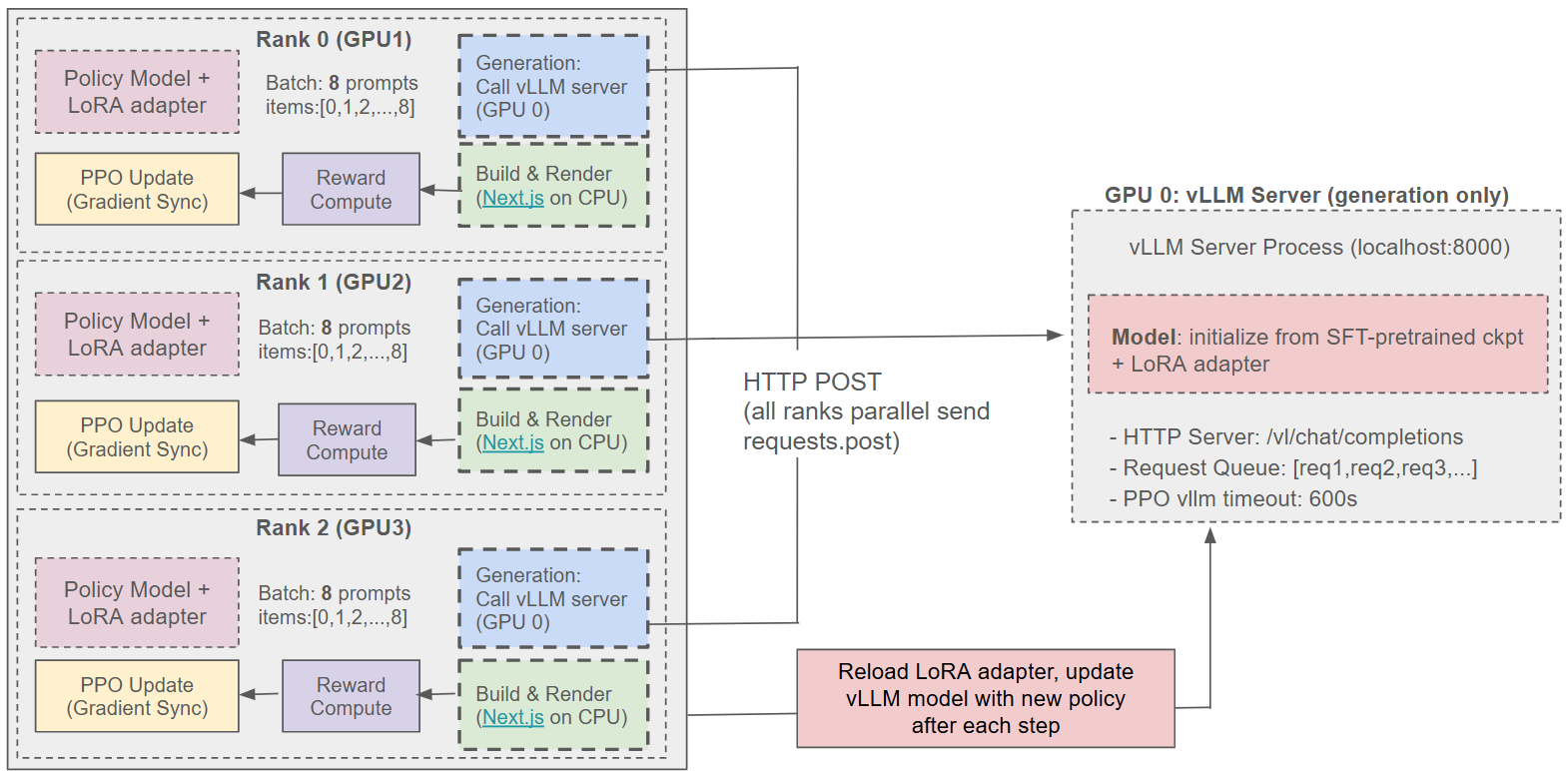}
    \caption{Training Infrastructure of VisRL.}
    \label{fig:visrl_infra_flow}
\end{figure}

After using SFT to train a cold-starting model, we start to use PPO with a hybrid on/off-policy approach that leverages vLLM for fast generation while maintaining near-on-policy learning through frequent adapter reloading.
Unlike prior SFT training, which distributes the model across 4 GPUs, we allocate 4 H100 GPUs: GPU-0 fully loads the model to serve as the vLLM inference server (generating code snippets), and other three GPUs run distributed PPO optimization via DeepSpeed ZeRO-2 for policy learning. 
Due to the memory constraint, we apply LoRA adapters to all attention and MLP projection layers for PPO tuning, where we set the rank-8 and $\alpha$-16 for general WebUI task. Since the 3D Simulation requires a higher learning capacity, we further set a larger rank-128 with $alpha$-256 for the 3D simulation task. To make the model loaded on GPU-0 synchronize with the on-policy model, we reload the latest adapter every training step to minimize off-policy lag. Training proceeds for 100 steps with a conservative learning rate of $2\times10^{-7}$, batch size is 8 per device (effective batch of 24 across 3 GPUs). In Fig.~\ref{fig:visrl_infra_flow}, we illustrate the training infrastructure of VisRL. In addition, we show the detailed training hyperparameters of VisRL in Table~\ref{tab:ppo_train_config}. Training completes in approximately 24-48 hours with checkpoints saved every step for reproducibility. In the later Sec.~\ref{app:d_1_training_curve}, we provide the training curve analysis for different PPO strategies.

\begin{table}[h]
\small
\centering
\caption{PPO Training Configuration and Resource Usage}
\label{tab:ppo_train_config}
\begin{adjustbox}{max width=\textwidth}
\begin{tabular}{cc}
\begin{tabular}{lc}
\toprule
\multicolumn{2}{c}{\textbf{Training Hyperparameters}} \\
\midrule
Base Model & SFT Qwen2.5-14B \\
Training Steps & 100 steps \\
LoRA Rank (General WebUI) & 8   \\
LoRA Alpha (General WebUI) & 16 \\
Trainable Params (General WebUI) & $\sim$4.2M (0.03\%) \\
LoRA Rank (3D WebUI) & 128   \\
LoRA Alpha (3D WebUI) & 256 \\
Trainable Params (3D WebUI) & $\sim$598M (4.3\%) \\
LoRA Dropout & 0.05 \\
LoRA Targets & q,k,v,o,gate,up,down \\
Learning Rate & $2 \times 10^{-7}$ \\
Batch Size (per-device) & 8 \\
\textbf{Effective Batch Size} & \textbf{24} (3 GPUs) \\
Mini-batch Size & 8 \\
PPO Clip Range & 0.05 \\
Entropy Coefficient & 0.005 \\
Value Loss Coefficient & 1.0 \\
Gradient Clipping & 1.0 \\
Advantage Clipping & $\pm$2.0 \\
Reward Normalization & Enabled \\
\midrule
\multicolumn{2}{c}{\textbf{Reward Components}} \\
\midrule
Vision Weight ($\alpha_v$) & 0.6 \\
UI-Agent Weight ($\alpha_u$) & 0.3 \\
Analyzer Weight ($\alpha_m$) & 0.1 \\
Invalid Penalty & $-0.2$ \\
External Image Penalty & $-0.05$ \\
Keyword Bonus & 0.20 \\
Auto-Fix Bonus & 0.3 \\
\midrule
\multicolumn{2}{c}{\textbf{Sampling Strategy}} \\
\midrule
Num Candidates & 3 (best-of-3) \\
Keep Top-K & K=1 \\
Generation Temp. & [0.7, 0.45, 0.3] \\
Max Gen Tokens & 7,000 \\
Max Context Tokens & 2,048 \\
\bottomrule
\end{tabular}
&
\begin{tabular}{lc}
\toprule
\multicolumn{2}{c}{\textbf{Hardware \& Resources}} \\
\midrule
Total GPUs & 4$\times$ H100 (80GB) \\
Training GPUs & 3$\times$ H100 (GPUs 1-3) \\
Inference GPU & 1$\times$ H100 (GPU 0) \\
GPU Memory (training) & $\sim$70GB/device \\
GPU Memory (vLLM) & $\sim$75GB \\
System RAM & 256GB \\
Storage & 1TB \\
Training Time & $\sim$24-48 hours \\
Time per Step & 15-30 minutes \\
\midrule
\multicolumn{2}{c}{\textbf{System Configuration}} \\
\midrule
RL Framework & TRL + Custom PPO \\
Inference Engine & vLLM (version 0.6+) \\
Precision & BF16 mixed \\
Parallelization & DeepSpeed ZeRO-2 \\
Distributed Strategy & DDP (3 GPUs) \\
Adapter Method & LoRA \\
Adapter Reload Freq. & Every 1 step \\
\midrule
\multicolumn{2}{c}{\textbf{Timeouts \& Limits for Execution Harness}} \\
\midrule
npm Install & 300s \\
Next.js Build & 360s \\
Dev Server Load & 60s \\
vLLM Request & 600s \\
Step Timeout & 7,200s (2 hours) \\
\bottomrule
\end{tabular}
\end{tabular}
\end{adjustbox}
\end{table}

\clearpage

\paragraph{$\bullet$ Training Infrastructure of VisRL}

We provide the implementation flow of the training infrastructure to ensure reproducibility in Fig.~\ref{fig:visrl_infra_flow}. 

\subsection{Sequence-Level Rollout and Probability}
\label{app:c_1_seq_ppo}

Given a prompt $p$, the policy generates a complete code sequence $c=\{c_t\}_{t=1}^{T}$.
We compute sequence log-probability via the chain rule:
\begin{equation}
\log \pi_\theta(c\mid p) \;=\; \sum_{t=1}^{T} \log \pi_\theta(c_t \mid c_{<t}, p).
\end{equation}
The PPO importance ratio is computed at the sequence level:
\begin{equation}
\rho(\theta)=\exp\!\left(\log \pi_\theta(c\mid p)-\log \pi_{\theta_{\text{old}}}(c\mid p)\right).
\end{equation}

\paragraph{Advantage and Loss Terms}

Because the environment returns a single outcome reward after execution, we use a prompt-conditioned baseline $V_\phi(c,p)$ and compute one advantage per sequence:
\begin{equation}
A(p,c) = r(c,p)-V_\phi(c,p).
\end{equation}
We normalize advantages within each batch $\mathcal{B}$:
\begin{equation}
\hat{A}(c,p)=\frac{A(c,p)-\mu_{\mathcal{B}}}{\sigma_{\mathcal{B}}+\epsilon_{\text{norm}}}.
\end{equation}
The clipped PPO objective is applied once per sequence:
\begin{equation}
\mathcal{L}^{\text{CLIP}}(\theta)=
\mathbb{E}_{(c,p)\sim\mathcal{B}}
\Big[
\min\big(\rho(\theta)\hat{A}(c,p),\;\text{clip}(\rho(\theta),1-\epsilon,1+\epsilon)\hat{A}(c,p)\big)
\Big].
\end{equation}
We optimize the standard combined loss:
\begin{equation}
\mathcal{L}(\theta,\phi)=
-\mathcal{L}^{\text{CLIP}}(\theta)
+ c_v\,\mathbb{E}_{(c,p)\sim\mathcal{B}}\big[(V_\phi(c,p)-r(c,p))^2\big]
- c_e\,\mathcal{H}[\pi_\theta],
\end{equation}
where $\mathcal{H}[\pi_\theta]$ is token-entropy computed from the model’s next-token distributions to encourage exploration.

\subsection{Reward Definition}
\label{app:c_2_reward_definition}

We compute the outcome reward only after generating $c$, building/running it, and rendering a screenshot $\mathcal{I}(c)$. The total reward aggregates complementary signals:
\begin{equation}
r(c,p) \;=\; \alpha_m\, r_{\text{metrics}}(c)
\;+\; \alpha_v\, r_{\text{vision}}(\mathcal{I}(c))
\;+\; \alpha_u\, r_{\text{UI agent}}(c,\mathcal{I}(c))
\;+\; r_{\text{penalty}}(c).
\end{equation}

\paragraph{Vision score $r_{\text{vision}} \in [0,1]$.}
A vision evaluator scores the rendered UI for overall layout quality and visual polish (e.g., hierarchy, spacing, typography, color coherence). This provides dense feedback even when task success is hard to verify.

\paragraph{UI-agent interaction reward $r_{\text{UI agent}}\ge 0$.}
A UI agent performs lightweight interaction tests (e.g., form filling, clicking, navigation) on the rendered page and returns a success rate. This signal directly captures functional correctness beyond static appearance.

\paragraph{Code metrics bonus $r_{\text{metrics}}\ge 0$.}
We award small bonuses for richer, more realistic implementations, such as non-trivial component structure and interactive behaviors (state/event handling). This discourages overly minimal outputs that ``look okay'' but are not functional.


\subsection{Additional Learning Strategies}
\label{app:c_3_additional_learning}

\paragraph{Multi-Temperature Candidate Sampling and Selection (Best-of-$N$)}

For each prompt, we generate a small set of candidates with different temperatures to balance exploration and quality:
\begin{equation}
c_i \sim \pi_{\theta_{\text{old}}}(\cdot\mid p,\tau_i),
\quad \tau_i \in \{\tau_{\text{low}}, \tau_{\text{high}}\}, \quad i=1,\ldots,N.
\end{equation}
We execute all candidates and keep the best one under a minimum-quality constraint:
\begin{equation}
c^* = \arg\max_{c_i} r(c_i,p)
\quad \text{s.t. } r(c_i,p)\ge r_{\min}.
\end{equation}
This increases the diversity of UI Generation, and also preserves the good enough samples such that the model can learn to generate more advanced UI templates without forgetting the old format of template it already learn before.

\paragraph{Fast-gate pre-build filter.}
Before expensive build/render, we skip candidates that are almost surely invalid (e.g., missing \texttt{export default}, clearly malformed structure, excessive length of code that highly possible truncated). This saves compute by avoiding costly failures.


\subsection{Pseudo Algorithm for Training Procedure}
\label{app:c_4_pseudo_algo}

\begin{algorithm}[h]
\caption{Sequence-Level PPO Training for WebUI Code Generation}
\label{alg:seqppo-summary}
\begin{algorithmic}[1]
\REQUIRE Prompts $\mathcal{D}$, initial policy $\pi_{\theta}$, value $V_\phi$, temperatures $\{\tau_i\}_{i=1}^N$
\FOR{each PPO iteration}
    \STATE Sample a batch of prompts $\mathcal{B}\subset\mathcal{D}$
    \FOR{each prompt $p\in\mathcal{B}$}
        \STATE Sample candidates $\{c_i\}_{i=1}^N$ with temperatures $\{\tau_i\}$
        \STATE Apply fast-gate filter; build, run, and render remaining candidates
        \STATE Compute rewards $r(c_i,p)$
        \STATE Select $c^*=\arg\max_i r(c_i,p)$ (optionally thresholded)
        \STATE Store $(p,c^*,r^*)$ for PPO update
    \ENDFOR
    \STATE Compute advantages $A(p,c)=r(c,p)-V_\phi(p)$ and normalize within batch
    \STATE Update $\theta,\phi$ using clipped PPO objective and value regression
\ENDFOR
\end{algorithmic}
\end{algorithm}

\clearpage

\subsection{UI Agent Task Execution in VisRL}
\label{app:c_5_uiagent_exec}
We integrate a vision-grounded UI agent that tests functional correctness by proposing and executing interactions (clicks, form fills) on rendered UIs, computing a success rate that contributes to the composite reward with weight $\alpha_u = 0.3$ alongside vision quality and code metrics. We describe every step as follows:

\begin{tcolorbox}[
  title={UI-Agent Task Execution (Playwright + Vision)},
  colback=cyan!4!white,
  colframe=cyan!75!black,
  coltitle=white,
  colbacktitle=cyan!75!black,
  boxrule=0.8pt,
  arc=3pt,
  left=6pt,right=6pt,top=6pt,bottom=6pt,
  fonttitle=\bfseries\small,
  fontupper=\ttfamily\footnotesize,
  enhanced,
  breakable
]

\textbf{Step 1 — Element Detection (DOM Scanning).}
Enumerate interactive elements via JS injection and assign stable element IDs.
\begin{itemize}[leftmargin=1.6em, itemsep=0.15em, topsep=0.15em]
  \item Candidate tags: \texttt{INPUT}, \texttt{TEXTAREA}, \texttt{SELECT}, \texttt{BUTTON}, \texttt{A}
  \item Also include nodes with \texttt{onclick} or \texttt{cursor:pointer}
  \item Filtering: remove nested clickables; enforce minimum area threshold
  \item Visualization: render a labeled screenshot with red dashed borders and IDs
\end{itemize}

\smallskip
\textbf{Scanning output (example):}
\begin{codebox}
\begin{lstlisting}
[0]: <input>  "Product Name"
[1]: <input>  "Price"
[2]: <button> "Submit"
\end{lstlisting}
\end{codebox}

\smallskip
\textbf{Step 2 — Vision-Guided Task Proposal.}
Send the labeled screenshot + element list to the vision model to propose executable test actions.

\begin{promptbox}
\textbf{Prompt to vision model:}
\begin{lstlisting}
Available elements:
[0]: <input>  "Product Name"
[1]: <input>  "Price"
[2]: <button> "Submit"

Suggest actions to test UI functionality.
Return a JSON action list with element_id and (optional) value.
\end{lstlisting}
\end{promptbox}

\smallskip
\textbf{Returned task proposal (example JSON):}
\begin{codebox}
\begin{lstlisting}
{
  "actions": [
    {"action": "fill",  "element_id": 0, "value": "Test Product"},
    {"action": "fill",  "element_id": 1, "value": "19.99"},
    {"action": "click", "element_id": 2}
  ]
}
\end{lstlisting}
\end{codebox}

\smallskip
\textbf{Step 3 — Action Execution by Element ID.}
Execute the proposed actions using stored DOM references:
\begin{itemize}[leftmargin=1.6em, itemsep=0.15em, topsep=0.15em]
  \item \textbf{Fill:} \texttt{INPUT}/\texttt{TEXTAREA} (\texttt{type()} + input event)
  \item \textbf{Select:} \texttt{SELECT} (match by text/value + change event)
  \item \textbf{Click:} \texttt{BUTTON}/\texttt{A} (check for navigation / DOM change)
\end{itemize}

\smallskip
\textbf{Step 4 — Multi-Strategy Verification.}
Determine success/failure using observable UI signals:
\begin{itemize}[leftmargin=1.6em, itemsep=0.15em, topsep=0.15em]
  \item Success if expected text appears, key elements change, or DOM count updates
  \item Failure if error keywords appear, exceptions occur, or no observable change is detected
\end{itemize}

\smallskip
\textbf{Step 5 — UI-Agent Scoring.}
Compute per-task success rate as:
\[
\text{score} = \frac{\text{successful\_actions}}{\text{total\_actions}} .
\]


\end{tcolorbox}
\label{fig:uiagent_rubric}

\clearpage

\subsection{Vision Agent -- Evaluation Rubric}
\label{app:c_6_visionrubric}
The vision evaluator assesses UI quality and the visual presence of interactive UI components (buttons, forms, charts, modals) based on screenshot analysis, providing a proxy signal for interface complexity. Complementary to this, the UI-Agent performs functional verification by executing interactions and measuring actual success rates, ensuring the model learns both visual appeal and operational correctness." We provide the rubric detail as follows:

\begin{tcolorbox}[
  title={Vision-Agent Rubric (0--5)},
  colback=cyan!5!white,
  colframe=cyan!75!black,
  coltitle=white,
  colbacktitle=cyan!75!black,
  boxrule=0.8pt,
  arc=3pt,
  left=6pt,right=6pt,top=6pt,bottom=6pt,
  fonttitle=\bfseries\small,
  fontupper=\ttfamily\footnotesize
]

\textbf{Step 0: Hard-gate failures (Score = 0).}
Assign \texttt{0} if any of the following occur:
\begin{itemize}[leftmargin=1.6em, itemsep=0.15em, topsep=0.15em]
  \item Build/runtime error messages are visible.
  \item Completely blank/white page (no rendered UI).
  \item Browser/system error page (e.g., navigation failure).
\end{itemize}

\smallskip
\textbf{Step 1: Base score (0.0--4.0) = Design (0--2) + Interactivity (0--2).}

\textbf{A) Visual design \& complexity (0--2).}
\begin{itemize}[leftmargin=1.6em, itemsep=0.1em, topsep=0.1em]
  \item \texttt{0.0--0.5}: Plain/text-only layout; minimal styling.
  \item \texttt{0.5--1.0}: Basic styled components (color, spacing) with simple layout.
  \item \texttt{1.0--1.5}: Clear hierarchy, multiple sections/cards, consistent typography and spacing.
  \item \texttt{1.5--2.0}: Polished UI with advanced styling (shadows/gradients), cohesive theme, animations where appropriate.
\end{itemize}

\textbf{B) Visual Presence of UI Components.(0--2).}
\begin{itemize}[leftmargin=1.6em, itemsep=0.1em, topsep=0.1em]
  \item \texttt{0.0--0.5}: Static content only; no executable UI components interaction.
  \item \texttt{0.5--1.0}: Basic UI components(e.g., 1--3 buttons/inputs) with simple layout.
  \item \texttt{1.0--1.5}: Multiple UI sections (4-8 components) with clear visual organization and variety.
  \item \texttt{1.5--2.0}: Rich component ecosystem (tables, charts, tabs, filters, modals, cards) with sophisticated visual structure.
\end{itemize}

\smallskip
\textbf{Step 2: Excellence tier (4.0--5.0).}
After computing the base score, assign a bonus within \texttt{[0, 1]} based on overall polish:
\begin{itemize}[leftmargin=1.6em, itemsep=0.15em, topsep=0.15em]
  \item \texttt{4.0--4.3}: \textbf{Good UI} --- professional palette, consistent spacing, feature-complete.
  \item \texttt{4.3--4.6}: \textbf{Outstanding UI} --- smooth transitions, advanced components, production-ready details.
  \item \texttt{4.6--5.0}: \textbf{Near-perfect UI (rare)} --- exceptional hierarchy, cohesive design language, flawless execution.
\end{itemize}

\smallskip
\textbf{Step 3: Mandatory deductions (apply after Steps 1--2).}
\begin{itemize}[leftmargin=1.6em, itemsep=0.15em, topsep=0.15em]
  \item Missing buttons/links: \texttt{-0.3}
  \item Inconsistent spacing or misalignment: \texttt{-0.3}
  \item Broken images / missing assets: \texttt{-0.3}
\end{itemize}

\end{tcolorbox}

\clearpage

\section{VisRL: Training Reward and Hyparameter Analyses}
\label{app:d_main_training_analysis}

\subsection{Training Curve for Different PPO Strategies}
\label{app:d_1_training_curve}

\begin{figure*}[h]
  \centering

  \begin{subfigure}[t]{0.65\textwidth}
    \centering
    \includegraphics[width=\linewidth]{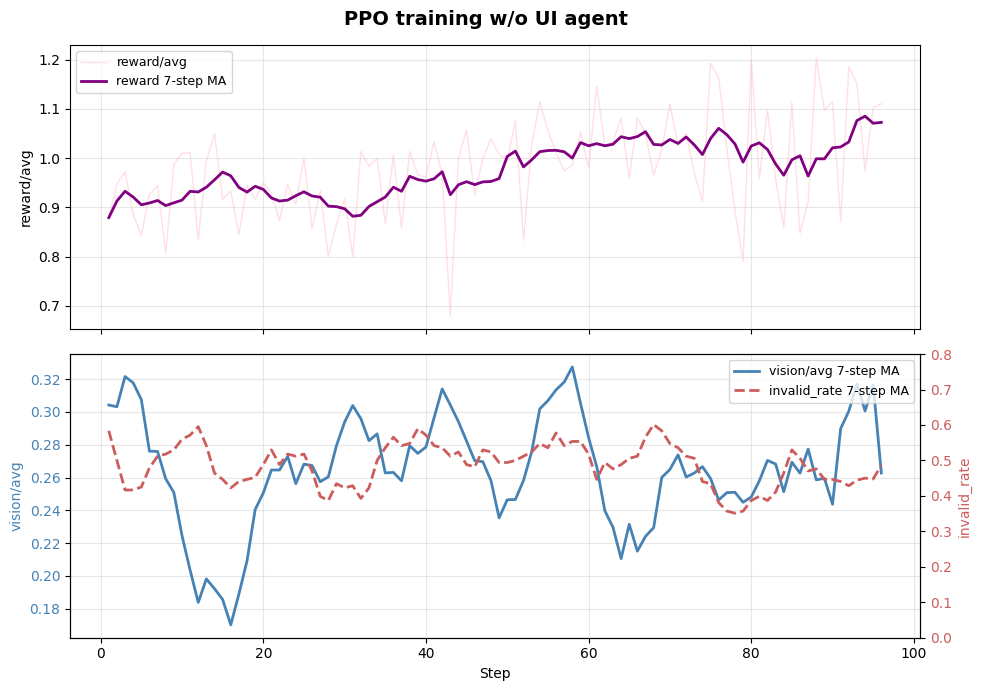}
    \caption{}
    \label{fig:ppo:no-ui}
  \end{subfigure}

  \begin{subfigure}[t]{0.65\textwidth}
    \centering
    \includegraphics[width=\linewidth]{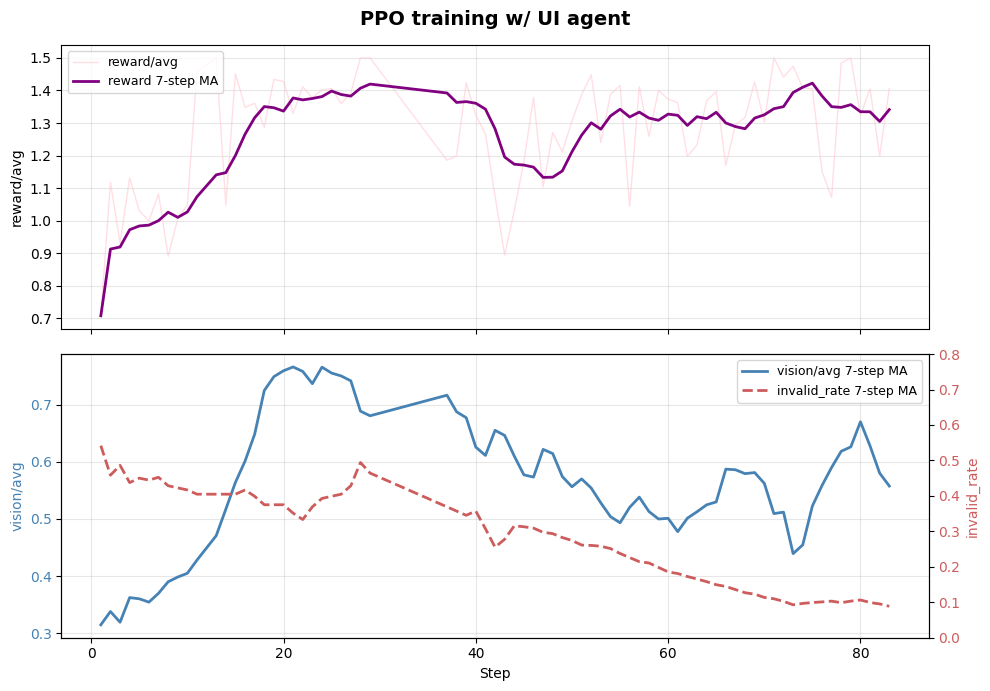}
    \caption{}
    \label{fig:ppo:with-ui}
  \end{subfigure}

  \caption{Comparison of PPO training dynamics with and without a UI agent. Top: average reward (with 7-step moving average). Bottom: vision score and invalid action rate.}
  \label{fig:ppo-learning-curve}
\end{figure*}

Fig.~\ref{fig:ppo-learning-curve} shows that incorporating the UI agent substantially improves PPO training stability and efficiency. In Fig.~\ref{fig:ppo-learning-curve} a, we show the training curve for PPO learning w/ vision agent only. The average reward increases only gradually and remains noisy; the vision score fluctuates with no clear upward trend, and the invalid action rate stays relatively high throughout training. In contrast, VisRL adopts the UI agent (Fig.~\ref{fig:ppo-learning-curve} b), the reward rises sharply in the early phase and maintains a consistently higher level. This improvement is accompanied by a strong increase in vision score and a steady reduction in invalid actions, indicating that agent-based interaction feedback provides a cleaner learning signal and guides the policy toward more executable, UI-consistent code.

\clearpage

\subsection{Reward Decomposition Analysis}
\label{app:d_2_reward_decompose_analysis}

\begin{figure*}[h]
  \centering

  \begin{subfigure}[t]{0.49\textwidth}
    \centering
    \includegraphics[width=\linewidth]{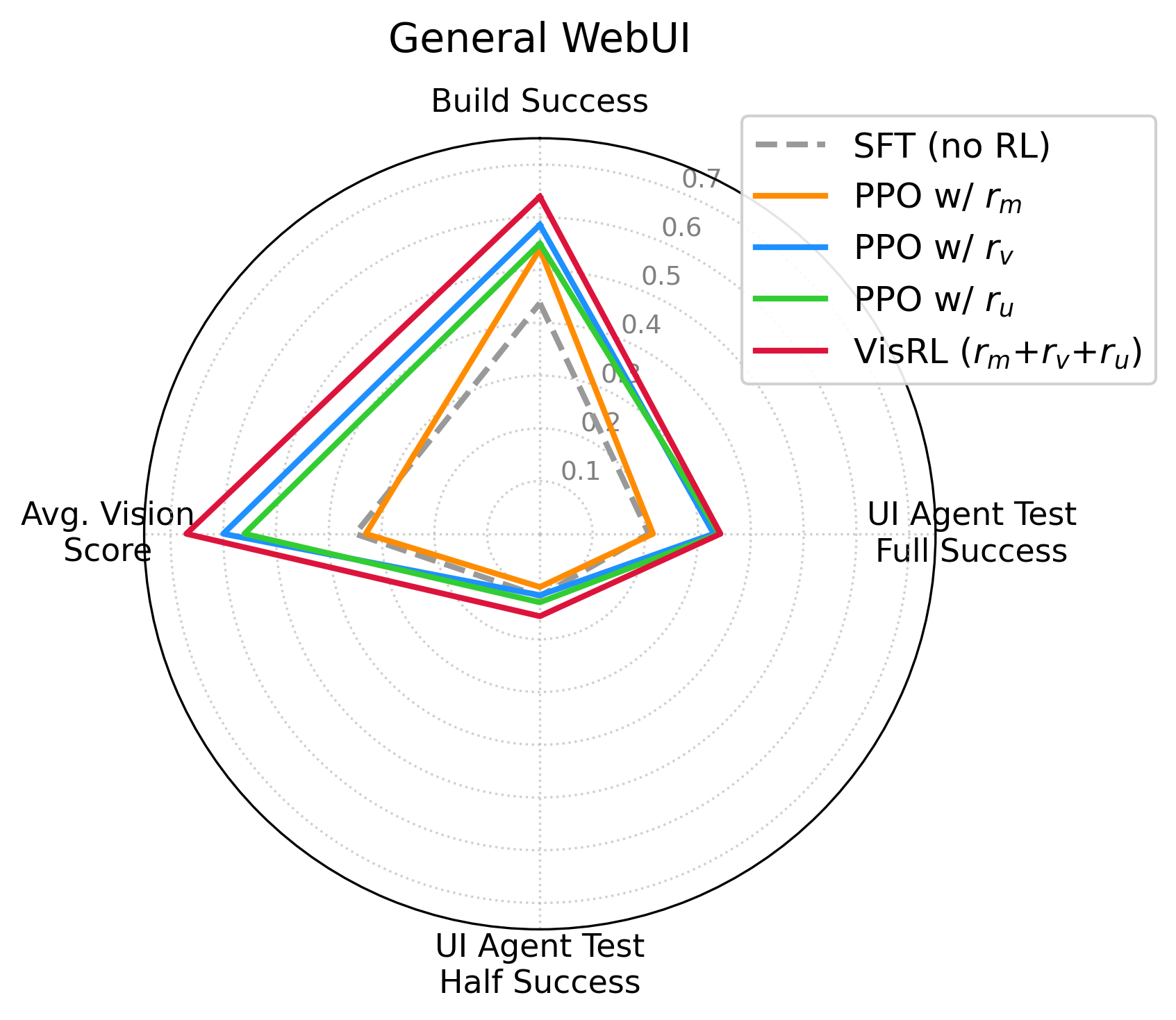}
    \label{fig:generalui-reward}
  \end{subfigure}
  \hfill
  \begin{subfigure}[t]{0.49\textwidth}
    \centering
    \includegraphics[width=\linewidth]{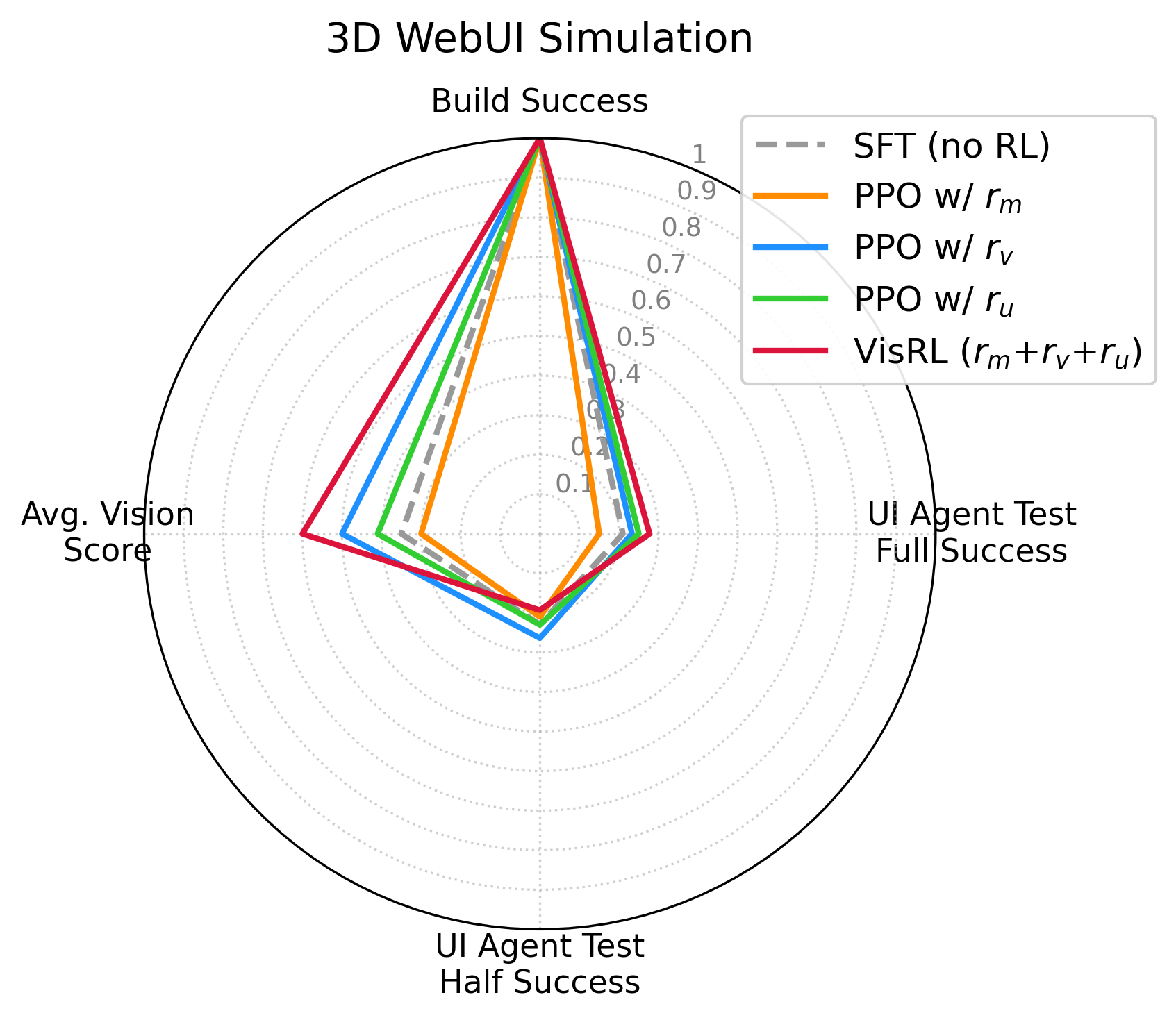}
    \label{fig:sim3d-reward}
  \end{subfigure}

  \caption{Radar plot for reward decompsition analysis}
  \label{fig:reward_decompose}
  \vspace{-3mm}
\end{figure*}

To better understand the effectiveness of each reward component, we present a reward decomposition analysis. Figure~\ref{fig:reward_decompose} shows a radar plot analysis comparing reward components across different training strategies for both General WebUI and 3D WebUI Simulation tasks. We evaluate five metrics: Build Success (compilation rate), Average Vision Score (aesthetic quality from GPT-4o-mini), UI Agent Test Full Success (all proposed interactions succeed), UI Agent Test Half Success (≥50\% interactions succeed). The analysis compares: (1) SFT baseline without RL (\textcolor{gray}{gray}), (2) PPO with analyzer-only reward ($r_m$, \textcolor{orange}{orange}), (3) PPO with vision-only reward ($r_v$, \textcolor{blue}{blue}), (4) PPO with UI-agent-only reward ($r_u$, \textcolor{green}{green}), and (5) VisRL with the full composite reward ($r_m + r_v + r_u$,\textcolor{red}{red}). 
Results demonstrate that single-component rewards lead to unbalanced optimization: vision-only PPO achieves high aesthetic scores but poor functional correctness, UI-agent-only PPO maximizes interaction success but produces visually simplistic interfaces, and analyzer-only PPO maintains build success but lacks both visual appeal and interactivity. In contrast, VisRL with the composite reward ($\alpha_v=0.6, \alpha_u=0.3, \alpha_m=0.1$) achieves the most balanced performance, excelling across all dimensions simultaneously. Notably, for 3D simulation tasks, the balanced reward maintains high build success (>0.9) while significantly improving both vision scores and UI agent test success rates compared to single-component baselines, validating the necessity of multi-component reward design for holistic web UI generation quality.

\clearpage
\subsection{Reward Weight Analysis -- $\alpha$ Sensitivity}
\label{app:d_3_alpha_sensitivity_analysis}
\begin{figure}[h]
    \centering
    \includegraphics[width=0.70\linewidth]{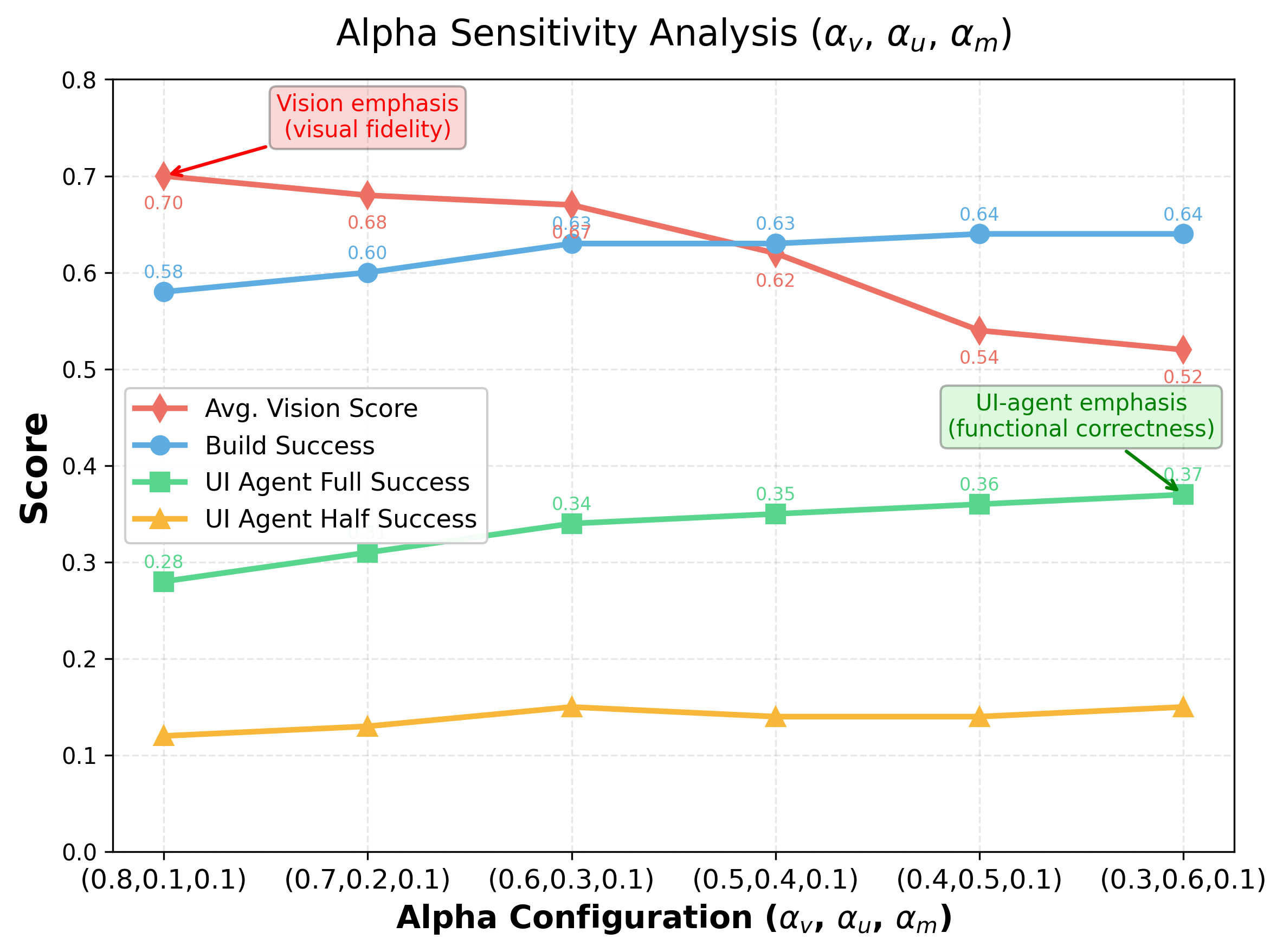}
    \caption{Sensitivity analysis of $\alpha$ Configuration}
    \label{fig:alpha_sensitivity}
    \vspace{-2mm}
\end{figure}

In Fig.~\ref{fig:alpha_sensitivity}, we study the sensitivity of VisRL to the composite reward weights
$(\alpha_v,\alpha_u,\alpha_m)$ by sweeping $\alpha_v$ (vision) and $\alpha_u$ (UI-agent).
We keep the analyzer weight fixed ($\alpha_m=0.1$), and evaluate four outcome metrics, including
average vision score, build success, UI-agent full success, and UI-agent half success.

The different combination of $\alpha$ shows a clear \emph{visual quality-functionality} trade-off. When placing more weight on the vision term yields higher
visual scores (e.g., $\sim$0.70 at $(0.8,0.1,0.1)$), and shifting weight toward the UI-agent term increases functional
correctness (UI-agent full success rises from $\sim$0.28 to $\sim$0.37). 
Notably, build success also improves with
larger $\alpha_u$ (from $\sim$0.58 to $\sim$0.64), suggesting that emphasizing executable, interaction-grounded rewards
encourages more compilable implementations 
rather than just changing surface-level UI appearance.

As a result, we set the weight at the point $\alpha=[0.6, 0.3, 0.1]$ that preserves strong visual
quality while maintaining build success and performance gains in UI-agent-testing correctness. 

\clearpage

\section{Extended Results and Diagnostics}
\label{app:e_main_extended_results}

\subsection{Additional Experimental Results (per-Test Category Results)}

\label{app:e_1_additional_results}

In Sec.~\ref{app:b_1-5_uicategory}, we provide the test case distribution of WebUIProof , including functionality testing, data display testing, UI design validation testing. Here, we present the per-test category results for both proprietary API models and our self-trained models. For each model, we show the No/Partial/Full completion scores and accuracy by test category (test-category distribution shown in Fig.~\ref{fig:webui-testcase-distribution}). For example, the accuracy of functionality testing (FT) for a certain model is computed as: 

\begin{equation}
\frac{\#\text{Full completion in FT} + 0.5*\#\text{Partial completion in FT}}{\text{Total}  \#\text{FT}}
\end{equation}
In Table~\ref{tab:cat_acc_generalui_apimodel} and ~\ref{tab:cat_acc_generalui_selftrained}, we show the category accuracy on the general WebUI task for proprietary API and self-trained models. In Table~\ref{tab:cat_acc_sim3d_apimodel} and ~\ref{tab:cat_acc_sim3d_selftrained}, we show the category accuracy on the 3D WebUI task.

\begin{table}[h]
\caption{Category accuracy of General WebUI task on Proprietary API.}
\centering
\small
\setlength{\tabcolsep}{2pt} 
\renewcommand{\arraystretch}{1} 
\begin{tabular}{lclcccc}
\toprule
Model Name                       & \multicolumn{1}{c}{\begin{tabular}[c]{@{}c@{}}Build Failure\\ Rate (\%)\end{tabular}} & \begin{tabular}[c]{@{}c@{}}Test \\ Category\end{tabular} & \multicolumn{3}{c}{\begin{tabular}[c]{@{}c@{}}Build Sucess\\ UI-Agent Verified Task (\%)\end{tabular}} & \multicolumn{1}{c}{\begin{tabular}[c]{@{}c@{}}Accuracy \\ (Cat. \%)\end{tabular}} \\ \cmidrule{4-6}
                                 &                                                                                  &                                                          & \multicolumn{1}{c}{\begin{tabular}[c]{@{}c@{}}No \\ Completion\end{tabular}}                          & \multicolumn{1}{c}{\begin{tabular}[c]{@{}c@{}}Partial \\ Completion\end{tabular}}                           & \multicolumn{1}{c}{\begin{tabular}[c]{@{}c@{}}Full \\ Completion\end{tabular}}                               &                                                                                    \\ \hline
\multirow{3}{*}{Claude Sonnet 4~\cite{Anthropic2025Sonnet4}} &                                                                          & Functionality Testing                                          &    10.15                             &   23.86                             &  30.28                               & 42.21                                                                                   \\
                                 &    30.37                                                                              & Data Display Testing                                            &   15.04                              &       15.76                         &    43.41                            &    51.29                                                                                \\
                                 &                                                                                  & UI Design Validation                                        &  18.52                               &   29.63                             &  37.03                              &    51.85                                                                                \\ \midrule
\multirow{3}{*}{Gemini 2.5 Pro~\cite{GoogleDeepMind2025Gemini2_5Pro}}                   &                                                                                  & Functionality Testing                                          &      16.15                                &   19.44                       &    22.47                            &    32.19                                                                              \\
                                 &                                                                  32.21                & Data Display Testing                                            &  12.89                              &   17.19                           &   51.58                             &  60.17                                                                                  \\
                                 &                                                                                  & UI Design Validation                                        &  14.81                               &    33.33                            &   35.19                             &   51.85                                                                                 \\ \midrule
\multirow{3}{*}{Qwen 3 Coder 480B~\cite{Alibaba2025Qwen3Coder}}                     &                                                                                  & Functionality Testing                                          &   15.78                              &   21.57                             &    21.77                            &     32.56                                                                               \\
                                 &                                                               33.05                   & Data Display Testing                                            &    12.89                             &   14.32                             &  50.14                              &   57.31                                                                                 \\
                                 &                                                                                 & UI Design Validation                                        &  16.67                               &   33.33                             &    33.33                            &   50.00                                                                                 \\ \midrule
\multirow{3}{*}{DeepSeek R1~\cite{DeepSeek2025R1}}                      &                                                                                  & Functionality Testing                                          &    15.96                             &           21.14                    &         26.24                       &      36.81 \\
                                 &                                                                 32.89                & Data Display Testing                                            &     14.32                            & 17.19                               &     42.98                           &      51.58                                                                              \\
                                 &                                                                                  & UI Design Validation                                        &         11.11                        & 25.93                               &      29.63                          &       42.59                                                                             \\ \midrule
\multirow{3}{*}{GPT 4.1~\cite{OpenAI2025GPT4_1}}                          &                                                                                  & Functionality Testing                                          &     7.48                           &    17.84                          &      29.96                          &    38.88                                                                                \\
                                 &                                                                36.58                 & Data Display Testing                                            &        18.62                         &   21.49                             &       35.82                         &    46.56                                                                                \\
                                 &                                                                                  & UI Design Validation                                        &   14.81                              &  29.63                              &     25.93                           &    40.74                                                                                \\ \midrule
\multirow{3}{*}{Llama 4 Maverick~\cite{MetaAI2025Llama4Maverick}}                         &                                                                                  & Functionality Testing                                          &     1.39                            &  15.87                              &        30.97                        &       38.91                                                                            \\
                                 &                                                              46.64                   & Data Display Testing                                            &   4.30                              &            24.36                    &       34.38                         &   46.56                                                                                 \\
                                 &                                                                                  & UI Design Validation                                        &    3.70                             &            18.52                    &     25.93                           &           35.19                                                                         \\ \midrule
\multirow{3}{*}{Kimi K2~\cite{MoonshotAI2025KimiK2}}                         &                                                                                  & Functionality Testing                                          &    2.39                             &   8.67                             &   34.99                             &      39.33                                                                              \\
                                 &    46.31                                                                              & Data Display Testing                                            &      10.02                           &          8.59                      &     35.82                           &   40.11                                                                                 \\
                                 &                                                                                  & UI Design Validation                                        &    37.04                             &           37.04                     &    18.52                            &     37.04                                                                               \\ \midrule
\multirow{3}{*}{Llama 3 405B~\cite{MetaAI2024Llama3_405B}}                     &                                                                                  & Functionality Testing                                          &      3.37                           &   15.45                             &               32.17                 &    39.89                                                                                \\
                                 &   47.32                                                                               & Data Display Testing                                            &   5.73                              &            14.33                    &     27.22                           &      34.38                                                                              \\
                                 &                                                                                  & UI Design Validation                                        &  55.56           & 7.40           & 18.51       &    22.22                                                                                \\ \bottomrule
\end{tabular}
\label{tab:cat_acc_generalui_apimodel}
\end{table}

\begin{table}[h]
\caption{Category accuracy of General WebUI task on Self-Trained Model.}
\centering
\small
\setlength{\tabcolsep}{2pt} 
\renewcommand{\arraystretch}{1} 
\begin{tabular}{lclcccc}
\toprule
\multicolumn{1}{l}{\begin{tabular}[l]{@{}l@{}}Training Variants of\\ Qwen 2.5 14B Instruct \end{tabular}}                     & \multicolumn{1}{c}{\begin{tabular}[c]{@{}c@{}}Build Failure\\ Rate (\%)\end{tabular}} & \begin{tabular}[c]{@{}c@{}}Test \\ Category\end{tabular} & \multicolumn{3}{c}{\begin{tabular}[c]{@{}c@{}}Build Sucess\\ UI-Agent Verified Task (\%)\end{tabular}} & \multicolumn{1}{c}{\begin{tabular}[c]{@{}c@{}}Accuracy \\ (Cat. \%)\end{tabular}} \\ \cmidrule{4-6}
                                 &                                                                                  &                                                          & \multicolumn{1}{c}{\begin{tabular}[c]{@{}c@{}}No \\ Completion\end{tabular}}                          & \multicolumn{1}{c}{\begin{tabular}[c]{@{}c@{}}Partial \\ Completion\end{tabular}}                           & \multicolumn{1}{c}{\begin{tabular}[c]{@{}c@{}}Full \\ Completion\end{tabular}}                                    &                                                                                    \\ \hline
\multirow{3}{*}{Baseline SFT} &                                                                                  & Functionality Testing                                          & 7.75                               &      10.71                 &      20.66                     &     26.01                                                                               \\
                                 &                                                                 56.25                 & Data Display Testing                                            &  14.33                               &           14.33                     &    21.49                            &    28.65                                                                                \\
                                 &                                                                                  & UI Design Validation                                        &  18.52                               &                11.11                &      18.52                          &       24.07                                                                             \\ \midrule
\multirow{3}{*}{PPO W/ Vision agent}                   &                                                                                  & Functionality Testing                                          &   12.87                              &    11.05                            &   34.38                             &     39.90                                                                               \\
                                 &     41.42                                                                             & Data Display Testing                                            &    14.33                             &                7.30                &   28.65                             &    32.31                                                                                \\
                                 &                                                                                  & UI Design Validation                                        &     18.52                            &              25.93                  &    37.04                            &      50.00                                                                              \\ \midrule
\multirow{3}{*}{VisRL W/ UI agent}                     &                                                                                  & Functionality Testing                                          &    11.70                             &         17.08                       &               35.80                 &        44.34                                                                            \\
                                 &     36.10                                                                             & Data Display Testing                                            &  15.76                               &         12.89                       &    28.65                            &    35.10                                                                                \\
                                 &                                                                                  & UI Design Validation                                        &  18.52                               &            11.11                    &       40.74                         &     46.30                                                                               \\ \bottomrule
\end{tabular}
\label{tab:cat_acc_generalui_selftrained}
\end{table}

\begin{table}[h]
\caption{Category accuracy of 3D WebUI task on Proprietary API.}
\centering
\small
\setlength{\tabcolsep}{2pt} 
\renewcommand{\arraystretch}{1} 
\begin{tabular}{lclcccc}
\toprule
Model Name                       & \multicolumn{1}{c}{\begin{tabular}[c]{@{}c@{}}Build Failure\\ Rate (\%)\end{tabular}} & \begin{tabular}[c]{@{}c@{}}Test \\ Category\end{tabular} & \multicolumn{3}{c}{\begin{tabular}[c]{@{}c@{}}Build Sucess\\ UI-Agent Verified Task (\%)\end{tabular}} & \multicolumn{1}{c}{\begin{tabular}[c]{@{}c@{}}Accuracy \\ (Cat. \%)\end{tabular}} \\ \cmidrule{4-6}
                                 &                                                                                  &                                                          & \multicolumn{1}{c}{\begin{tabular}[c]{@{}c@{}}No \\ Completion\end{tabular}}                          & \multicolumn{1}{c}{\begin{tabular}[c]{@{}c@{}}Partial \\ Completion\end{tabular}}                           & \multicolumn{1}{c}{\begin{tabular}[c]{@{}c@{}}Full \\ Completion\end{tabular}}                           &                                                                                    \\ \hline
\multirow{3}{*}{Claude Sonnet 4~\cite{Anthropic2025Sonnet4}} &                                                                                  & Functionality Testing                                          &        53.15                         & 16.79                               &                30.08                &      38.47                                                                              \\
                                 &      0.00                                                                            & Data Display Testing                                            &    21.43                             &                    42.86            &     35.71                           &      57.14                                                                              \\
                                 &                                                                                  & UI Design Validation                                        &   68.35                              &                 6.33               &                                &        25.32                                                                            \\ \midrule
 \multirow{3}{*}{Qwen 3 Coder 480B~\cite{Alibaba2025Qwen3Coder}}                     &                                                                                  & Functionality Testing                                          &     55.25                            &    18.87                            &        27.27                        &       36.71                                                                             \\
                                 &    1.43                                                                              & Data Display Testing                                            &   40.00                              &            25.71                    &    34.29                            &      47.14                                                                              \\
                                 &                                                                                  & UI Design Validation                                        &      54.43                           &         6.33                       &    27.85                            &    31.01                                                                                \\ \midrule
\multirow{3}{*}{Gemini 2.5 Pro~\cite{GoogleDeepMind2025Gemini2_5Pro}}                   &                                                                                  & Functionality Testing                                          &      62.93                           &                   15.38             &      21.67                          &        29.36                                                                          \\
                                 &     0.00                                                                             & Data Display Testing                                            &    35.71                             &           28.57                     &   35.71                             &         50.00                                                                           \\
                                 &                                                                                  & UI Design Validation                                        &    56.96                             &         15.19                       &        27.85                        &     35.44                                                                               \\ \midrule
\multirow{3}{*}{Kimi K2~\cite{MoonshotAI2025KimiK2}}                         &                                                                                  & Functionality Testing                                          &     56.54                         &    14.69                            &         28.67                       &     36.02                                                                               \\
                                 &      0.00                                                                            & Data Display Testing                                            &        42.86                       &         40.00                      &    17.14                            &         37.14                                                                       \\
                                 &                                                                                  & UI Design Validation                                        &    62.03                             &             12.66                   &      25.32                          &     31.65                                                                               \\ \midrule
 \multirow{3}{*}{Llama 4 Maverick ~\cite{MetaAI2025Llama4Maverick}}                         &                                                                                  & Functionality Testing                                          &     67.13                            &    7.69                            &     25.41                           &   29.25                                                                                 \\
                                 &     0.00                                                                             & Data Display Testing                                            &     50.00                            &                   28.57             &      21.43                          &     35.71                                                                               \\
                                 &                                                                                  & UI Design Validation                                        &           55.70                      &        18.99                        &      25.32                          &               34.81                                                                     \\ \midrule
\multirow{3}{*}{DeepSeek R1~\cite{DeepSeek2025R1}}                      &                                                                                  & Functionality Testing                                          &    34.27                             &        18.87                        &         22.37                       &      31.81                                                                              \\
                                 &                                                                     11.43             & Data Display Testing                                            &       57.14                          &              21.43                  &      21.43                          &    32.14                                                                                \\
                                 &                                                                                  & UI Design Validation                                        &    70.89                             &                10.13                &    18.99                            &      24.05                                                                              \\ \midrule
\multirow{3}{*}{GPT 4.1~\cite{OpenAI2025GPT4_1}}                          &                                                                                  & Functionality Testing                                          &   67.83                              &    12.58                            &      11.89                          &      18.18                                                                              \\
                                 &      2.85                                                                            & Data Display Testing                                            &     50.00                            &              28.57                  &     21.43                           &     35.71                                                                               \\
                                 &                                                                                  & UI Design Validation                                        &        74.68                         &               10.13                 &     15.19                           &    20.25                                                                                \\ \midrule

\multirow{3}{*}{Llama 3 405B~\cite{MetaAI2024Llama3_405B}}                     &                                                                                  & Functionality Testing                                          &     65.03                            &    13.99                            &     13.28                           &     20.28                                                                               \\
                                 &     2.86                                                                             & Data Display Testing                                            &        74.29                         &                      8.57          &      17.14                          &          21.43                                                                         \\
                                 &                                                                                  & UI Design Validation                                        & 74.68           & 17.72          & 7.59           &    16.46                                                                                \\ \bottomrule
\end{tabular}
\label{tab:cat_acc_sim3d_apimodel}
\end{table}

\begin{table}[h]
\caption{Category accuracy of 3D WebUI task on Self-Trained Model.}
\centering
\small
\setlength{\tabcolsep}{2pt} 
\renewcommand{\arraystretch}{1} 
\begin{tabular}{lclcccc}
\toprule
\multicolumn{1}{l}{\begin{tabular}[l]{@{}l@{}}Training Variants of\\ Qwen 2.5 14B Instruct \end{tabular}}                              & \multicolumn{1}{c}{\begin{tabular}[c]{@{}c@{}}Build Failure\\ Rate (\%)\end{tabular}} & \begin{tabular}[c]{@{}c@{}}Test \\ Category\end{tabular} & \multicolumn{3}{c}{\begin{tabular}[c]{@{}c@{}}Build Sucess\\ UI-Agent Verified Task (\%)\end{tabular}} & \multicolumn{1}{c}{\begin{tabular}[c]{@{}c@{}}Accuracy \\ (Cat. \%)\end{tabular}} \\ \cmidrule{4-6}
                                 &                                                                                  &                                                         & \multicolumn{1}{c}{\begin{tabular}[c]{@{}c@{}}No \\ Completion\end{tabular}}                          & \multicolumn{1}{c}{\begin{tabular}[c]{@{}c@{}}Partial \\ Completion\end{tabular}}                           & \multicolumn{1}{c}{\begin{tabular}[c]{@{}c@{}}Full \\ Completion\end{tabular}}                            &                                                                                    \\ \midrule
\multirow{3}{*}{Baseline SFT} &                                                                                  & Functionality Testing                                          &    52.16                          &    12.82                          &   19.74                            &   26.16                                                                                 \\
                                 &        0.00                                                                          & Data Display Testing                                            &     48.57                          &      25.71                          &       25.71                         &    38.57                                                                                \\
                                 &                                                                                  & UI Design Validation                                        &         68.35                        &              12.66                  &    18.99                            &        25.32                                                                            \\ \midrule
\multirow{3}{*}{PPO w/ Vision agent}                   &                                                                                  & Functionality Testing                                          &     59.38                            &    27.97                            &    20.96                            &          34.95                                                                          \\
                                 &                                                                 0.00                 & Data Display Testing                                            &  21.43                               &                  35.71              &   28.57                             &         46.43                                                                           \\
                                 &                                                                                  & UI Design Validation                                        &    62.03                             &        15.19                        & 22.78                               &      30.38                                                                              \\ \midrule
\multirow{3}{*}{VisRL w/ UI agent}                     &                                                                                  & Functionality Testing                                          &        53.53                         &        14.95                        &               30.05                 &   37.53                                                                                 \\
                                 &                                                                0.00                  & Data Display Testing                                            &        42.86                         &               28.57                 &       28.57                        &     42.86                                                                               \\
                                 &                                                                                  & UI Design Validation                                        &     58.23                            &                   18.99             &      22.78                          &      32.28                                                                              \\ \bottomrule
\end{tabular}
\label{tab:cat_acc_sim3d_selftrained}
\end{table}

\clearpage

\subsection{Visualization of UI Agent Traces}
\label{app:e_2_additional_visualization}

\begin{figure}[h]
  \centering
    \begin{subfigure}[t]{\textwidth}
    \centering
    \includegraphics[width=\textwidth]{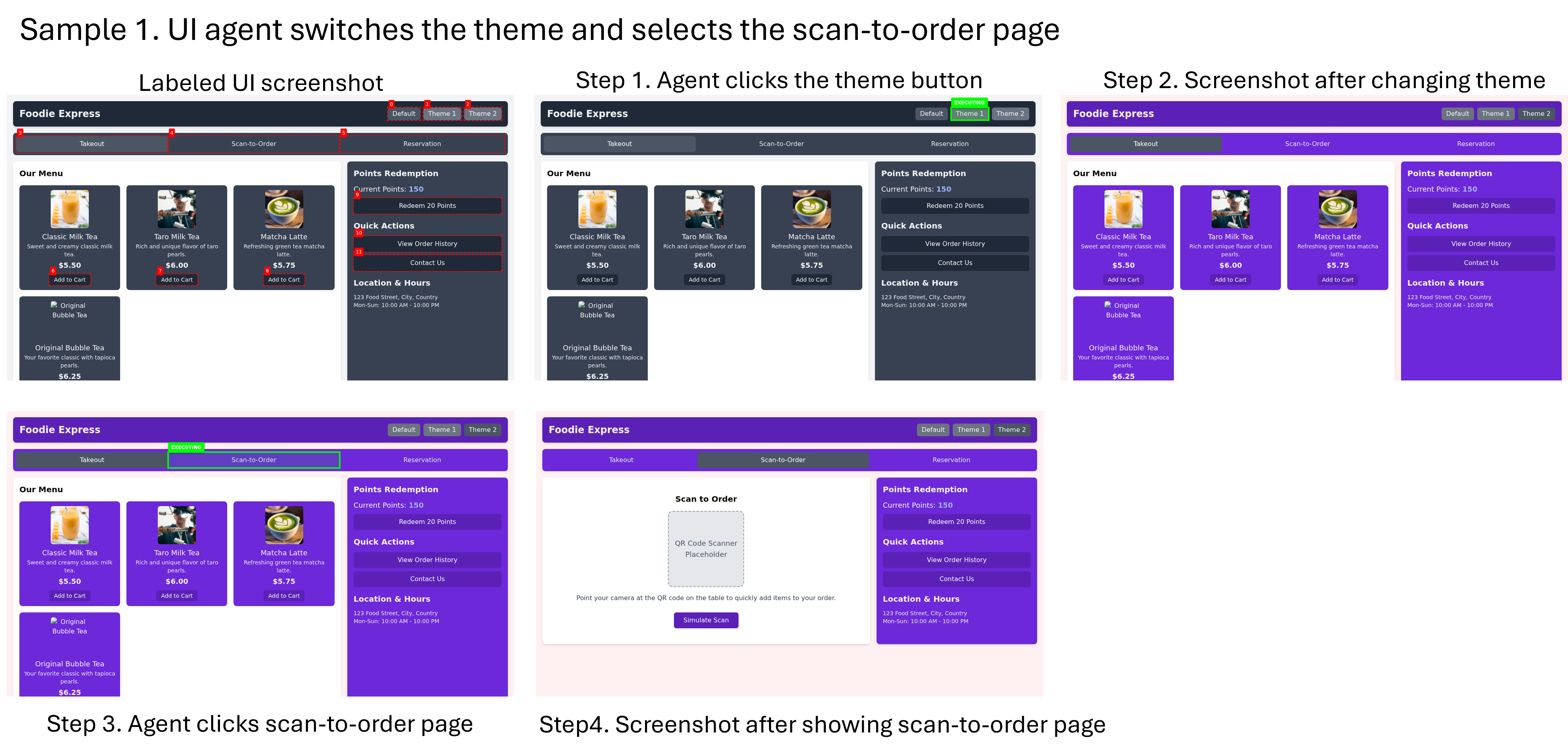}
    \label{fig:webui:movie}
  \end{subfigure}

   \begin{subfigure}[t]{\textwidth}
    \centering
    \includegraphics[width=\textwidth]{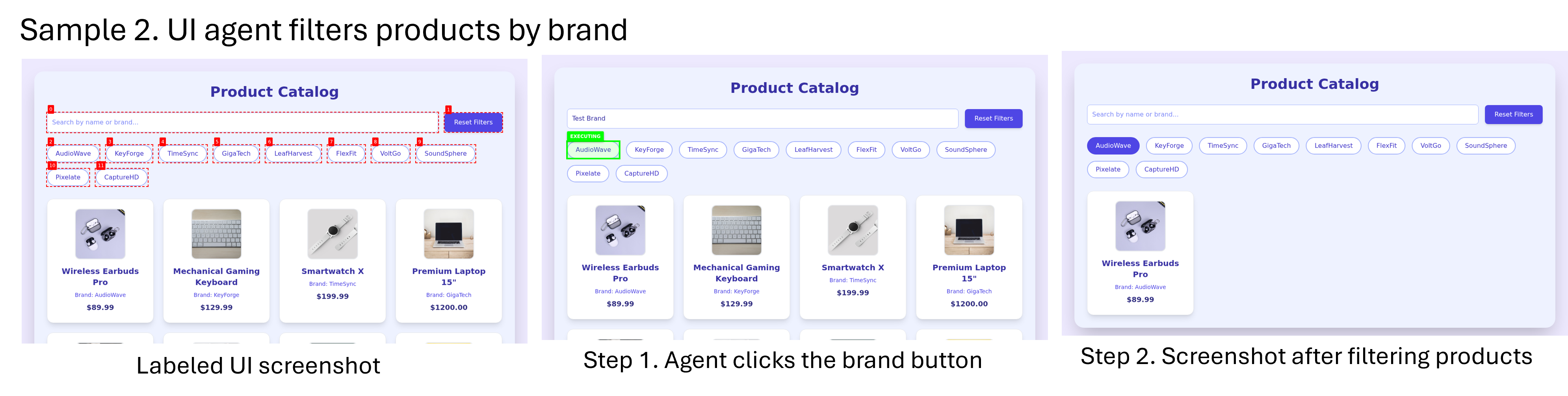}
    \label{fig:webui:catalog}
  \end{subfigure}
\caption{WebUI interaction examples (sample 1 and 2). Each row shows a full UI state transition with agent-labeled elements and executed actions.}
\end{figure}


\begin{figure*}[t]
  \centering


  \begin{subfigure}[t]{\textwidth}
    \centering
    \includegraphics[width=\textwidth]{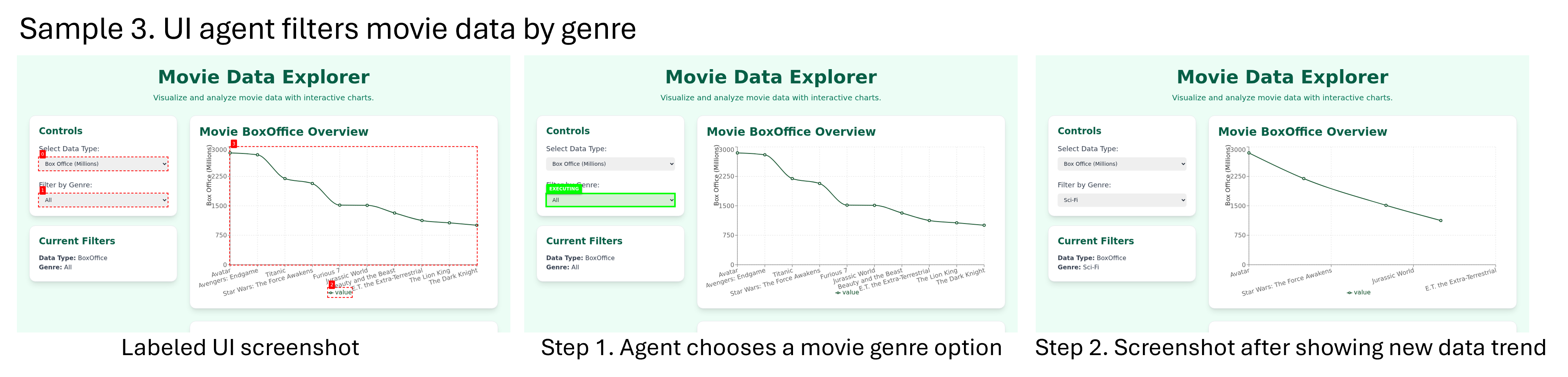}
    \label{fig:webui:movie}
  \end{subfigure}

  \begin{subfigure}[t]{\textwidth}
    \centering
    \includegraphics[width=\textwidth]{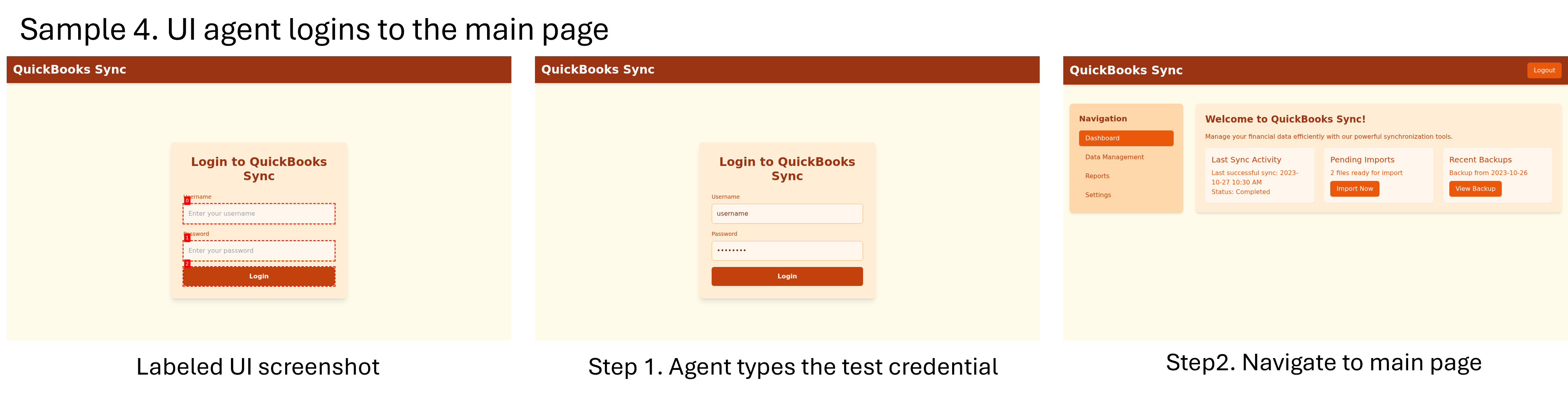}
    \label{fig:webui:quickbooks}
  \end{subfigure}

  \begin{subfigure}[t]{\textwidth}
    \centering
    \includegraphics[width=\textwidth]{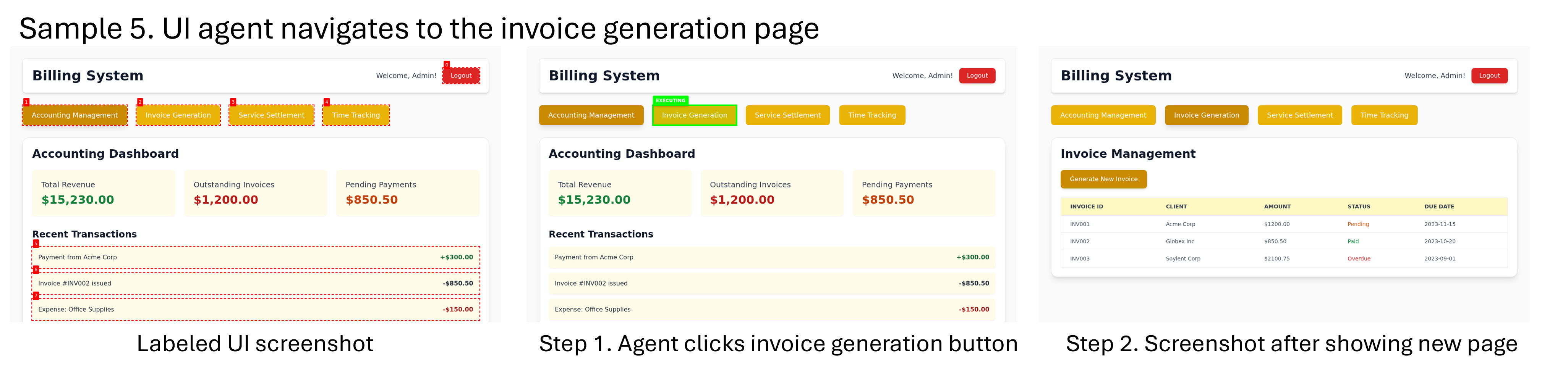}
    \label{fig:webui:billing}
  \end{subfigure}

  \caption{WebUI interaction examples (sample 3 to 5). Each row shows a full UI state transition with agent-labeled elements and executed actions.}
  \label{fig:webui-samples}
\end{figure*}

\begin{figure*}[t]
  \centering

  \begin{subfigure}[t]{0.99\textwidth}
    \centering
    \includegraphics[width=\linewidth]{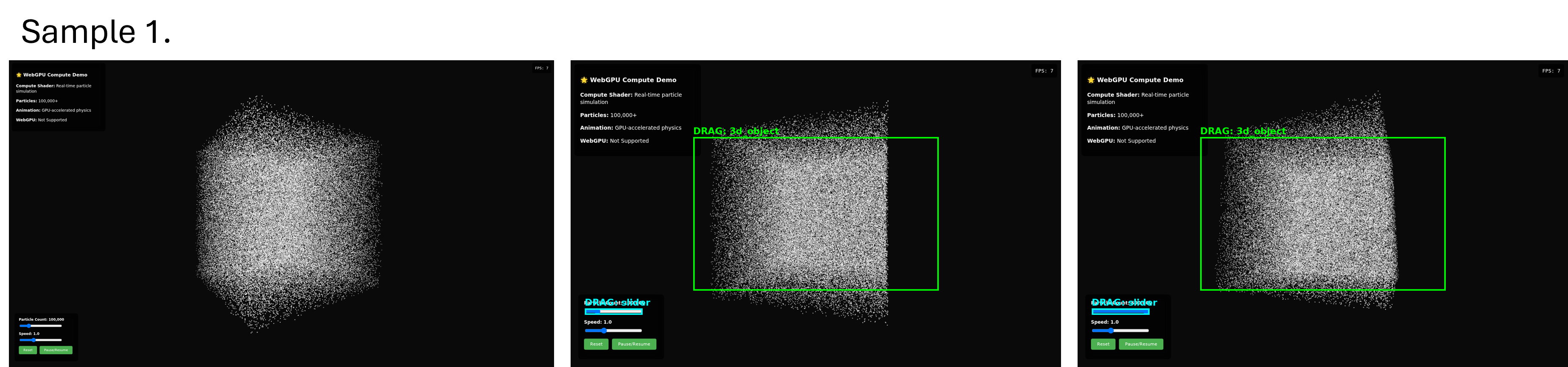}
    \caption{WebGPU Compute}
    \label{fig:s1}
  \end{subfigure}\hfill
  \begin{subfigure}[t]{0.99\textwidth}
    \centering
    \includegraphics[width=\linewidth]{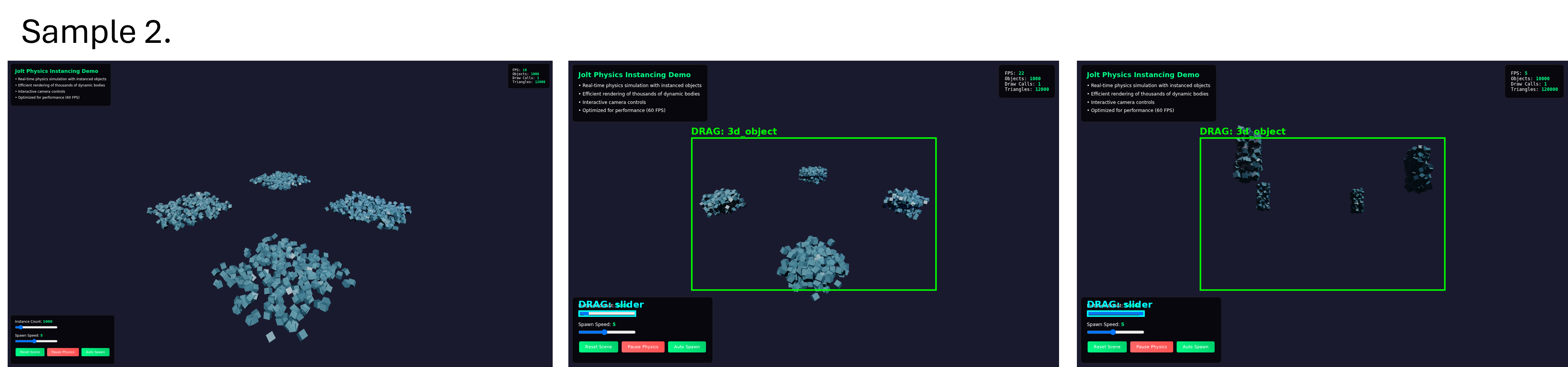}
    \caption{Jolt Physics Instancing}
    \label{fig:s2}
  \end{subfigure}\hfill
  \begin{subfigure}[t]{0.99\textwidth}
    \centering
    \includegraphics[width=\linewidth]{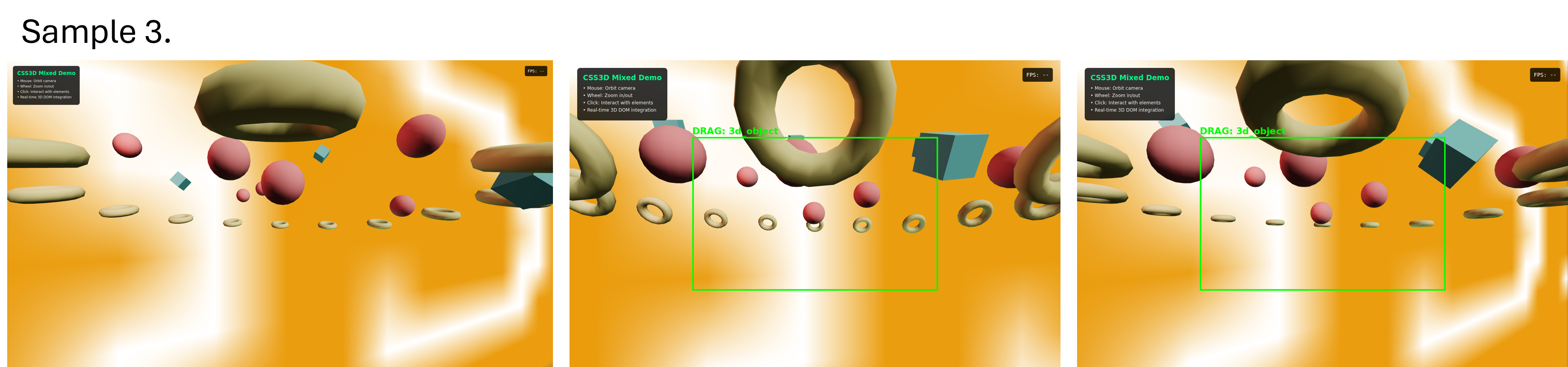}
    \caption{CSS3D Mixed}
    \label{fig:s3}
  \end{subfigure}

  \begin{subfigure}[t]{0.99\textwidth}
    \centering
    \includegraphics[width=\linewidth]{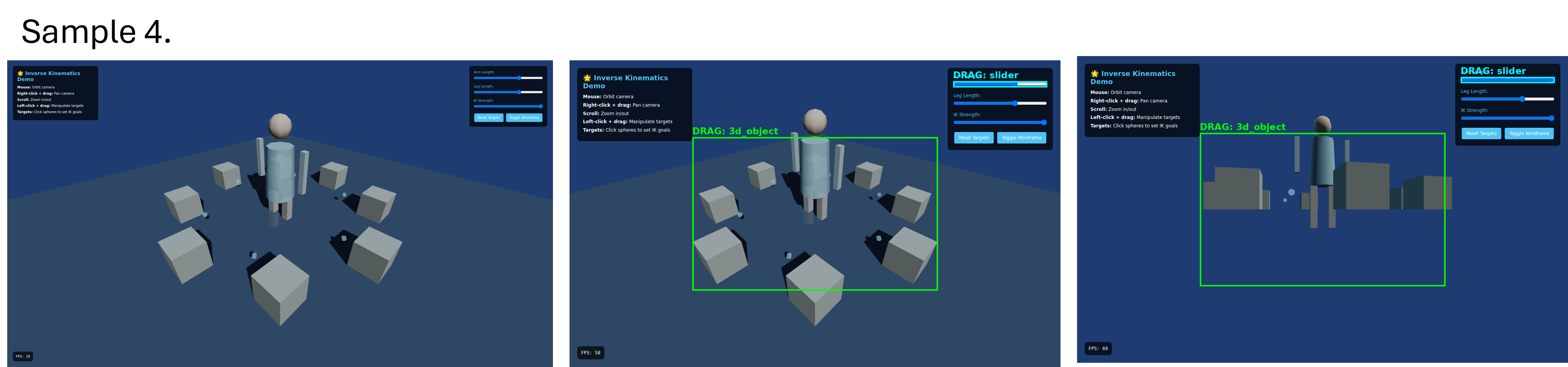}
    \caption{Inverse Kinematics}
    \label{fig:s4}
  \end{subfigure}\hfill
  \begin{subfigure}[t]{0.99\textwidth}
    \centering
    \includegraphics[width=\linewidth]{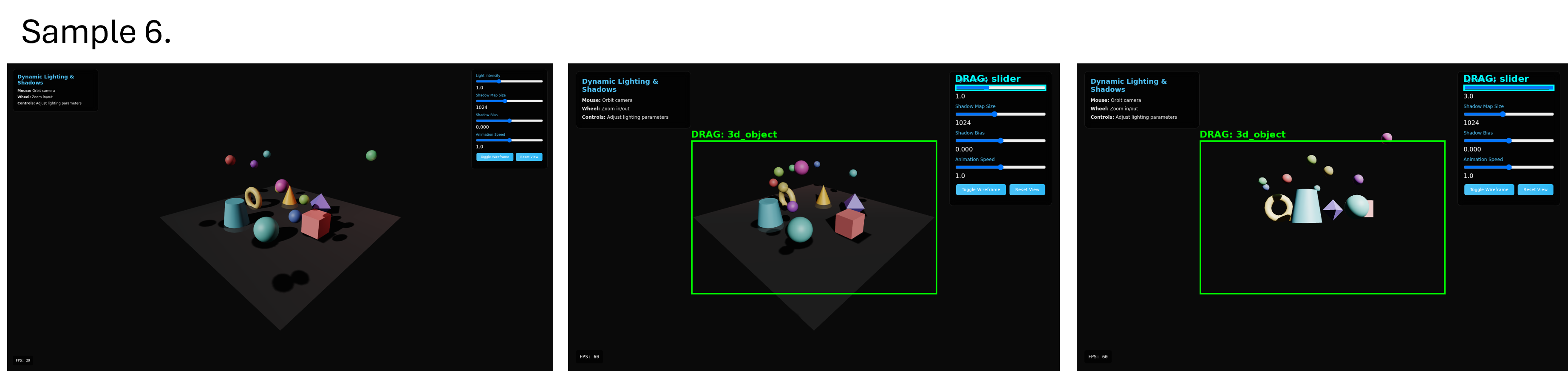}
    \caption{Dynamic Lighting}
    \label{fig:s6}
  \end{subfigure}

  \caption{Six simulation samples. Each subfigure shows the UI state and interaction annotation for one sample.}
  \label{fig:sim3d-samples}
\end{figure*}

\clearpage

\subsection{More Visualization}
\label{app:e_3_moremore_visualization}
\begin{figure}[h]
    \centering
    \includegraphics[width=0.90\linewidth]{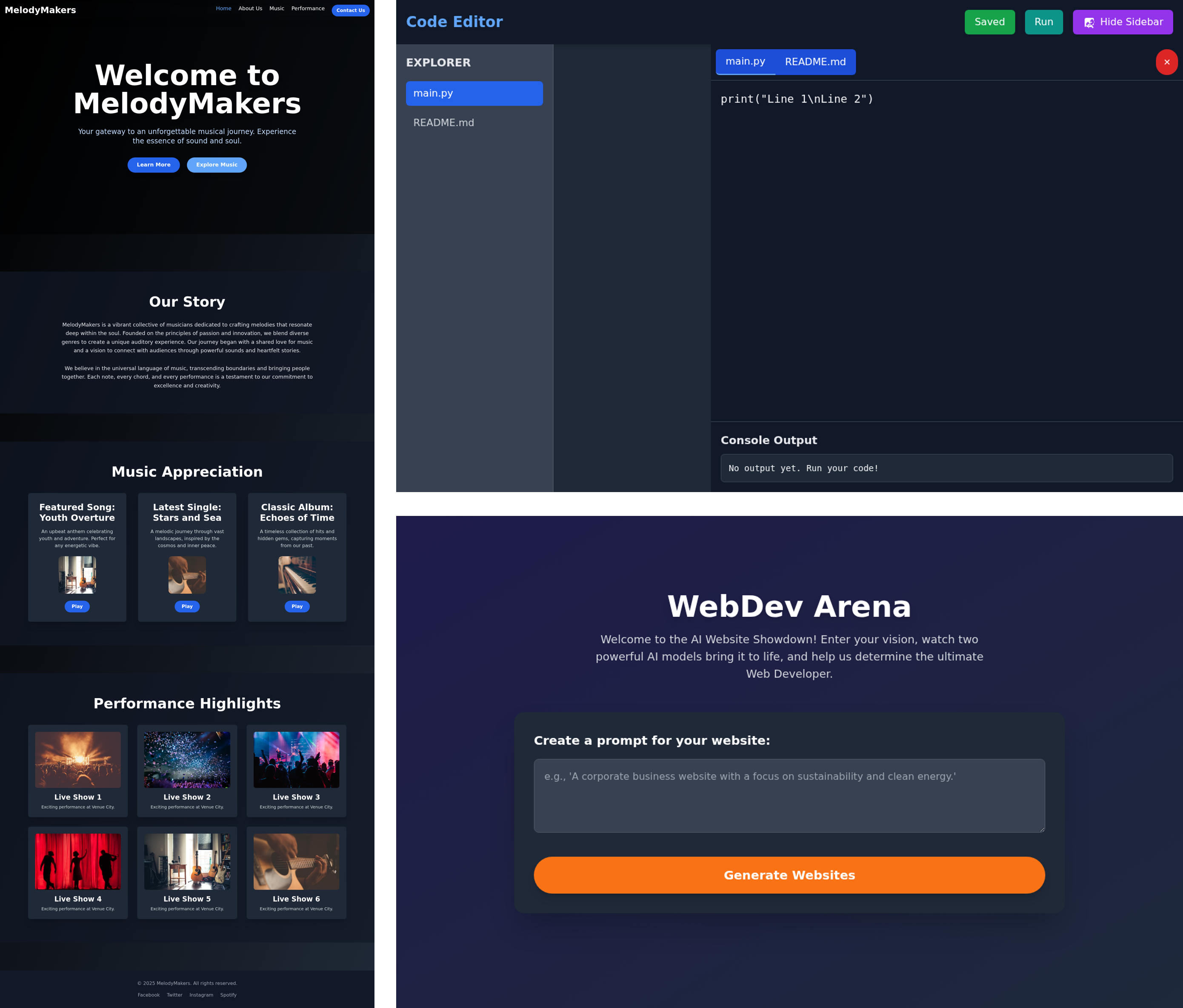}
    \caption{We present three webpages generated from self-trained model Qwen 2.5 14B w/ VisRL. The left-most webpage is a music band showcase portal. The upper-right webpage is a codecraft editor. The lower-right webpage is a AI website showdown.}
    \label{fig:morevis1}
    \vspace{-2mm}
\end{figure}

\begin{figure}[h]
    \centering
    \includegraphics[width=0.90\linewidth]{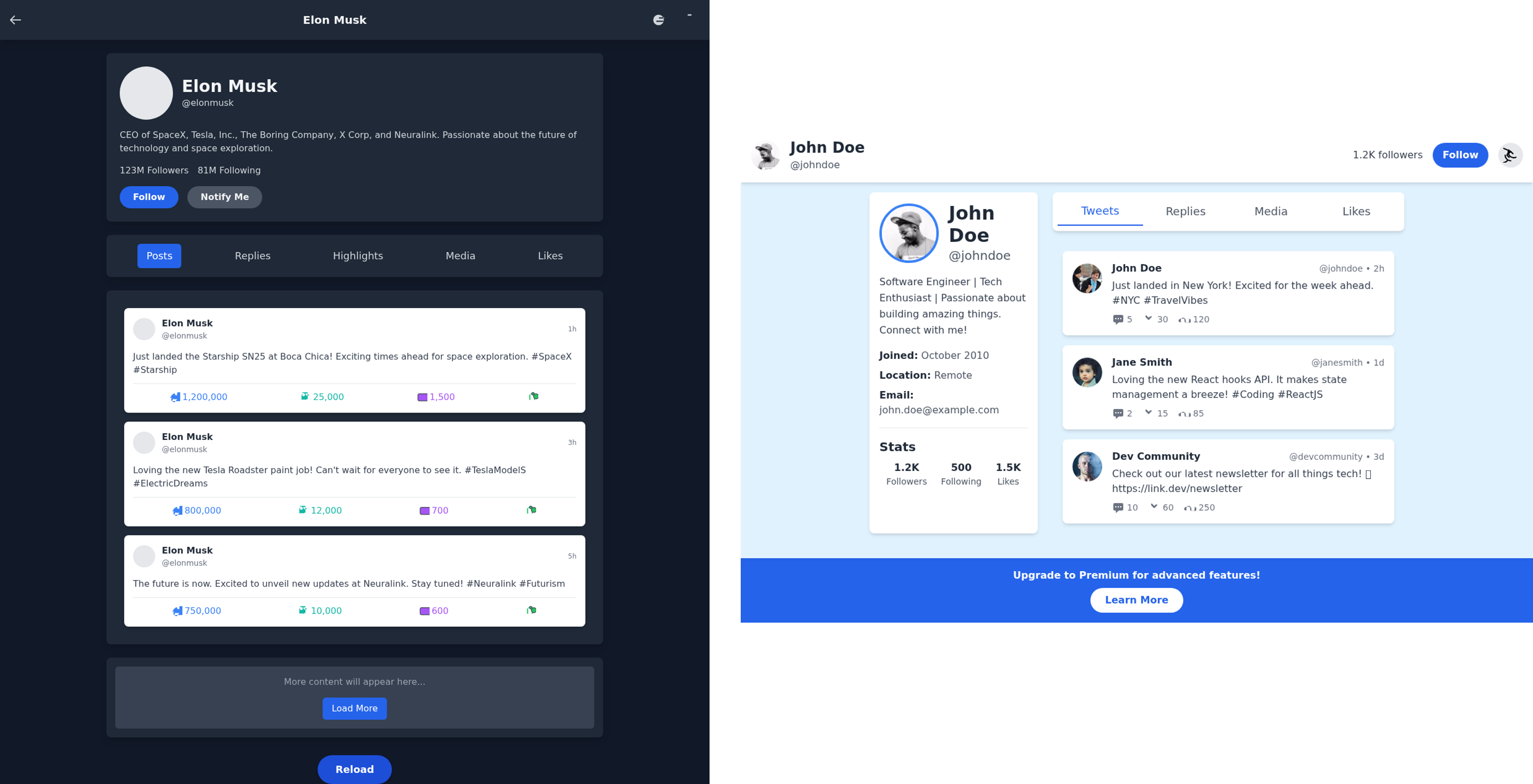}
    \caption{Two webpages generated from self-trained model Qwen 2.5 14B w/ VisRL. Both of them are the social media viewer.}
    \label{fig:morevis2}
    \vspace{-2mm}
\end{figure}

\begin{figure}[h]
    \centering
    \includegraphics[width=0.90\linewidth]{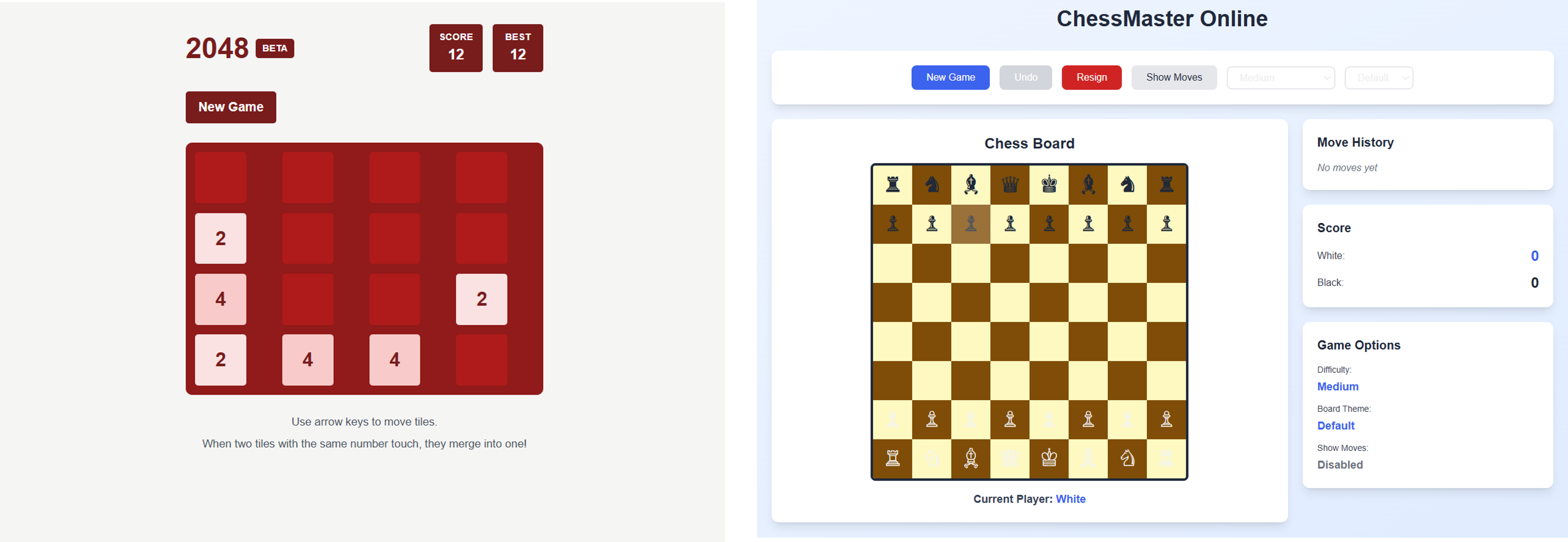}
    \caption{Two webpages generated from self-trained model Qwen 2.5 14B w/ VisRL. The left side is a 2048 game interface and the right side is a chess game simulator.}
    \label{fig:morevis3}
    \vspace{-2mm}
\end{figure}

\begin{figure}[h]
    \centering
    \includegraphics[width=0.90\linewidth]{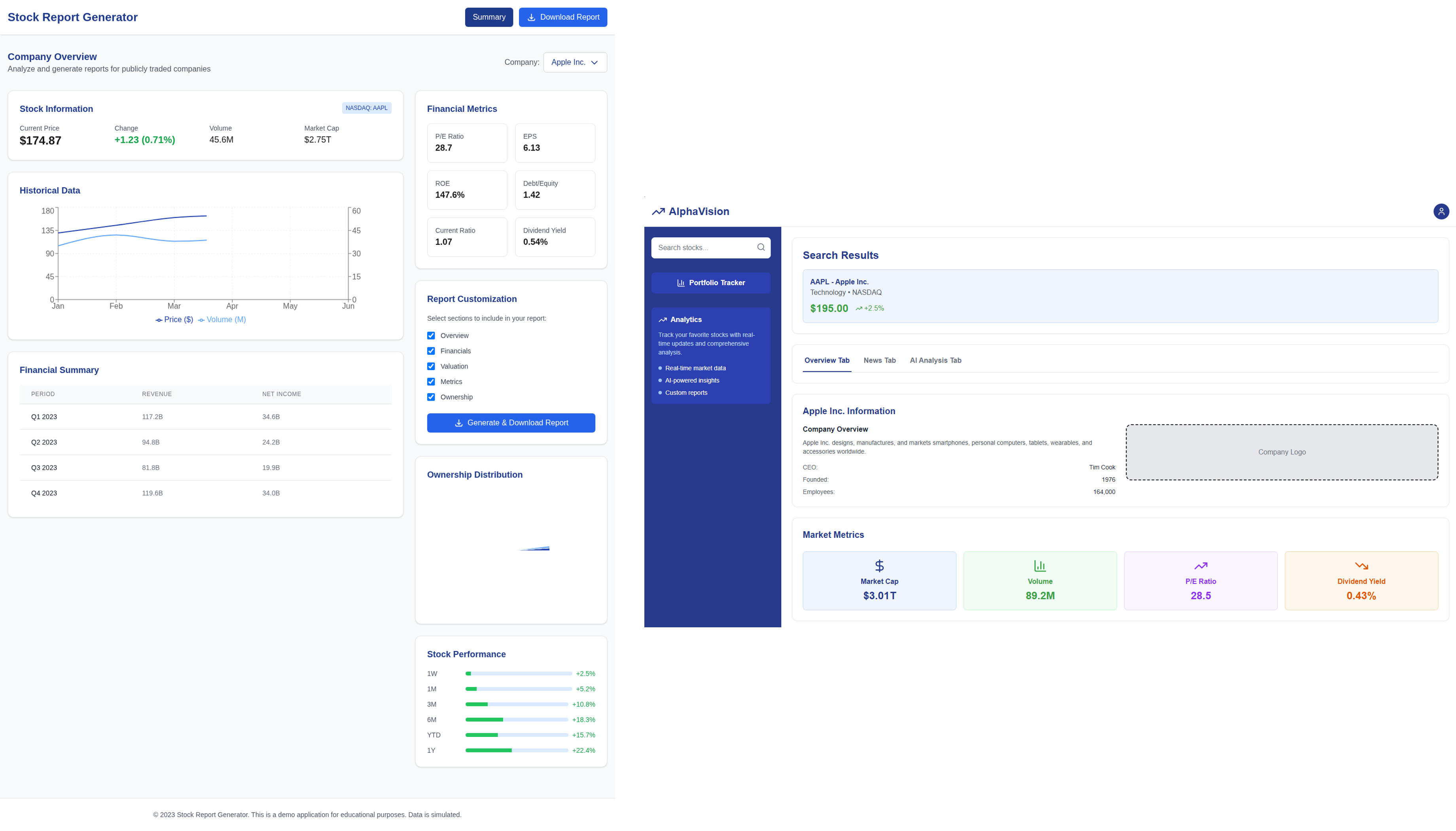}
    \caption{Two webpages generated from self-trained model Qwen 2.5 14B w/ VisRL. Both of them are the stock report generator.}
    \label{fig:morevis4}
    \vspace{-2mm}
\end{figure}

\begin{figure}[h]
    \centering
    \includegraphics[width=0.90\linewidth]{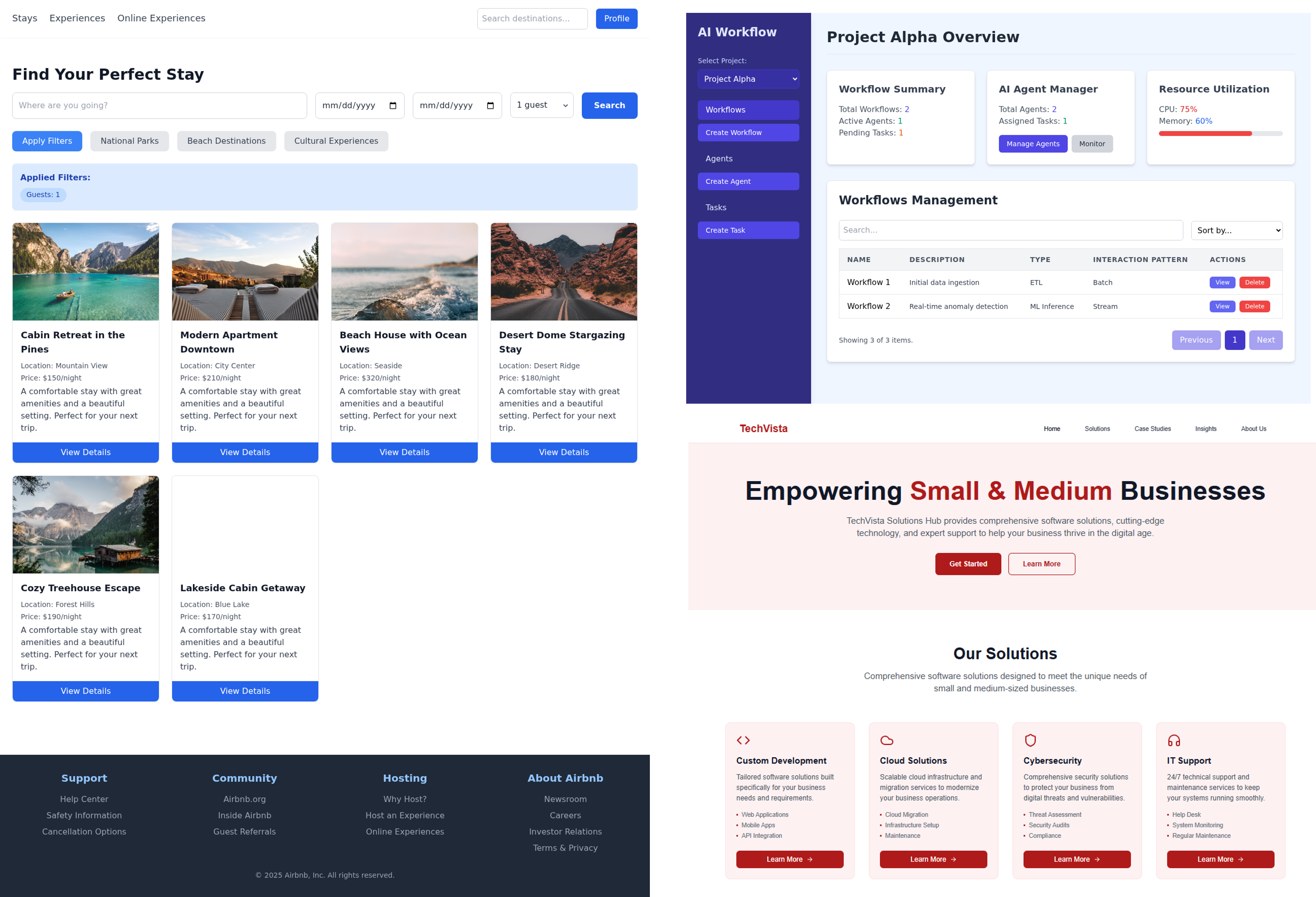}
    \caption{Three webpages generated from self-trained model Qwen 2.5 14B w/ VisRL. The left side is a travel accommodation finder, the upper-right side is an AI agent workflow manager, and the lower-right side is a landing page for tech. company.}
    \label{fig:morevis5}
    \vspace{-2mm}
\end{figure}

\paragraph{Sample Visualization for general WebUI}
Fig.~\ref{fig:morevis1},~\ref{fig:morevis2},~\ref{fig:morevis3},~\ref{fig:morevis4},~\ref{fig:morevis5} show the general WebUI samples under different UI categories, including creative media, data management, games, social interaction, and e-commerce platforms.

\clearpage

\begin{figure}[h]
    \centering
    \includegraphics[width=0.90\linewidth]{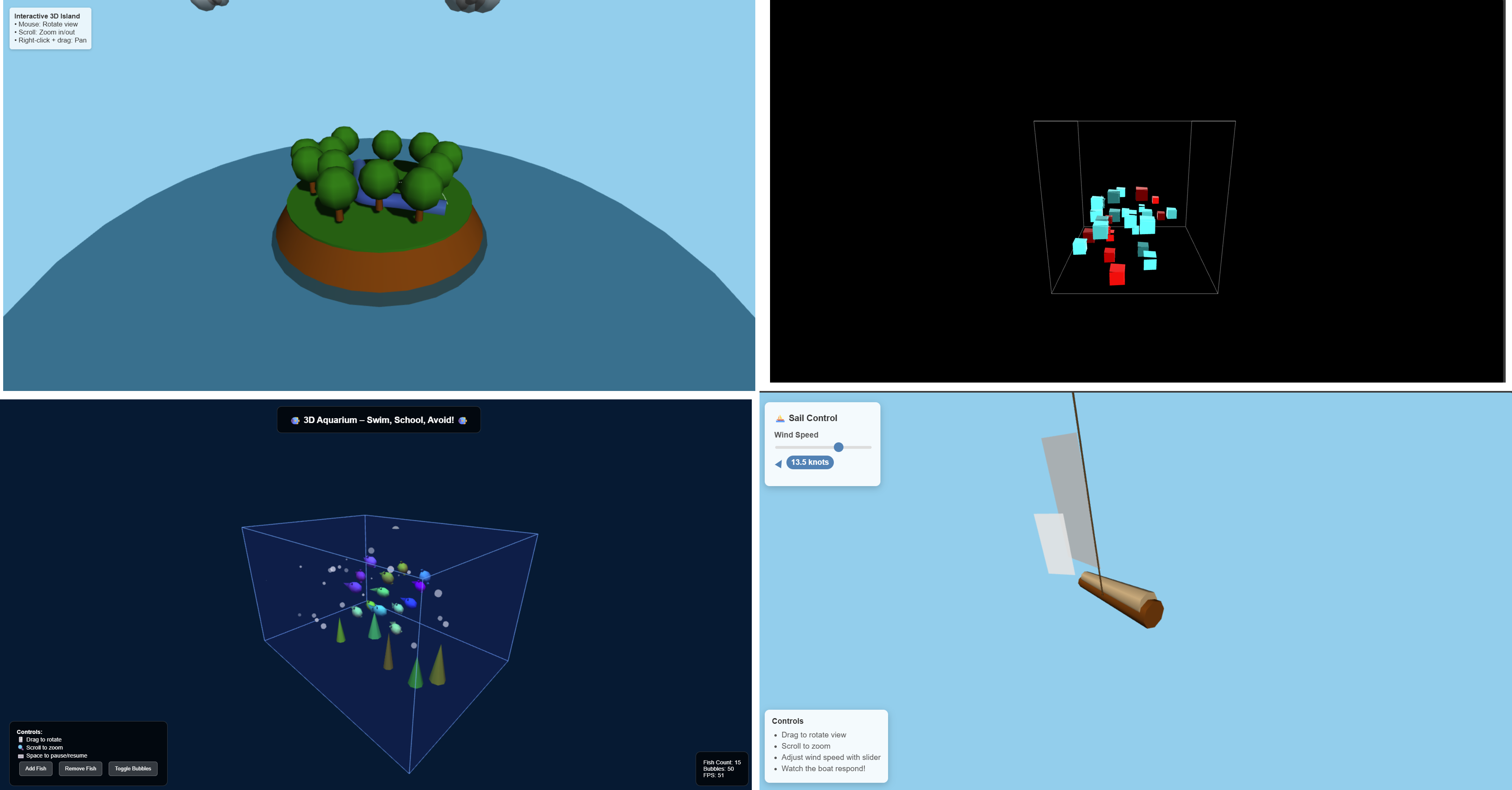}
    \caption{Four webpage demo generated from self-trained model Qwen 2.5 14B w/ VisRL. We provide simple prompt for each to showcase how they are generated. (1.) The upper-left demo: Create a 3D animation of a circular island that has multiple trees on it and a river flowing within. (2.) The upper-right demo: Create a 3D simulation of jelly cubes bouncing in a box. (3.) The lower-left demo: Create a 3D aquarium with fishes that swim around and avoid collisions with the tank walls. (4.) The lower-right demo: Create a 3D animation of a sail boat floating in the sea. }
    \label{fig:morevis6}
    \vspace{-2mm}
\end{figure}

\begin{figure}[h]
    \centering
    \includegraphics[width=0.90\linewidth]{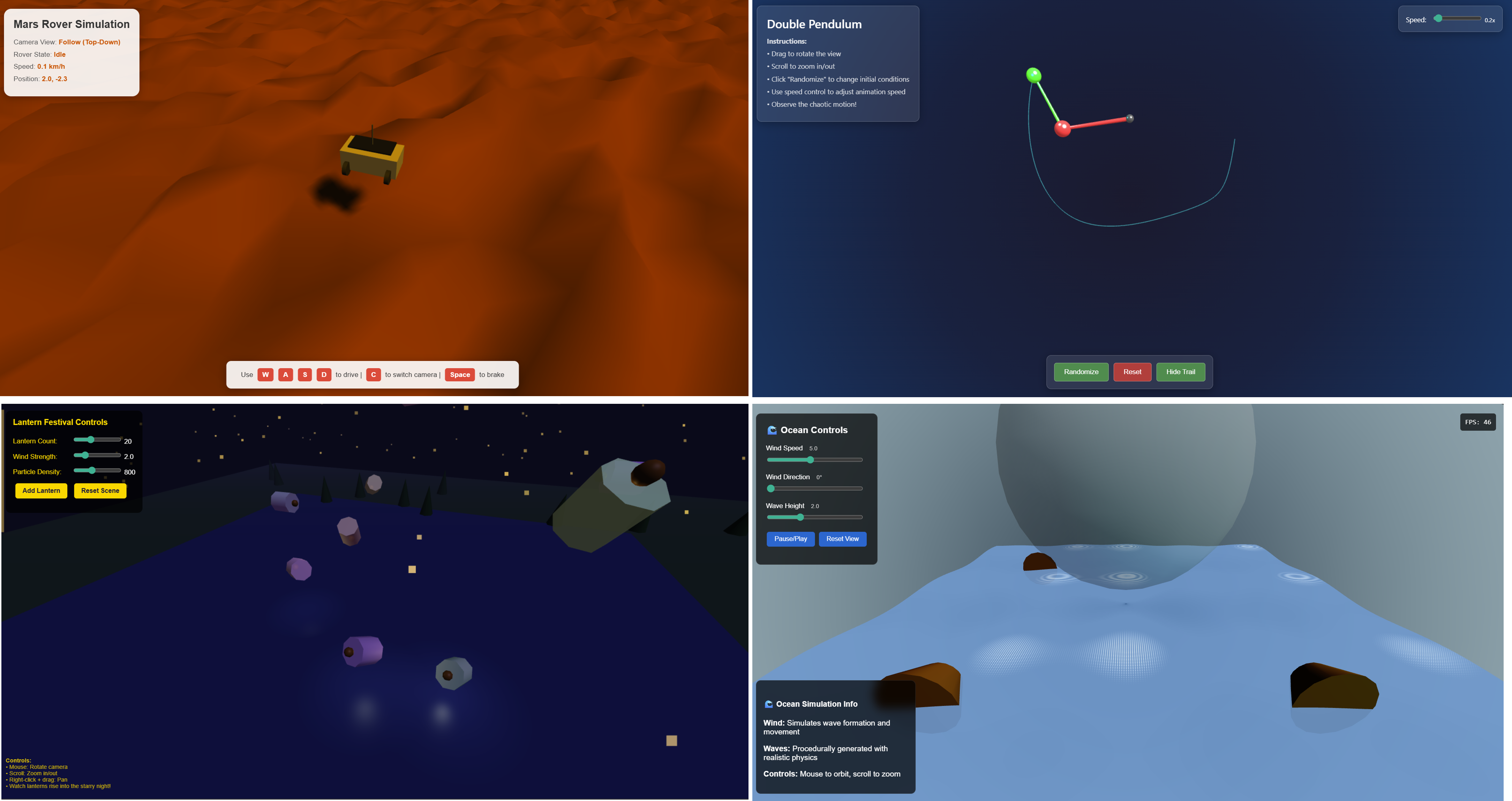}
    \caption{Same as prior figure, we provide simple prompt for each to showcase how they are generated. (1.) The upper-left demo: Create a 3D simulation of a mars rover moving on the surface of Mars. Users should be able to accelerate and control the direction of the car. Incorporate accurate suspension dynamics as the car moves on the rugged Mars surface. 
    (2.) The upper-right demo: Create a visually engaging animation of a double pendulum using JavaScript. The animation should demonstrate the chaotic motion of two pendulums connected together. (3.) The lower-left demo: Simulate 3D paper lanterns rising into the night sky during a festival, with buoyancy and wind physics, above a reflective lake. (4.) The lower-right demo: Create a realistic 3D ocean scene with: dynamic water movement, accurate wave simulations, and reflective surface effects. }
    \label{fig:morevis7}
    \vspace{-2mm}
\end{figure}

\begin{figure}[h]
    \centering
    \includegraphics[width=0.90\linewidth]{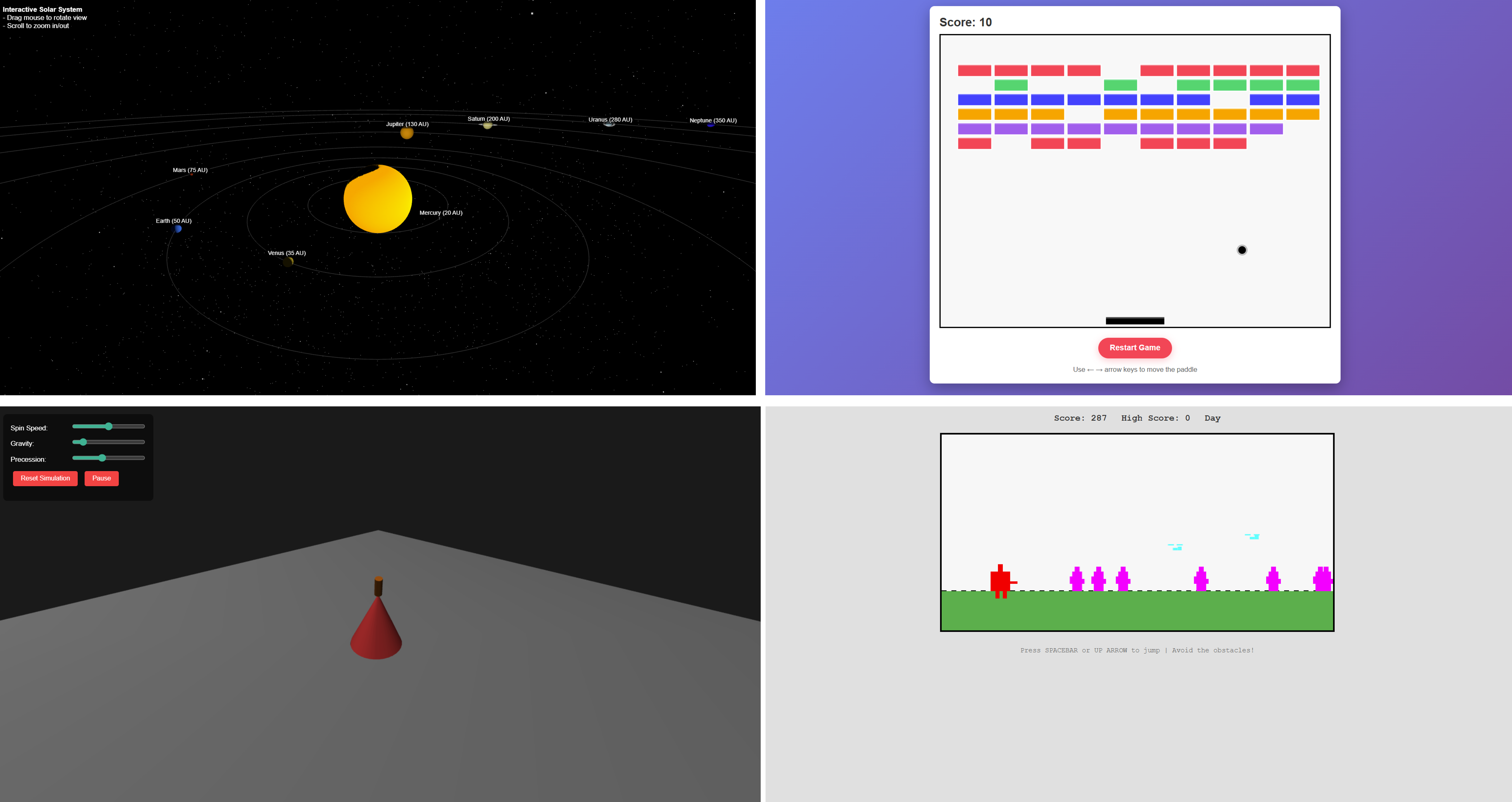}
    \caption{Same as prior figure, we provide simple prompt for each to showcase how they are generated. (1.) The upper-left demo: Create a 3D solar system simulation using Three.js with realistic planet orbits, textures, and lighting effects.
    (2.) The upper-right demo: Create a breakout-style game where a paddle at the bottom of the screen reflects a bouncing ball toward a grid of colored blocks above. The paddle moves left and right with arrow keys, and blocks disappear when hit by the ball. Each block is worth different points based on its color, and the game ends when all blocks are cleared or the ball falls below the paddle. (3.) The lower-left demo: Create a 3D animation of a top spinning. Use accurate physics to demonstrate its movements. (4.) The lower-right demo: Create a dinosaur running game where: (i) A dinosaur character can jump using spacebar/up arrow, (ii) Cacti and birds appear as obstacles from the right side, (iii) Ground moves continuously to create running effect (iv) Game ends when dinosaur hits an obstacle, (v) Simple pixel art style, and (vi) Day/night cycle every 500 points.}
    \label{fig:morevis8}
    \vspace{-2mm}
\end{figure}

\paragraph{Sample Visualization for 3D WebUI Simulation}
Fig.~\ref{fig:morevis6}, \ref{fig:morevis7}, and \ref{fig:morevis8} show the 3D WebUI samples under different categories.

\clearpage

\subsection{Analysis of Build Failure Taxonomy}

\label{app:e_3_build_failure}

\begin{figure}[h]
  \centering
  \includegraphics[width=0.70\linewidth]{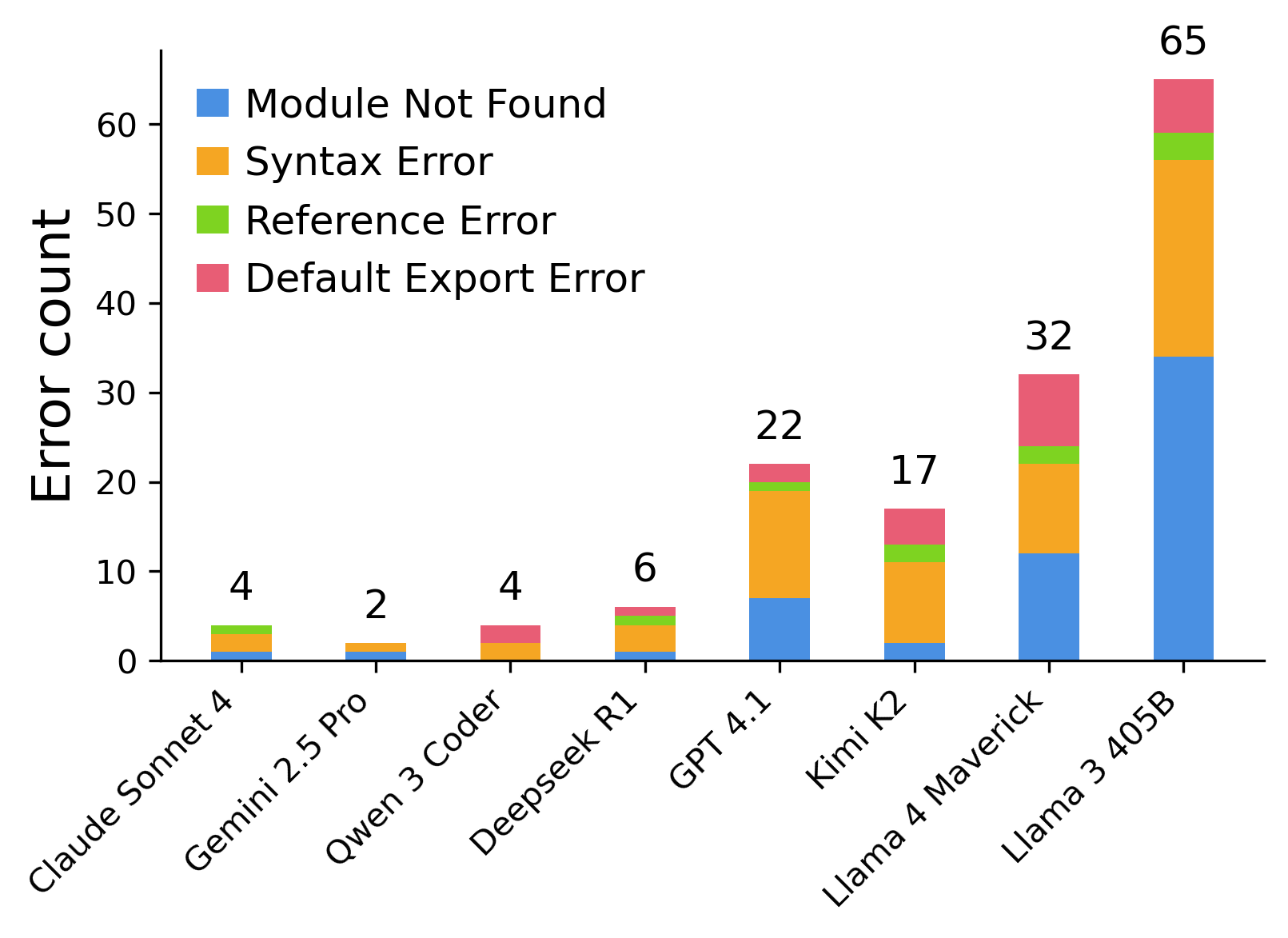}
  \caption{Breakdown of build-time errors across models in our WebUI generation benchmark. Bars show the number of failures by error type, including missing dependencies, syntax mistakes, reference issues, and invalid default exports. Higher stacks indicate more frequent build failures.}
  \label{fig:buildfailanalysis}
\end{figure}

To investigate the build failure reason, we categorize build-time errors into four primary types based on failure patterns observed during Next.js execution logs. \textit{Module Not Found errors} occur when Webpack cannot resolve imported packages (e.g., \texttt{@heroicons/react}, \texttt{react-icons}, or \texttt{react-dnd}), typically due to missing installations, incorrect import paths, or renamed files. \textit{Syntax errors} arise when the TypeScript/JavaScript parser encounters malformed code (e.g., missing braces, parentheses, or template literal errors), preventing the compilation stage from completing. \textit{Default Export errors} (happen when a Next.js page compiles but fails during the static generation stage because some certain file do not contain a valid default React component export (e.g., \texttt{pages/modelOutputPage.tsx}). Finally, \textit{Reference errors} occur at runtime during server-side rendering or static generation, often due to accessing variables before initialization or circular imports. In Table~\ref{fig:buildfailanalysis}, we show the four dominant failure counts on eight models, providing insight into the generation ability of each model.






\end{document}